\documentclass[showpacs,prl,onecolumn,aps,superscriptaddress,preprintnumbers,nofootinbib]{revtex4}
\usepackage[T1]{fontenc}
\usepackage[latin9]{inputenc}
\usepackage{amsmath,amssymb}
\usepackage{epsfig}
\usepackage{graphicx}
\usepackage{amsmath}
\usepackage{amsfonts}
\def\slashchar#1{\setbox0=\hbox{$#1$}     		
   \dimen0=\wd0                                 	
   \setbox1=\hbox{/} \dimen1=\wd1               	
   \ifdim\dimen0>\dimen1                        	
      \rlap{\hbox to \dimen0{\hfil/\hfil}}      	
      #1                                        	
   \else                                        	
      \rlap{\hbox to \dimen1{\hfil$#1$\hfil}}   	
      /                                         	
   \fi}

\renewcommand{\vec}{\boldsymbol}
\newcommand{\beq}{\begin{equation}}
\newcommand{\eeq}{\end{equation}}
\newcommand{\bea}{\begin{eqnarray}}
\newcommand{\eea}{\end{eqnarray}}
\newcommand{\ba}{\begin{array}}
\newcommand{\ea}{\end{array}}

\def\eq#1{{Eq.~(\ref{#1})}}
\def\fig#1{{Fig.~\ref{#1}}}
\newcommand{\bas}{\bar{\alpha}_S}
\newcommand{\as}{\alpha_S}
\newcommand{\nn}{\nonumber}

\newcommand{\h}{\frac{1}{2}}

\newcommand{\ga}{\gamma}

\newcommand{\Lb}{\left(}
\newcommand{\Rb}{\right)}
\def\pom{{I\!\!P}}

\begin{document}

\title{CGC/saturation  approach for high energy soft interactions:  `soft'
  Pomeron structure and $\mathbf{v_{n}}$ in  hadron and nucleus collisions
  from Bose-Einstein correlations}
\author{E. ~Gotsman}
\email{gotsman@post.tau.ac.il}
\affiliation{Department of Particle Physics, School of Physics and Astronomy,
Raymond and Beverly Sackler
 Faculty of Exact Science, Tel Aviv University, Tel Aviv, 69978, Israel}
 \author{ E.~ Levin}
\email{leving@post.tau.ac.il, eugeny.levin@usm.cl}
\affiliation{Department of Particle Physics, School of Physics and Astronomy,
Raymond and Beverly Sackler
 Faculty of Exact Science, Tel Aviv University, Tel Aviv, 69978, Israel}
 \affiliation{Departemento de F\'isica, Universidad T\'ecnica Federico
 Santa Mar\'ia, and Centro Cient\'ifico-\\
Tecnol\'ogico de Valpara\'iso, Avda. Espana 1680, Casilla 110-V,
 Valpara\'iso, Chile} 
 \author{  U.~ Maor}
\email{maor@post.tau.ac.il}
\affiliation{Department of Particle Physics, School of Physics and Astronomy,
Raymond and Beverly Sackler
 Faculty of Exact Science, Tel Aviv University, Tel Aviv, 69978, Israel}
\date{\today}


\begin{abstract}
In the framework of our model of soft interactions at high
 energy based on CGC/saturation approach,
we show that Bose-Einstein correlations of identical gluons lead to large
 values of $v_n$.
  We demonstrate how three
 dimensional scales of high  energy interactions: hadron radius,
 typical size of the wave function in diffractive production of small
 masses  (size of the constituent quark), and the saturation momentum,
  influence  the values of BE correlations, and in particular,
 the values of $v_n$. Our calculation shows that the structure of
 the `dressed' Pomeron leads to  values of $v_n$ which are close to 
experimental
 values for proton-proton scattering,  20\% smaller than  the observed
 values for proton-lead collisions, and close to lead-lead collisions
 for 0-10\% centrality. Bearing this result in mind, we conclude that
 it is premature to consider,  that 
the appearance of long range rapidity  azimuthal correlations
  are   due only to the hydrodynamical behaviour of the quark-gluon 
plasma.
 \end{abstract}
 
 \preprint{TAUP-3007/16}

\maketitle


\section{Introduction}

In our previous paper\cite{GLMBE} we  showed that  Bose-Einstein 
correlations
 lead to  strong azimuthal angle correlations, which do not depend on 
the
 difference in rapidity of the two produced hadrons  (long range rapidity 
LRR
 correlations). The mechanism suggested by us, has a general origin, and
thus
 manifests itself in   hadron-hadron, hadron-nucleus and  nucleus-nucleus
 interactions, and  generates  the correlation that  has been observed
 experimentally
 \cite{CMSPP,STARAA,PHOBOSAA,STARAA1,CMSPA,CMSAA,ALICEAA,ALICEPA,
ATLASPP,ATLASPA,ATLASAA}. The fact that Bose-Einstein correlations
 lead to strong  LRR azimuthal angle  correlations, was found  long
 ago in the framework of Gribov Pomeron calculus\cite{PION}, and it
 has been  re-discovered recently  in Refs.\cite{KOWE,KOLUCOR} in
 CGC/saturation approach\cite{KOLEB}.   In  Ref.\cite{GLMBE}
it was noticed,
 that these correlations give rise to $v_n$ for odd and even $n$, while all
 other mechanisms in CGC/saturation approach, including the
 correlations observed in \cite{KOWE,KOLUCOR}, generate only
 $v_n$ with even $n$.

\begin{figure}[h]
    \centering
  \leavevmode
      \includegraphics[width=12cm]{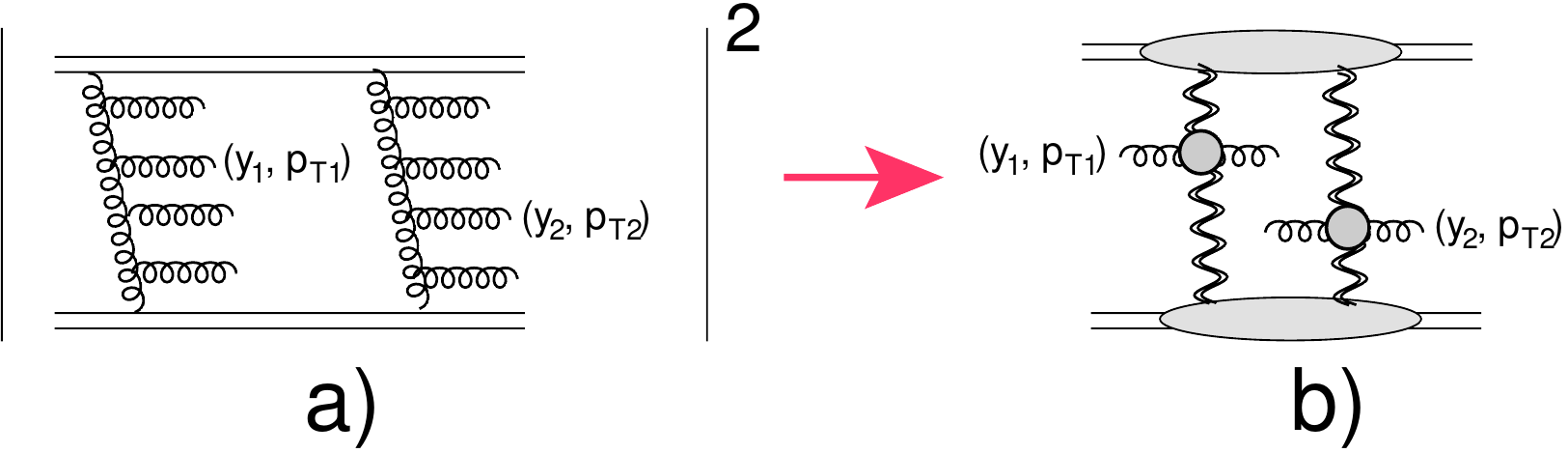}  
      \caption{ Production of two gluons with $(y_1,\vec{p}_{T1})$ and
 $(y_2,\vec{p}_{T2})$ in
 two parton showers (\protect\fig{2sh}-a).   \protect\fig{2sh}-b shows
  the double inclusive
 cross section in the Mueller diagram technique \protect\cite{MUDI}. The 
wavy
 lines denote the BFKL Pomerons\cite{BFKL,LI}.}
\label{2sh} 
   \end{figure}

 The LRR correlations in CGC/saturation approach originate  from
 the production of two parton
 showers (see  \fig{2sh}).  The double inclusive cross section is
 described by the Mueller diagram of \fig{2sh}-b, in which the production
 of gluons   from the parton cascade, 
 is described by the exchange of the BFKL Pomeron (wavy double line in
 \fig{2sh}-b),  while, due to our 
poor theoretical
 knowledge of the confinement of quarks and gluons, the upper and lower
 blobs in \fig{2sh}-b require  modeling.
. 

\begin{figure}[ht]
    \centering
  \leavevmode
      \includegraphics[width=12cm]{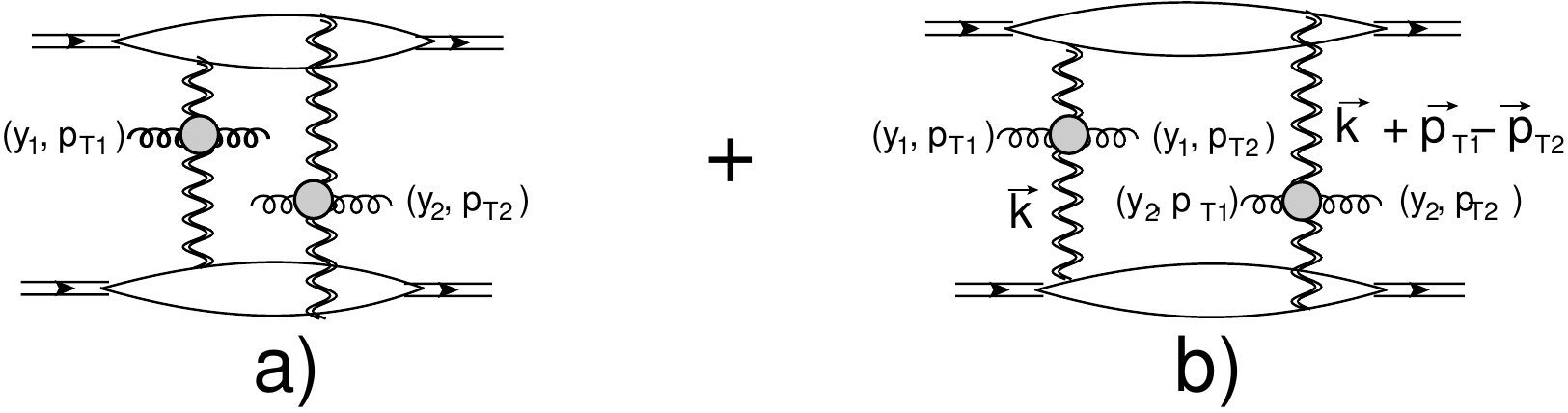}  
      \caption{ Production of two identical  gluons with $(y_1,\vec{p}_{T1})$
 and $(y_2,\vec{p}_{T2})$ in two parton showers.  The diagrams in the
 Mueller diagram technique
 \protect\cite{MUDI} are shown in \fig{2shiden}-a and \fig{2shiden}-b.
 The wavy lines denote the BFKL Pomerons\cite{BFKL,LI}.} 
      
      \label{2shiden}
       \end{figure}

 
  If  the two
 produced gluons have the same quantum numbers,  one can see that
 in addition to the Mueller diagram for different gluons (see 
\fig{2sh}-b),
  we need to take into
 account a second Mueller diagram of \fig{2shiden}-b, in which two
 gluons with
 $(y_1,\vec{p}_{T2})$ and  $(y_2,\vec{p}_{T1})$ are produced.  
 When $\vec{p}_{T1} \,\to\,\vec{p}_{T2}$, the two  production processes 
become
 identical, leading to the cross section $\sigma\Lb \mbox{two identical
 gluons}\Rb = 2 \sigma\Lb \mbox{two different  gluons}\Rb$, as one expects.
  When $|\vec{p}_{T2} -   \vec{p}_{T1}| \gg 1/R$, where $R$ is the
 size of the emitter\cite{HBT}, the interference diagram becomes small
 and can be neglected.

 The angular correlation emanates from the diagram of \fig{2shiden}-b,  in 
which
 the upper BFKL Pomerons carry momentum
 $\vec{k} - \vec{p}_{T,12}$ with $\vec{p}_{T, 12}\,=\,\vec{p}_{T1}\,-\,
\vec{p}_{T2}$, while the lower BFKL   Pomerons have  momenta $\vec{k}$.
   The Mueller diagrams for the correlation between two
gluons are shown in \fig{2shiden}.

\begin{figure}[ht]
    \centering
  \leavevmode
      \includegraphics[width=16cm]{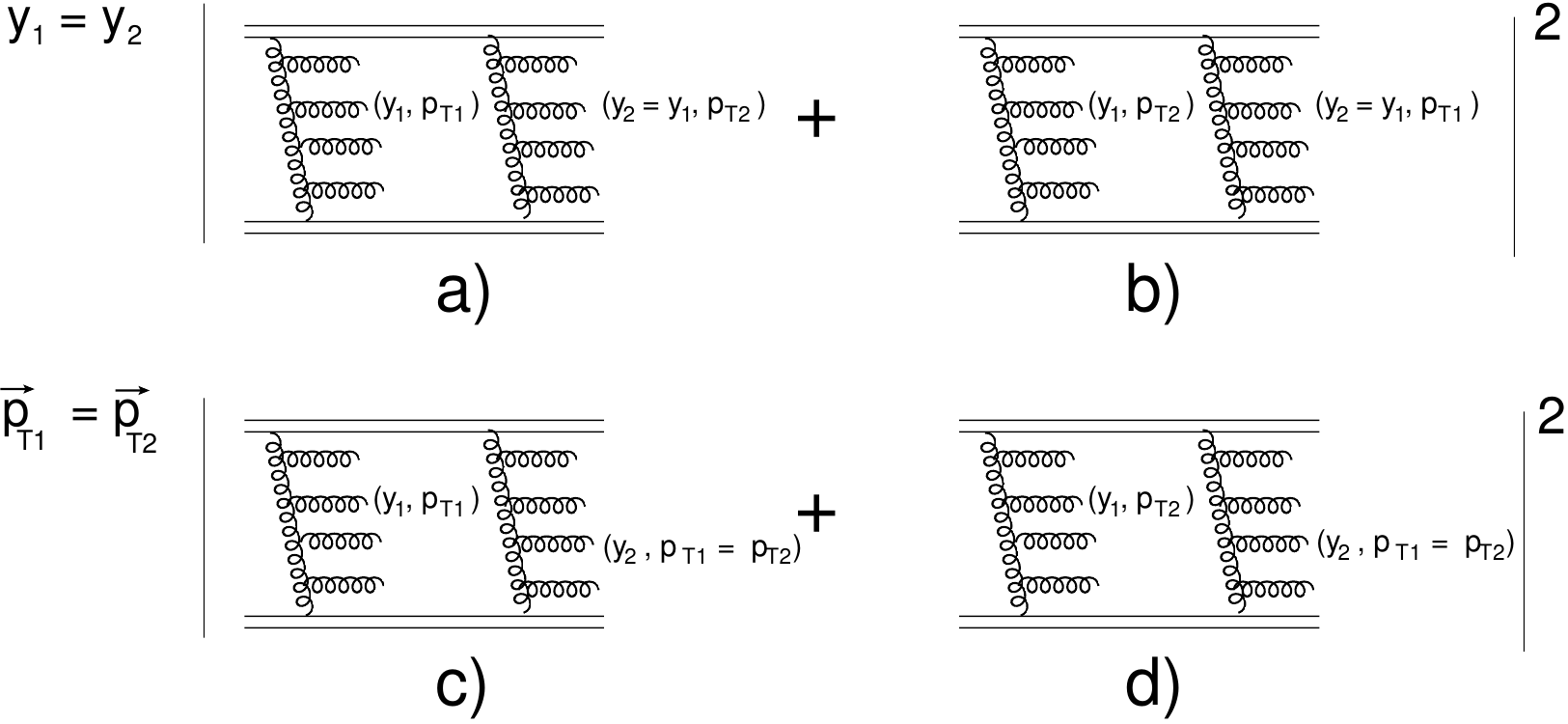}  
      \caption{ The interferences between two states  for the production
 of two identical  gluons in two specific cases: $y_1=y_2$
 (\fig{bec2shy1y2}-a and      \fig{bec2shy1y2} -b)  and $\vec{p}_{T1}
 = \vec{p}_{T2}$(\fig{bec2shy1y2}-c and      \fig{bec2shy1y2} -d).     
} 
  \label{bec2shy1y2}
       \end{figure}

 After integration over $ k_T$, the sum of diagrams \fig{2shiden}-a  and 
\fig{2shiden}-b can be written as
 \beq \label{I1}
 \frac{d^2 \sigma}{d y_1 \,d y_2  d^2 p_{T1} d^2 p_{T2}}\Lb \rm identical\,\,
 gluons\Rb\,\,=\,\,
  \, \frac{d^2 \sigma}{d y_1 \,d y_2  d^2 p_{T1} d^2 p_{T2}}\Lb \rm different
 \,\, gluons\Rb \Big( 1 \,+\,C\Lb R |\vec{p}_{T2} -   \vec{p}_{T1}|\Rb\Big)
  \eeq
  
  \eq{I1} coincides with the general formula for the Bose-Einstein
 correlations \cite{HBT,IPCOR}
  \beq \label{I11}
  \frac{d^2 \sigma}{d y_1 \,d y_2  d^2 p_{T1} d^2 p_{T2}}\Lb \rm identical\,\,
 gluons\Rb\,\,\propto\,\,\Big{ \langle}  1\,\,+\,\,e^{i
 r_\mu Q_\mu}\Big{\rangle}
 \eeq
 where  averaging $\langle \dots \rangle$ includes the integration
 over $r_\mu = r_{1,\mu} - r_{2,\mu}$. There  is only one difference:  
  $Q_\mu = p_{1.\mu} \,-\,p_{2,\mu}$ degenerates to $\vec{Q}\,\equiv
 \,\vec{p}_{T,12}$, due to the fact that the production of two gluons
 from the two parton showers do not depend on rapidities. 
 Note,  that the contribution of \fig{2shiden}-b does not
 depend  on the rapidity difference $y_1 - y_2$ nor  on $y_1$
 and $y_2$.   For $y_1=y_2$  \eq{I1}  follows directly  from   the 
general
 \eq{I11},  and the interference diagram of  \fig{bec2shy1y2}-b  leads
 to \eq{I1}, and allows us to  calculate the 
typical
 correlation radius and the correlation function $C\Lb R |\vec{p}_{T2} -
   \vec{p}_{T1}|\Rb$. On the other hand, for $y_1 \neq y_2$  but for
 $\vec{p}_{T1} = \vec{p}_{T,2}$ \eq{I1}, 
 gives a constant which does not depend on $y_1$ and $y_2$. However,
 in general case $y_1 \neq y_2$  and  $\vec{p}_{T1} \neq \vec{p}_{T,2}$
  the diagram of \fig{2shiden}-b looks problematic \footnote{We thank Alex
 Kovner for vigorous discussions on this subject.}, since it seems to 
describe the 
interference between two different   final states. In appendix A we
 demonstrate that the contribution of \fig{2shiden}-b does not vanish
 even in this general case.  Note, that for $y_1 = y_2$,
 the sum of two Mueller diagrams , indeed, relates to the interference
 between two diagrams, as  is shown in \fig{bec2shy1y2}-a and
 \fig{bec2shy1y2}-b.  For these kinematics, as we have mentioned 
 \beq \label{I12}
 C\Lb R |\vec{p}_{T2} -   \vec{p}_{T1}|\Rb\,\,=\,\, \Big{ \langle} e^{i 
 \vec{r}_T\cdot \vec{Q}_T}\Big{\rangle} 
 ~~~~~~~~~\mbox{where}~~~~~~ \vec{Q}_T\,=\,\vec{p}_{T,12}
 \eeq
 For $\vec{p}_{T1} = \vec{p}_{T2}$, the sum of two Mueller diagrams 
can also
  be viewed as the interference of two diagrams of \fig{bec2shy1y2}-c
 and  \fig{bec2shy1y2}-d, leading to
 \beq \label{I13}
  C\Lb| 0 |\Rb\,\,=\,\, \Big{ \langle} e^{i  \vec{r}^+\,  \vec{Q}_-\,+\,i
  \vec{r}^-\,  \vec{Q}_+}\Big{\rangle}   
  \eeq
 The calculation of the Mueller diagram shows that this average does not
 depend on $y_1$ and $y_2$.
 
 Remembering that for two parton showers in each order of perturbative 
QCD,
 (or in other words at fixed multiplicity of the produced gluons)  the
 amplitude can be written in the factorized form $ A = A_L\Lb r_+,
 r_-\Rb\,A_T\Lb \vec{r}_T\Rb$ leading to 
 \beq \label{I14}
 \Big{ \langle} e^{i  r_\mu\,Q_\mu}\Big{\rangle} \,\,=\,\,\,\underbrace{\Big{
 \langle} e^{i  \vec{r}_T\cdot \vec{Q}_T}\Big{\rangle}}_{\mbox{\small
 averaging over } r_T} \times \underbrace{\Big{ \langle} e^{i  \vec{r}^+\,
   \vec{Q}_-\,+\,i  \vec{r}^-\, \vec{Q}_+} \Big{\rangle}}_{\mbox{\small
 averaging over} \,  r_+, r_-}  
 \eeq 
  
\begin{figure}[ht]
    \centering
  \leavevmode
      \includegraphics[width=10cm]{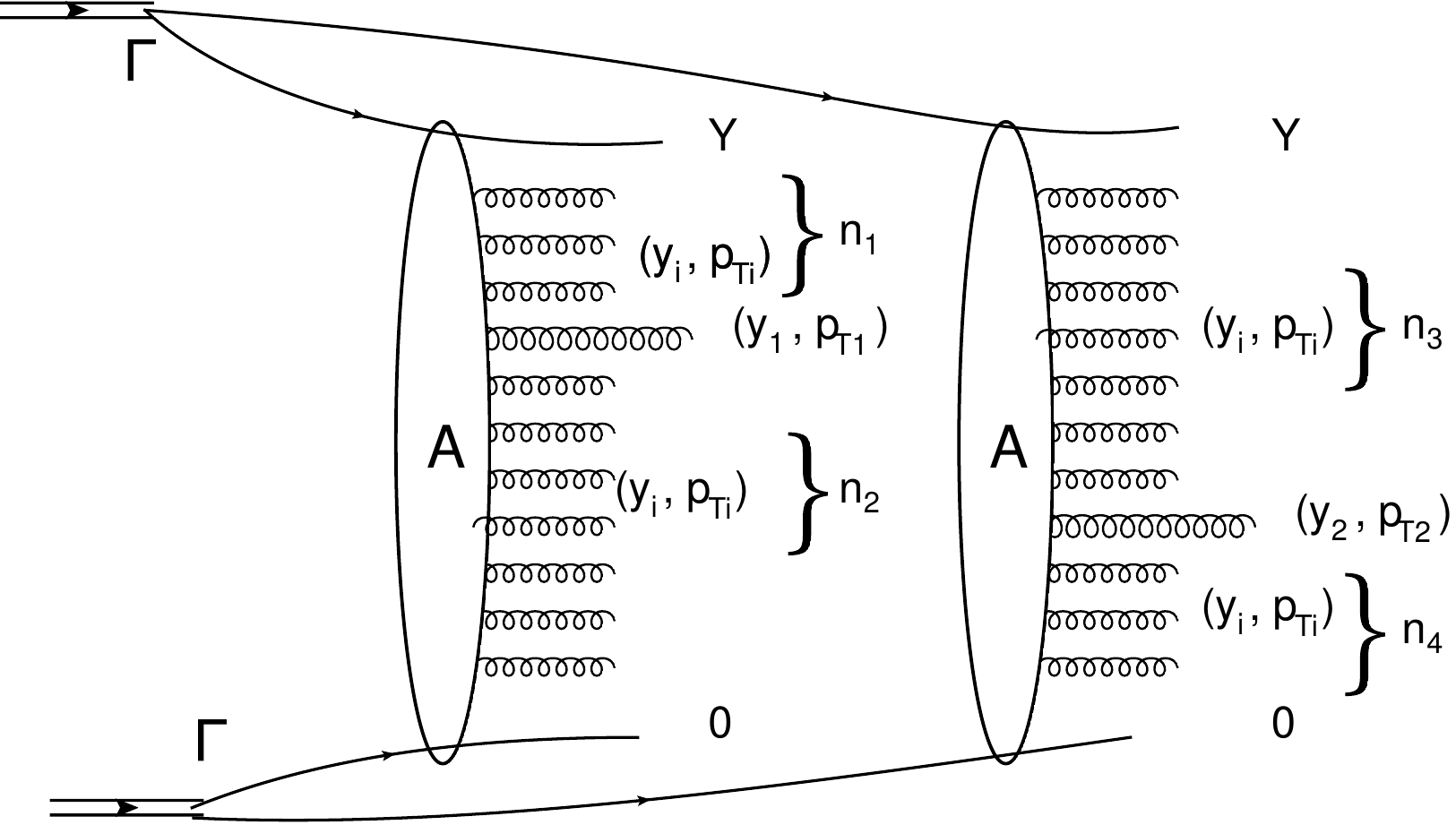}  
      \caption{ The amplitude of production of $n=n_1+n_2+n_3+n_4$ particles, $ A^{\rm different\,gluons}\Lb 2 \to n | \{ y_i, p_{Ti}\}; y_1, \vec{p}_{T1} ; y_2, \vec{p}_{T2} \Rb$ (see \protect\eq{LLA1})} 
  \label{bella}
       \end{figure}

 In our opinion, the above discussion shows  that the Mueller diagram 
 of \fig{2shiden}-b, does not characterize the interference between
 two orthogonal states, but is  an economical way to describe the 
independence 
of
  identical gluon production on rapidities,  providing the smooth
 analytical description of the cross section from $y_1=y_2$ to the 
general case $y_1 \neq y_2$.   Since this point is not obvious  we
 would like to recall the main features of the leading log(1/x) approximation
 (LLA).  In the LLA  we account for the following kinematic
 region\cite{BFKL,LI} for the production of two parton showers
 (see \fig{bella}):
   \bea \label{LLA1}
 \mbox{first parton shower} & \rightarrow & Y\,>\,\dots \,> \,y_i\,>\,\dots\,>\,y_{n_1} \,>\,y_1\,>\,y_{n_2}\,>\,\dots \,> \,y_i\,>\,\dots\,>0;\nn\\
   \mbox{second parton shower} & \rightarrow& Y\,>\,\dots \,> \,y_i\,>\,\dots\,>\,y_{n_3} \,>\,y_2\,>\,y_{n_4}\,>\,\dots \,> \,y_i\,>\,\dots\,>0;\\
   \mbox{parameters of LLA}:   &\bas\,\ll\,1& \bas\Lb y_{i+1} - y_i\Rb\,\geq\,1;~~\bas\Lb Y - y_i\Rb\,\geq\,1;~~\bas\Lb  y_i - 0 \Rb\,\geq\,1;~~\bas\Lb Y - y_1\Rb\,\geq\,1;\nn\\
  &&\bas\Lb Y - y_2\Rb\,\geq\,1;\bas\Lb  y_1 - 0\Rb\,\geq\,1;  ~~ \bas\Lb  y_2 - 0 \Rb\,\geq\,1; ~~\bas\Lb y_1 - y_2\Rb\,\geq\,1;\nn
   \eea

  The cross sections of double inclusive productions can be calculated in LLA  for the production of two parton showers in the following way:
  \begin{subequations} 
   \bea \label{LLA1}
   \frac{d^2 \sigma^{\rm different\,gluons}}{d y_1 \,d y_2  d^2 p_{T1} d^2 p_{T2}} \,&=&\,\,\,\sum^\infty_{n_1+n_2-2>2}\,\, \sum^\infty_{n_3+n_4-2>2}  \,\int d\Phi^{(1)}_{n_1 + n_2} d \Phi^{(2)}_{n_3+n_4}\,|A^{\rm different \,gluons}\Lb \{y_i,p_{Ti}\}; y_1,p_{T1}; y_2,p_{T2}\Rb|^2 \nn \\   
   &=&\sum_{n_1+n_2 =n-2>2}^\infty\,\,\,\,\prod^{n_1}_{Y> y_i > y_1} \int^{y_{i+1}}_{y_{i-1}} d y_i  d^2 p_{T,i}  \prod^{n_2}_{y_1 > y_i > 0 } \int^{y_{i+1}}_{y_{i-1}} d y_i  d^2 p_{T,i} \label{lla11}\\
   &\times& \,\sum_{n_3+n_4 =n'-2>2}^\infty\,\,\,\,\prod^{n_3}_{Y> y_i > y_2} \int^{y_{i+1}}_{y_{i-1}} d y_i  d^2 p_{T,i}  \prod^{n_4}_{y_2 > y_i > 0 } \int^{y_{i+1}}_{y_{i -1}} d y_i  d^2 p_{T,i}   \,\nn\\
   &\times & 
  |  \Gamma^2  A_{n_1 n_2}\Lb 2 \to n | \{ y_i, p_{Ti}\}; y_1, \vec{p}_{T1} \Rb A_{n_3 n_4}\Lb 2 \to n | \{ y_i, p_{Ti}\}; y_2, \vec{p}_{T2} \Rb]^2  \nn\\
     &\xrightarrow{\rm LLA}&\sum_{n_1+n_2 =n-2>2}^\infty\,\,\,\,\prod^{n_1}_{Y> y_i > y_1} \int^{y_{i+1}}_{y_{i-1}} d y_i  d^2 p_{T,i} \,\,\, \prod^{n_2}_{y_1 > y_i > 0 } \int^{y_{i+1}}_{y_{i-1}} d y_i  d^2 p_{T,i} \label{lla12} \\
   &\times& \,\sum_{n_3+n_4 =n'-2>2}^\infty\,\,\,\,\prod^{n_3}_{Y> y_i > y_2} \int^{y_{i+1}}_{y_{i-1}} d y_i  d^2 p_{T,i}  \prod^{n_4}_{y_2 > y_i > 0 } \int^{y_{i+1}}_{y_{i -1}} d y_i  d^2 p_{T,i}   \,\nn\\
   &\times & 
   | \Gamma^2  A_{n_1 n_2}\Lb 2 \to n | \{ y_i=0, p_{Ti}\}; y_1=0, \vec{p}_{T1} \Rb A_{n_3 n_4}\Lb 2 \to n | \{ y_i=0, p_{Ti}\}; y_2=0, \vec{p}_{T2} \Rb|^2 \nn
       \eea
       \end{subequations} 
         where $d \Phi^{(1)}_{n_1+n_2}$ and $d \Phi^{(2)}_{n_3+n_4}$
 are the phase spaces of the produced gluons in the first and second 
parton
 showers. $A^{\rm different \,gluons}\Lb \{y_i,p_{Ti}\}; y_1,p_{T1};
 y_2,p_{T2}\Rb$ = $  \Gamma^2  A_{n_1 n_2}\Lb 2 \to n | \{ y_i, p_{Ti}\};
 y_1, \vec{p}_{T1} \Rb A_{n_3 n_4}\Lb 2 \to n | \{ y_i, p_{Ti}\}; y_2,
 \vec{p}_{T2} \Rb$ (see \fig{bella}) and all other      notations are
 shown in \fig{bella}.
       
       The transition from \eq{lla11} to \eq{lla12}, occurs due to the fact that we want to obtain  the log contribution $ \propto (y_{i+1}  - y_{i -1})$ for each $d y_i$, These logarithms stem from the integration of the phase space of produced particles, while we can neglect the $y_i$ dependence of the production amplitude. In other words, the production amplitudes are   functions   {\it only} of the transverse momenta and \eq{lla12} shows that the longitudinal degrees of freedom can be factorize  out \cite{BFKL,LI}.

 \eq{LLA1} after integrations over $y_i$, can be re-written in a  more efficacious  form , viz.
 
 \bea \label{LLA2}
        \frac{d^2 \sigma}{d y_1 \,d y_2  d^2 p_{T1} d^2 p_{T2}} &=&\sum_{n_1+n_2 =n-2>2}^\infty 
        \,\sum_{n_1+n_2 =n'-2>2}^\infty \underbrace{
   \frac{1}{n_1!} \Lb Y - y_1\Rb^{n_1} \, \frac{1}{n_2!} \Lb y_1 - 0\Rb^{n_2\,}  \frac{1}{n_3!} \Lb  Y - y_2  \Rb^{n_3}\, \frac{1}{n_4!} \Lb   y_2  -0 \Rb^{n_4}}_{\rm integral \,\,over\,\,the \,longitudinal\,\,phase\,\,space} \\
   &\times& \int \prod_i d^2 p_{Ti}\,
    |\Gamma^2  A_{n_1 n_2}\Lb 2 \to n | \{ y_i=0, p_{Ti}\}; y_1=0, \vec{p}_{T1} \Rb A_{n_3 n_4}\Lb 2 \to n | \{ y_i=0, p_{Ti}\}; y_2=0, \vec{p}_{T2} \Rb|^2  \nn
     \eea
     
     Summing over $n_i$ we obtain the Mueller diagram of \fig{2sh}.

    For identical particles we need to replace 
    \bea \label{LLA3}
    && A^{\rm different \,gluons}\Lb \{y_i=0,p_{Ti}\}; y_1=0,p_{T1}; y_2=0,p_{T2}\Rb \,=\nn\\
     &&  \Gamma^2  A_{n_1 n_2}\Lb 2 \to n | \{ y_i=0, p_{Ti}\}; y_1=0, \vec{p}_{T1} \Rb A_{n_3 n_4}\Lb 2 \to n | \{ y_i=0, p_{Ti}\}; y_2=0, \vec{p}_{T2} \Rb    \,\,\to\,\,\\
     &&A^{\rm identical  \,gluons}\Lb \{y_i=0,p_{Ti}\}; y_1=0,p_{T1}; y_2=0,p_{T2}\Rb\,\,= \nn\\
     &&A^{\rm different \,gluons}\Lb \{0,p_{Ti}\}; y_1=0,p_{T1}; y_2=0,p_{T2}\Rb\,+\,A^{\rm different \,gluons}\Lb \{0,p_{Ti}\}; y_2=0,p_{T2};y_1=0,p_{T1}\Rb\,=\nn\\
      &&   \hspace{3cm} \Gamma^2  A_{n_1,n_2}\Lb 2 \to n | \{ y_i=0, p_{Ti}\}; y_1=0, \vec{p}_{T1} \Rb A_{n_3 n_4}\Lb 2 \to n | \{ y_i=0, p_{Ti}\}; y_2=0, \vec{p}_{T2} \Rb\,\,\nn\\
 &&\hspace{3cm}+\,\,  \Gamma^2  A_{n_1 n_2}\Lb 2 \to n | \{ y_i=0, p_{Ti}\}; y_2=0, \vec{p}_{T2} \Rb A_{n_3 n_4}\Lb 2 \to n | \{ y_i=0, p_{Ti}\}; y_1=0, \vec{p}_{T1} \Rb \nn
    \eea   
We wish to  stress, that in \eq{LLA3} we use the Bose-Einstein symmetry for  the production amplitudes,  which are   {\it only }     functions of   the transverse momenta of produced particles.
       
  Such a replacement leads to the sum of the diagrams of  \fig{2shiden}-a and  \fig{2shiden}-b.

  The goal of this paper is to calculate the function $\,C\Lb R
 |\vec{p}_{T2} - 
  \vec{p}_{T1}|\Rb$ which tends to 1 at $\vec{p}_{T2}  \to    \vec{p}_{T1} 
 $, and vanishes for $R |\vec{p}_{T2} -   \vec{p}_{T1}| \,\gg\,1$.
  To estimate $ C\Lb R |\vec{p}_{T2} -  
 \vec{p}_{T1}|\Rb$, it is     sufficient  to know the double inclusive cross
 section for $y_1 = y_2$ , where    \fig{2shiden}-b   contributes  significantly.

 To obtain the double inclusive cross section,  
  we need to add the cross section  for two different gluon production 
  which has the form
  \beq \label{I2}
   \frac{d^2 \sigma}{d y_1 \,d y_2  d^2 p_{T1} d^2 p_{T2}}\,\,=\,\,\frac{d^2
 \sigma}{d y_1 \,d y_2  d^2 p_{T1} d^2 p_{T2}}\Lb \rm different \,\, 
gluons\Rb\Bigg( 1\,+\,\frac{1}{N^2_c - 1}\,\,C\Lb R |\vec{p}_{T2} -  
 \vec{p}_{T1}|\Rb\Bigg)
   \eeq
   In \eq{I2} we take  into account, that we have $N^2_c  - 1$ pairs of
 the identical gluons, where $N_c $ is the number of colours, and that
 the polarizations of the identical gluons should be the same. The latter
 leads to a  suppression of $\h$ of the second term in \eq{I2}.  Using 
\eq{I2}
 we can find $v_n$, since
   \beq \label{I3}
     \frac{d^2 \sigma}{d y_1 \,d y_2  d^2 p_{T1} d^2 p_{T2}}\,\,\propto\,\,1
 \,\,+\,\,2 \sum_n V_{ n \Delta } \Lb p_{T1}, p_{T2}\Rb \,\cos\Lb \Delta\,
\varphi\Rb
     \eeq
     where $\Delta \varphi$ is the angle between    $\vec{p}_{T1}$ and
 $ \vec{p}_{T2} $.
     $v_n$ is determined  from  $V_{n \Delta } \Lb p_{T1}, p_{T2}\Rb $
     \beq \label{vn}  
 1.~~    v_n\Lb p_T\Rb\,\,=\,\,\sqrt{V_{n \Delta }\Lb p_T, p_T\Rb}\,;\,
~~~~~~~~~~~~~2.~~~~  v_n\Lb p_T\Rb\,\,=\,\,\frac{V_{n \Delta }\Lb p_T,
 p^{\rm Ref}_T\Rb}{\sqrt{V_{n \Delta }\Lb p^{\rm Ref}_T, p^{\rm Ref}_T\Rb}}\,;
     \eeq
 \eq{vn}-1 and \eq{vn}-2  depict   two methods  of how the  values
 of $v_n$ have been extracted from the experimentally measured
  $V_{n \Delta } \Lb p_{T1}, p_{T2}\Rb$. $ p^{\rm Ref}_T$ denotes the
 momentum of the reference trigger.
     These two definitions are equivalent if  $V_{n \Delta }\Lb p_{T1},
 p_{T2}\Rb $ can be factorized as $V_{n \Delta }\Lb p_{T1}, p_{T2}\Rb\,=\,
     v_n\Lb p_{T1}\Rb\,v_n\Lb p_{T2}\Rb $. We will show below that in our approach this is
 the case 
  for the restricted  kinematic region $R\,p_{Ti}\,\ll\,1$.


     The first problem that we face in calculating $C\Lb R |\vec{p}_{T2} -  
 \vec{p}_{T1}|\Rb$, is to estimate the value of $R$, which increases with
 energy (see for example LHC data of Ref.\cite{TOTEM}). On the other hand,
 the BFKL Pomeron\cite{BFKL,LI}  does not lead to the shrinkage of the
 diffraction peak, as it  has no slope for the Pomeron trajectory.   The only
 way  to obtain a size which increases with energy, is to use the
 unitarity constraints , $A^{\rm BFKL} \Lb Y,b\Rb \propto e^{\Delta_{\rm BFKL}
 \,Y} 
    a(b)\,<\,1$ \cite{FROI},  where  $\Delta_{\rm BFKL}$  is the intercept
 of the BFKL Pomeron and $b$ is the impact factor. However, in QCD $a\Lb
 b\Rb$ decreases as a power of $b$ and the unitarity constraints  lead
 to $R\,\, \propto\,\,\exp\Lb \Delta_{\rm BFKL}\,Y\Rb$ \cite{KOWE}.
 Therefore, to obtain the energy behaviour of $R$,  we need to introduce
    a non-perturbative correction  at large $b$, which assures $a(b)
 \,\,\propto\,\exp\Lb - \mu_{\rm soft} b\Rb$, and also to take into 
account
 the multi Pomeron interactions which satisfy the unitarity constraints.
 Fortunately, the second part of the problem has been solved in the
 CGC/saturation approach\cite{KOLEB}, but the first  needs
 modelling of the unknown confinement of quarks and gluons. Hence,
  we are doomed to build a model which includes everything that we
 know theoretically regarding the CGC/saturation approach, but
in addition, 
one needs to introduce some
 phenomenological descriptions of the hadron structure, and the large
 $b$ behaviour of the BFKL Pomeron.

     Such a model for hadron-hadron interactions at high energy has been
 developed in Refs.\cite{GLM2CH,GLMNI,GLMINCL,GLMCOR}, and it successfully
 describes the experimental data on total, inelastic and diffractive cross
 sections, as well as the inclusive production and LRR correlations. The 
goal
 of this paper is to show that the structure of the `dressed' Pomeron in 
this
 model leads to strong BE correlations, and generates $v_n$ both for even 
and
 odd $n$, in hadron and nucleus interactions. In the next section we 
consider
 the contribution to $C\Lb R |\vec{p}_{T2} -   \vec{p}_{T1}|\Rb$ from the  
first
 Mueller diagram,  and discuss the different sources of BE-correlations. 
In
 section 3 
    we   give a brief review of the structure of the Pomeron in  our 
 model,  in which we  incorporate  the solution to the 
CGC/saturation 
equations
 with additional  non-perturbative assumptions: the large $b$ behaviour for the
 saturation momentum,  and the structure of  hadrons.
  It has been known for a long time \cite{OLDREG,PION} in the framework
 of Gribov Pomeron calculus, and has been re-considered in CGC/saturation
 approach \cite{RIDGE}, that the LRR correlations stem from the production 
of
 gluon jets from two different parton showers (see \fig{2sh}). In section
 4 we evaluate the BE  correlations that result from the dressed Pomeron 
of
 our model, and  show that they are able to describe the main features
 of the experimental data.
 
 
 \section{Calculation of the first diagram}
 
 
 \subsection{Proton-proton scattering}
  
\begin{figure}[ht]
    \centering
  \leavevmode
      \includegraphics[width=12cm]{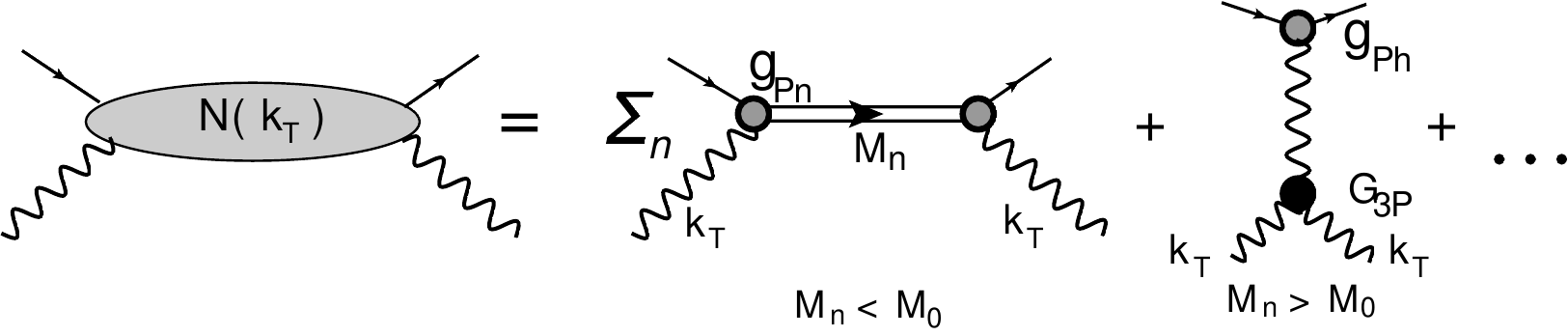}  
      \caption{ The structure of $N_{\pom h} \Lb k^2_T \Rb $.} 
    \label{n}
       \end{figure}

 The first Mueller diagram which contributes to   $C\Lb R |\vec{p}_{T2} - 
  \vec{p}_{T1}|\Rb$ and which we need to calculate, is shown in 
\fig{2shiden}-e and
 can be written in the form \cite{LERE}:
 \bea \label{1D1}
&&   \frac{d^2 \sigma}{d y_1 \,d y_2  d^2 p_{T1} d^2 p_{T2}}\,\,=\,\,
\frac{d^2 \sigma}{d y_1 \,d y_2  d^2 p_{T1} d^2 p_{T2}}\Lb \rm different
 \,\, gluons\Rb\Bigg( \frac{1}{N^2_c - 1}\,\,C\Lb R |\vec{p}_{T2} -  
 \vec{p}_{T1}|\Rb\Bigg)\\
&&=\,
 \Lb   \frac{\bas\,C_F}{2 \pi}\Rb^2 \int d^2 k_T \,N_{\pom h}\Lb k^2_T\Rb\,
N_{\pom h}\Lb \Lb \
 \vec{k_T} + \vec{p}_{T,12}\Rb^2\Rb  \,    \frac{d   \sigma}{d y_1 d^2
 p_{T1}}\Lb k_T, |  \vec{k_T} + \vec{p}_{T,12}|\Rb
 \frac{d   \sigma}{d y_2 d^2 p_{T2}}\Lb k_T, |  \vec{k_T} + \vec{p}_{T,12}|
\Rb \nn
    \eea   
where $\vec{p}_{T,12} = \vec{p}_{T1} \,-\,\vec{p}_{T2}$ and
\bea \label{1D2}
 &&   \frac{d   \sigma}{d y_1 d^2 p_{T1}}\Lb k_T, |  \vec{k_T} +
 \vec{p}_{T,12}|\Rb\,=\\
 &&\,\int d^2 q_T\, \phi^{\rm BFKL}\Lb q_T,\vec{k}_T - \vec{q}_T\Rb
 \,\Gamma_\mu \Lb q_T, p_{T1}\Rb \, \Gamma_\mu \Lb \vec{k}_T - \vec{q}_T, p_{T2}\Rb\,\phi^{\rm BFKL}\Lb q_T,\vec{k}_T + 
    \vec{p}_{T,12} - \vec{q}_T\Rb\nn
    \eea
    In \eq{1D2}   $\phi^{\rm BFKL}$ denotes the parton density of the
 BFKL Pomeron, with momentum transferred by the Pomeron $\vec{k}_T$
 or $\vec{k}_T + \vec{p}_{T,12}$. The Lipatov vertex $\Gamma_\mu$,
 as well as the equations for $\phi^{\rm BFKL}$ will be discussed
 in the appendix A.  Generally speaking, $N_{\pom h}$ has a structure
  which is shown in \fig{n}:
    \beq \label{N}
    N_{\pom h}\Lb k^2_T\Rb\,\,=\,\,\sum^{M_0}_{M_n}\,g^2_{\pom n}\Lb k^2_T\Rb
 \delta\Lb M^2 - M^2_{n}\Rb\,+\,g_{\pom h}\Lb 0\Rb\,G_{3 \pom}\Lb k^2_T\Rb\,
e^{\Delta_{\rm BFKL} Y}
    \eeq
    where $M_n$ denotes the mass of resonances, $\Delta_{\rm BFKL}$
      the intercept of the BFKL Pomeron, and $G_{3 \pom}$  the triple
 Pomeron vertex. 
 Considering the contribution of the first term to $N_{\pom h}$,  
 we can neglect,  in the first approximation,  the dependence of
 $\phi^{\rm BFKL} $ on the momentum transferred, since $Q_T$  
  turns out to be of the order of the saturation momentum 
$Q_s\,\gg\,1/R_h$,
 where $R_h$ is the hadron size  incorporated in $N_{\pom h}$.    


 This is not the case for the second term in \eq{N}, which has $Q_T
 \sim Q_s$.
It leads to the BFKL Pomeron calculus which takes  the Pomeron
 interactions into account. We will discuss this contribution in sections 
3 and 4. In
 this section
 we restrict ourselves  to the first term in the sum in \eq{n}. Collecting
 all formulae, we obtain that in the first diagram
 
 \beq \label{D1C}
 C\Lb R |\vec{p}_{T2} -   \vec{p}_{T1}|\Rb\,\,\propto\,\,\int d^2 k_T
 \,g^2_{\pom,{\rm pr}}  \Lb k^2_T\Rb\,  g^2_{\pom,{\rm tr}}  \Lb \Lb
 \vec{k} - \vec{p}_{T,12}\Rb^2\Rb\Bigg{/}  \int d^2 k_T \,g^2_{\pom,{\rm pr}}
  \Lb k^2_T\Rb\,  g^2_{\pom,{\rm tr}}  \Lb k^2_T\Rb \eeq
 
To obtain the first estimates
  for the vertices of the soft Pomeron interaction with the projectile
 and target,    we use the following
 parameterizations:
 \beq \label{G}
 g_{pr}\Lb k^2\Rb \,=\,g^0_{pr}\,e^{- \h B_{pr}\,k^2_T};~~~~~~~~ g_{tr}\Lb k^2\Rb \,=
\,g^0_{tr}\,e^{- \h B_{tr}\,k^2_T}; 
 \eeq 
 For proton-proton  collisions  we take $B_{pr} = B_{tr} = B$.
 
 In this case \cite{GLMBE,PION}
  \beq \label{D1CPP}
    C\Lb R |\vec{p}_{T2} -   \vec{p}_{T1}|\Rb\,\,=\,\,\exp\Lb
 - B_R\,\Lb p^2_{T1} - 2 p_{T1} p_{T2} \cos\Lb \Delta \varphi\Rb\,+\,
p^2_{T2}\Rb\Rb
\eeq
with $B_R = B_{pr} B_{tr}/\Lb B_{pr}\,+\, B_{tr}\Rb$. 
 $B_R = \h B$ for proton-proton scattering.

  In Ref.\cite{GLMBE} it is shown that \eq{D1CPP}      leads to $V_{\Delta
 n}$ of \eq{I3} which is equal to
  \beq \label{VDNPP}
  V_{\Delta\, n}\,\,=\,\,\
  I_n\Lb 2 B_R p_{T1} p_{T2}\Rb \frac{e^{-B_R\Lb p^2_{T1}\,+
\,p^2_{T2}\Rb}}{N^2_C - 1 \,\,+ \,\, I_0\Lb 2 B_R p_{T1}
 p_{T2}\Rb\,e^{-B_R\Lb p^2_{T1}\,+\,p^2_{T2}\Rb } }
  \eeq
  where $I_n$ is the modified Bessel function of the first kind.
  
\begin{figure}[ht]
    \centering
    \begin{tabular}{ccc}
  \leavevmode
      \includegraphics[width=6cm]{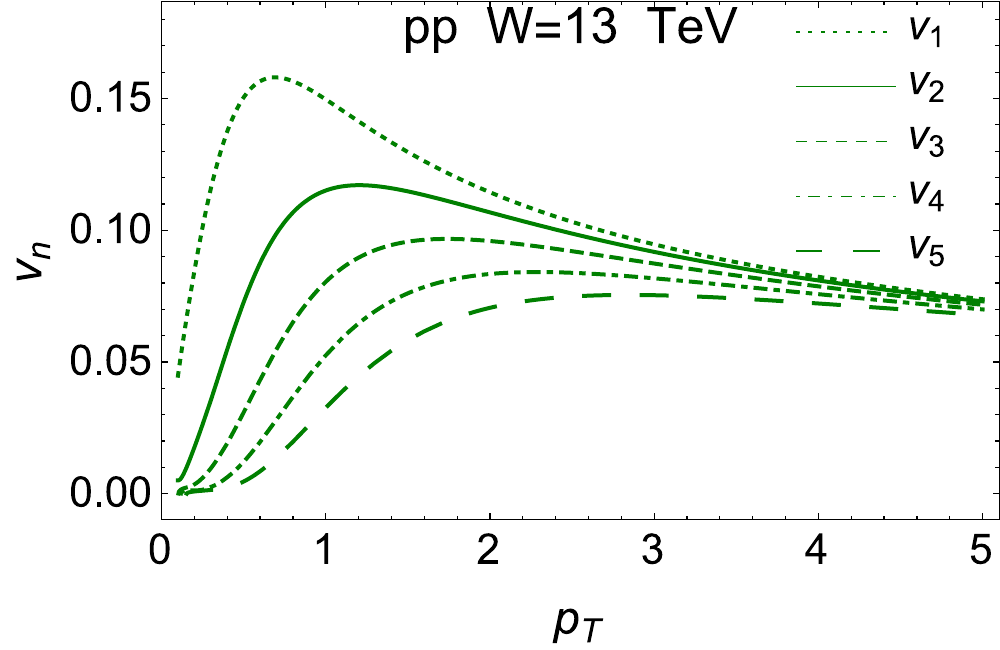} & 
 \includegraphics[width=6cm]{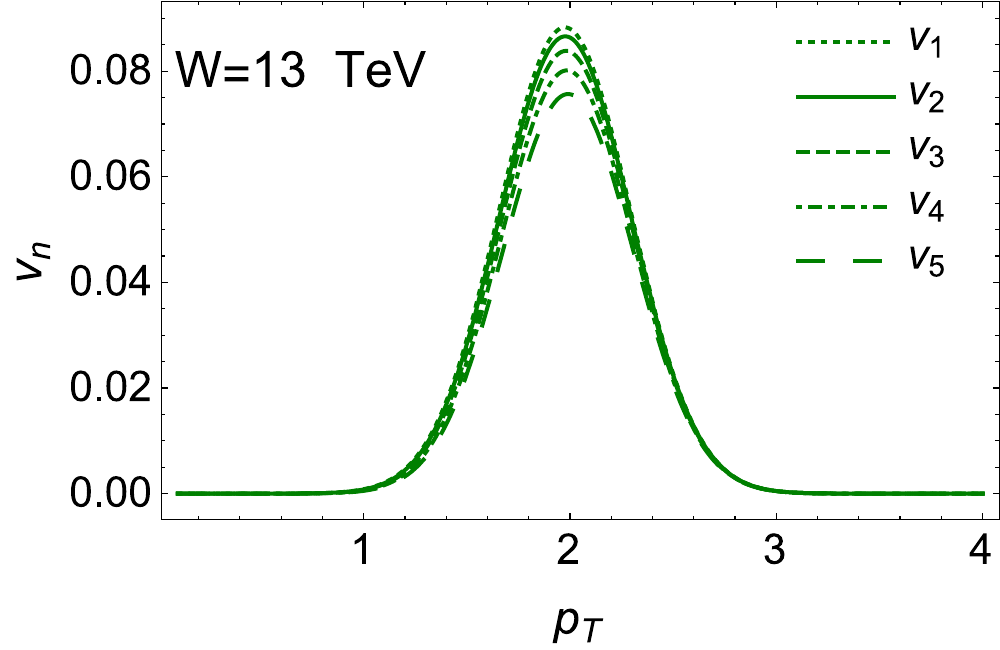} &
       \includegraphics[width=6cm]{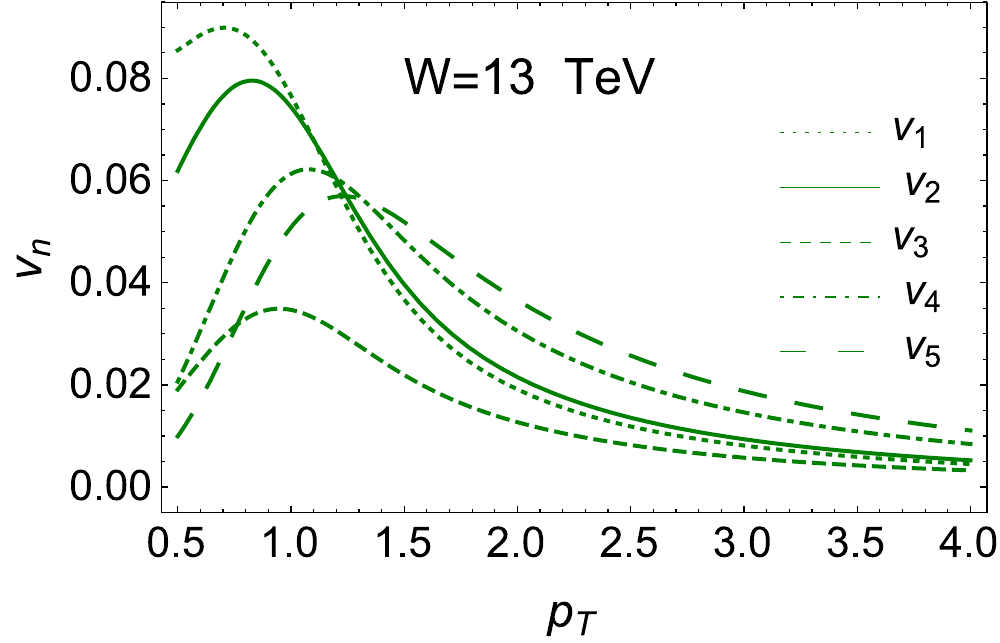}\\
       \fig{ppmod}-a & \fig{ppmod}-b & \fig{ppmod}-c\\
       \end{tabular}     
           \caption{ $v_n$ versus $p_T$ for proton-proton scattering at
 $W = 13\,TeV$, using  \eq{VDNPP} and \eq{vn}. \fig{ppmod}-a shows $v_n$
 that stem from \eq{vn}-1. In \fig{ppmod}-b the estimates from \eq{vn}-2
 for $p^{\rm Ref}_T = 2 \,GeV$ are plotted.  \fig{ppmod}-c describes the 
same
 $v_n$ as in \fig{ppmod}-b, where, $ p^{\rm Ref}_T$ is taken in the 
interval
 $0.5 - 5 \,GeV$, as  is measured in Ref.\cite{ATLASPP}.}           
      \label{ppmod}
       \end{figure}

  In \fig{ppmod}  taking $B_R = 5 \,GeV^{-2}$,
 we plot the prediction for $v_n$ using \eq{VDNPP}
 and \eq{vn}.  This value of $B_R$, corresponds to
 the slope of the elastic cross section for proton-proton scattering
 at $W = 13\,GeV$.
  One can see that \eq{vn}-1 and \eq{vn}-2 give different predictions,
 demonstrating that we do not have factorization $V_{n \Delta
 }\Lb p_{T1}, p_{T2}\Rb\,\neq\,     v_n\Lb p_{T1}\Rb\,v_n\Lb p_{T2}\Rb $. 
\fig{ppmod}-c shows $v_n$ for $p^{\rm min}_T \leq p_{T2} = p^{\rm Ref}_{T}
 \leq p^{\rm max}_T$ with $p^{\rm min}_T =0.5  \,GeV$ and $p^{\rm max}_T =
 5 \,GeV$,  as  done in Ref.\cite{ATLASPP}.
 To calculate such a $v_n$,
we need to know the dependence
 of the cross section on $p_{T2}$. Indeed, we need
 to take \eq{I2} and integrate it over $p_{T2}$: viz.
  \beq \label{1D3}
 H\Lb p_{T1},\Delta \varphi\Rb\,=\,  \int^{p^{\rm max}_{T2}}_{p^{\rm min
 }_{T2}} d p^2_{T2}\,\frac{d^2 \sigma}{d y_1 \,d y_2  d^2 p_{T1} d^2 p_{T2}}
\Lb \protect\eq{I2}\Rb \,\,\propto\,\,1 \,\,+\,\,2 \sum_n V_{ n \Delta } \Lb
 p_{T1}\Rb \,\cos\Lb \Delta\,\varphi\Rb
   \eeq 
For \fig{ppmod}-c,
we need to know the behaviour of the double inclusive cross section on
 $p_{T1} $ and $p_{T2}$.
  We assume   that $ \frac{d^2 \sigma}{d y_1
 \,d y_2\,  d^2 p_{T1} d^2 p_{T2}}  \,\propto\,1\Big{/}\Lb p^2_{T1}\,
p^2_{T2}\Rb$ for the cross sections given by \fig{2sh}-b and by
 \fig{2shiden}-c and \fig{2shiden}-d. In the appendix A   we show that
 the cross section for \fig{2shiden}-e $ \frac{d^2 \sigma}{d y_1
 \,d y_2\,  d^2 p_{T1} d^2 p_{T2}} \Lb \protect\fig{2shiden}-e\Rb
 \,\propto\, \Big( 1/p^2_{T1} \,+\,1/p^2_{T2}\Big)^2$.
  
 We took  the energy dependence into account, by calculating
 the $B_R$ from the slope of the elastic scattering at given
 energy $W$,   which was taken from Ref.\cite{GLM2CH}.
 
       \begin{figure}[ht]
    \begin{tabular}{cc}
  \leavevmode
      \includegraphics[width=6cm]{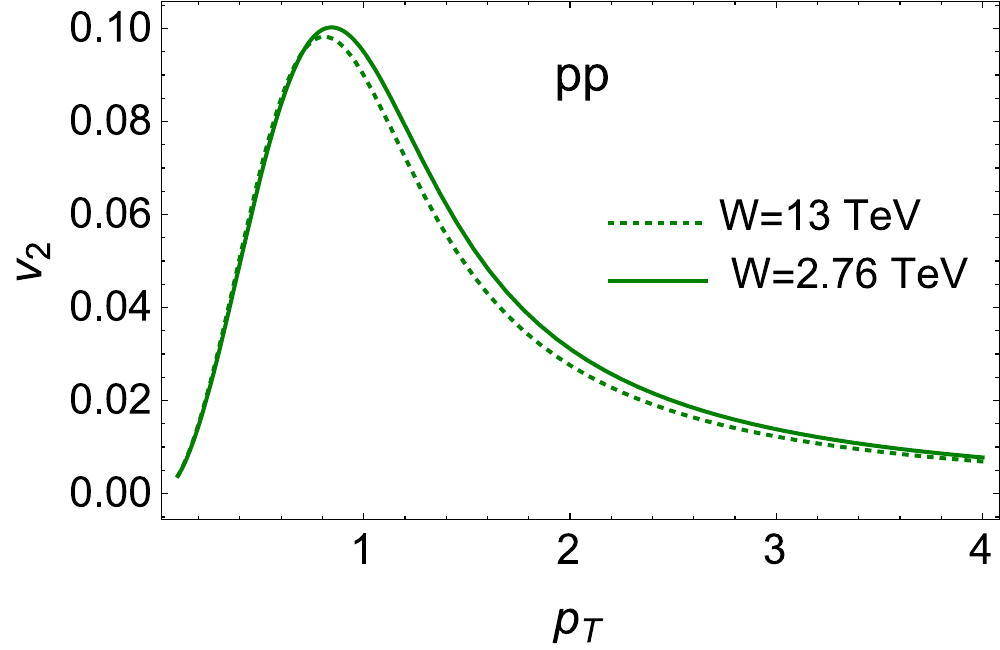} & 
 \includegraphics[width=8cm]{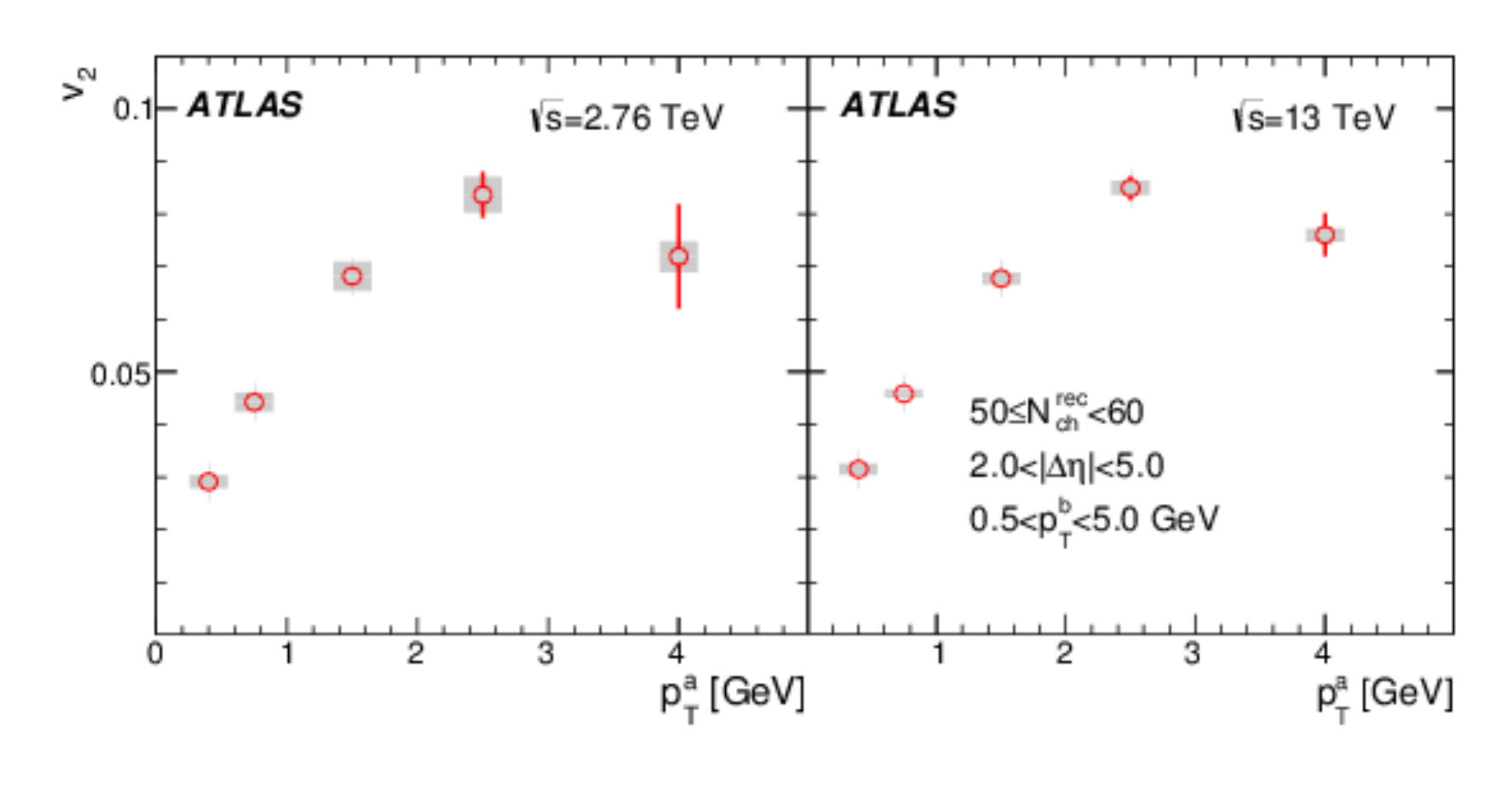} \\
       \fig{CppW}-a & \fig{CppW}-b \\
       \end{tabular}     
           \caption{ $v_2$ versus $p_T$ for proton-proton scattering at
 $W\,=\,2.76\,TeV$ and at
            $W\, =\, 13\,TeV$. \fig{CppW}-a shows $v_n$ that stem 
 from \eq{vn}-2 for $ p^{\rm Ref}_T$ which is taken in the interval
 0.5 to 5 GeV,   as  is measured in Ref.\protect\cite{ATLASPP}.
 \fig{CppW}-b exhibits the experimental data  taken from 
 Ref.\protect\cite{ATLASPP}.}           
      \label{CppW}
       \end{figure}

 
 One can see that the calculated values, as well as energy dependence are  
close
 to the experimental data of Ref.\cite{ATLASPP}.  The main difference
 is in the $p_T$  dependence, which suggests the necessity to include 
the diffractive
 dissociation process or, in other words, the entire sum in \eq{N}, as
 well as the enhanced diagrams that are generated by the BFKL Pomeron
 calculus (see \fig{n}).
 
 We can estimate  
the sum over resonances or, in other words, the diffraction production
 of states with low mass, by using our model (see
 appendix B for necessary formulae).  In \fig{Cpp} we plot the
 correlation function $C\Lb p_{T,12}\Rb$ as  defined in
 \eq{I2} for $|\vec{p}_{T1}|=|\vec{p}_{T2}|$, which is the result
 of these calculations.  One can see that the effective 
  $p_{T,12}$ dependence of the slope, turns out to be much smaller than 
 our estimates from the first diagrams that we obtained above.
 The slope that we used   for the calculation shown in \fig{ppmod}
 was estimated as $\frac{1}{4}B_{el}$,  where $B_{el} = 20\,GeV^{-2}$
 is the slope of the elastic cross section at $W = 13 \,TeV$. We see
 two reasons for such a drastic change in the $p_{T,12}$ dependence: 
first,
 we took into account the diffractive production processes which were
 neglected in \fig{ppmod}; and second, in our model the effective
 shrinkage of the diffraction peak originates from the shadowing
 corrections, as the BFKL Pomeron has no inherent shrinkage.
 Such corrections are stronger in net-diagrams of \fig{amp}-b that
 are responsible for elastic scattering, than for the fan diagrams
 of \fig{inclgen} that contribute to inclusive production.
 Recall that  $B_{\rm shr}\,\approx 10\,GeV^{-2}$ at $W = 13 \,TeV$,
 comes from the shrinkage of the diffraction peak. 
   
     \begin{figure}[ht]
    \centering
  \leavevmode
      \includegraphics[width=10cm]{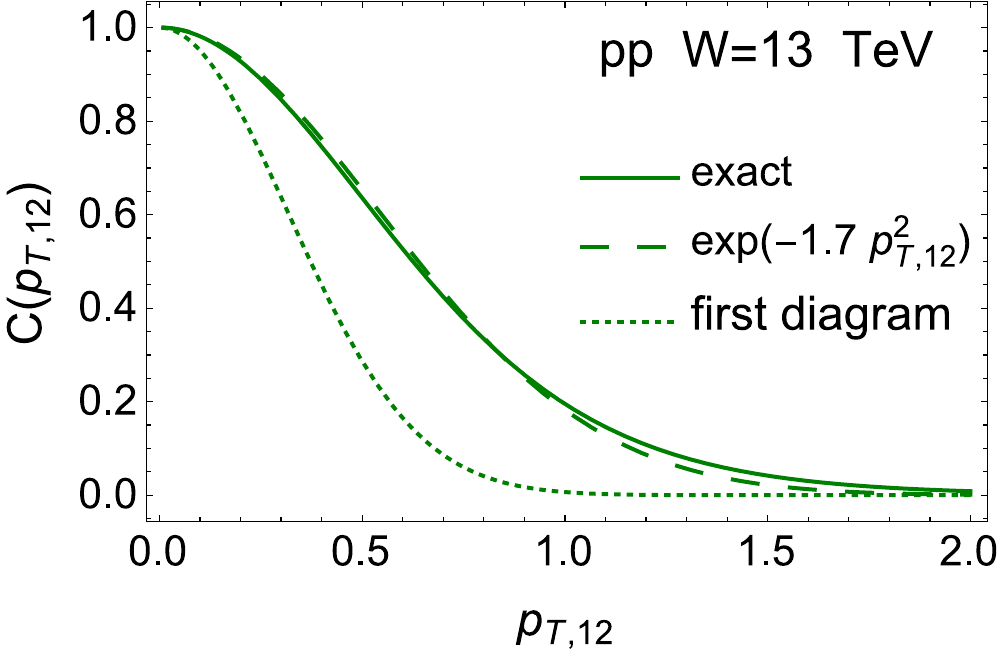}  
      \caption{ Correlation function $C\Lb p_{T,12}\Rb$ as it
 is defined in \eq{I2}, versus $p_{T,12} = |\vec{p}_{T1} - \vec{p}_{T2}|$.
      Dashed line corresponds to $\exp\Lb - B p^2_{T,12}\Rb$
 with $B = 1.7 \,GeV^{-2}$, while the dotted line shows the
 dependence that we used in section 2 to calculate the first
 diagram: $\exp\Lb - B\,p^2_{T,12}\Rb$  with $B = 5 \,GeV^{-2}$.}
\label{Cpp}
   \end{figure}

\begin{figure}[ht]
    \centering
    \begin{tabular}{ccc}
  \leavevmode
      \includegraphics[width=6cm]{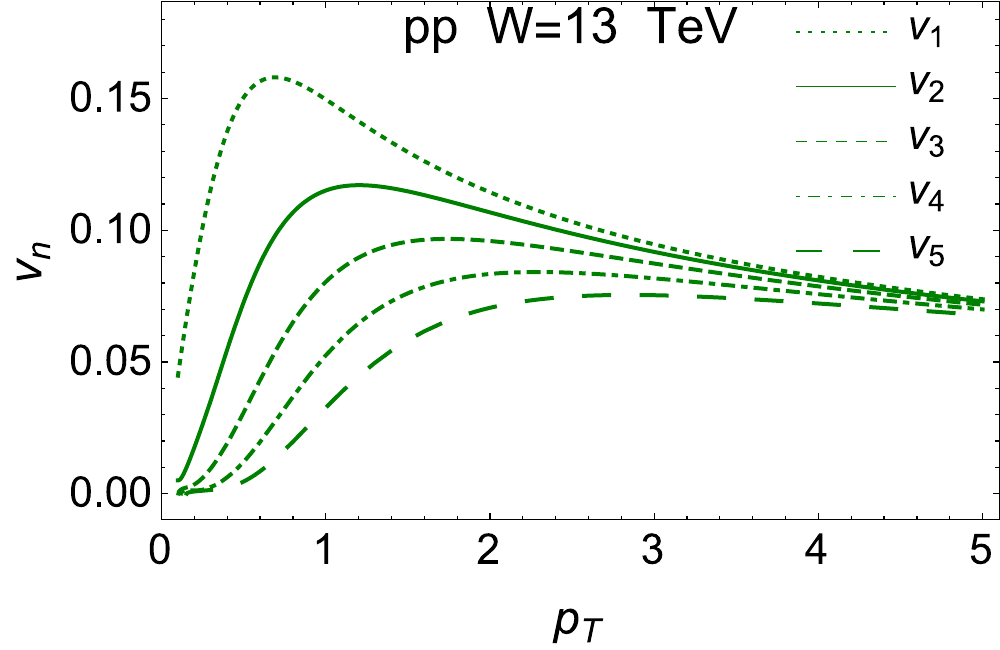} & 
 \includegraphics[width=6cm]{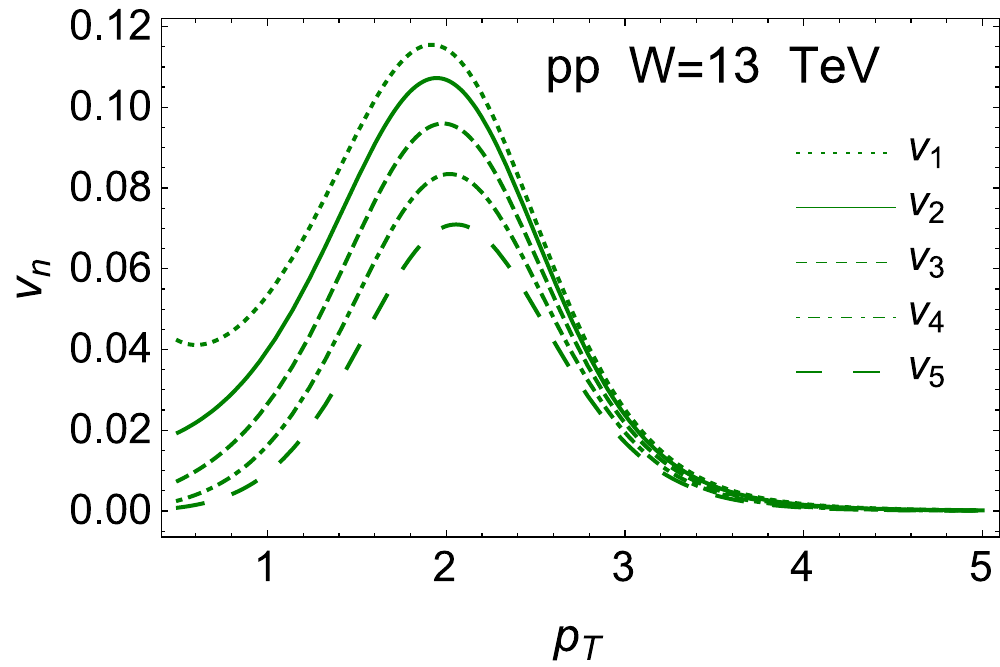} &
       \includegraphics[width=6cm]{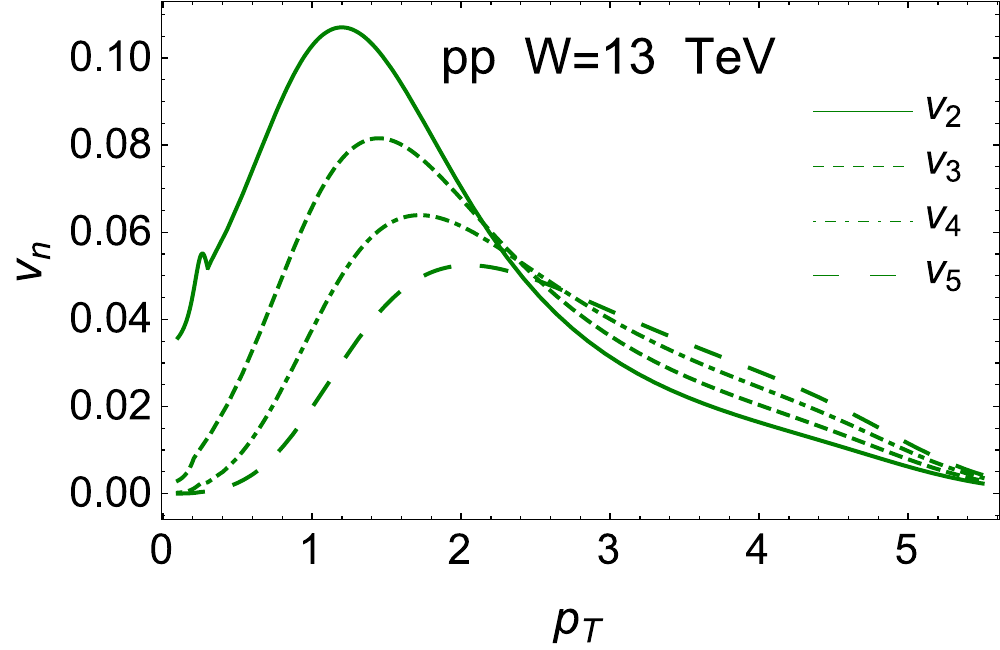}\\
       \fig{pp13}-a & \fig{pp13}-b & \fig{pp13}-c\\
       \end{tabular}     
           \caption{ $v_n$ versus $p_T$ for proton-proton
 scattering at $W = 13\,TeV$, using  \eq{PP1} and \eq{vn}.
 \fig{pp13}-a shows $v_n$ that stem from \eq{vn}-1. In \fig{pp13}-b
 the estimates from \eq{vn}-2 for $p^{\rm Ref}_T = 2 \,GeV$ are plotted.
\fig{pp13}-b describes the same $v_n$ as in \fig{ppmod}-c but
 $ p^{\rm Ref}_T$ is taken in the interval 0.5 to 5 GeV, as
  is measured in Ref.\cite{ATLASPP}.}           
      \label{pp13}
       \end{figure}

  The calculation of $v_n$ are shown in \fig{pp13}.
  One can see that we obtain large $v_n$ for both
 odd and even $n$. The value of $v_2$ from \fig{pp13}-c is
 about 10\% larger, than the experimental one from Ref.\cite{ATLASPP}
 (see \fig{CppW}-b). However, our calculations lead to narrower
 distributions in $p_T$ than the experimental one. The factorization
   $V_{n \Delta }\Lb p_{T1}, p_{T2}\Rb\,=\,     v_n\Lb p_{T1}\Rb\,v_n\Lb
 p_{T2}\Rb $  is strongly violated,   as  in the case of estimates of the 
first
 diagram.

      ~


  \fig{CppW} illustrates the energy dependence of $v_n$ for proton-proton 
scattering, showing $v_2$ for two energies $W=2.56 \,TeV$
 and $W=13 \,TeV$. Note, that $v_2$ does not depend
 on energy, in accord with the experimental data of 
Ref.\cite{ATLASPP}.
 
 Therefore, we can conclude that the first term in \eq{N} leads to a
 value of $v_n$, which is large and of the order of the experimental one;
 the inclusion of  diffraction in the region of small mass (sum over
 resonances in \eq{N}) leads to a decrease of the interaction volume,
 but cannot reproduce the experimental $p_T$ distributions of $v_n$,
 and BE correlations show the experimentally observed independence on 
energy.
 
 \subsection{Hadron-nucleus and nucleus nucleus interaction}
 
 
 For a nucleus we can simplify the calculation, considering  cylindrical 
nuclei
 which have a form factor
 \beq \label{1DSA}
 S_A\Lb k_T\Rb\,=\,\frac{R_A}{k_T}\,J_1\Lb k_T R_A\Rb
 \eeq
 where $J_1$ is the Bessel function. Taking \eq{1DSA} into account one
 can see that
 \bea \label{1DCpA}
 C_{pA}\Lb R |\vec{p}_{T2} -   \vec{p}_{T1}|\Rb\,\,&\propto&\,\,\int
 d^2 k_T \,  g^2_{\pom,{\rm tr}}  \Lb \Lb \vec{k}_T - \vec{p}_{T,12}\Rb^2\Rb
 \,S_A^2\Lb k^2_T\Rb\Bigg{/}     \int d^2 k_T \,  g^2_{\pom,{\rm tr}}  \Lb 
\Lb k^2_T\Rb^2\Rb \,S_A^2\Lb k^2_T\Rb \nn\\
 &=&\int d^2 k_T \,  e^{-B\Lb  k^2_T  + p^2_{T,12}\Rb} \,I_0\Lb 2 B \,k_T\,
  p_{T,12}\Rb \,S_A^2\Lb k^2_T\Rb \Bigg{/}     \int d^2 k_T \,\,  e^{ - B
 k^2_T} \,\,S_A^2\Lb k^2_T\Rb \eea
 
\begin{figure}[ht]
    \centering
    \begin{tabular}{cc}
  \leavevmode
      \includegraphics[width=8cm]{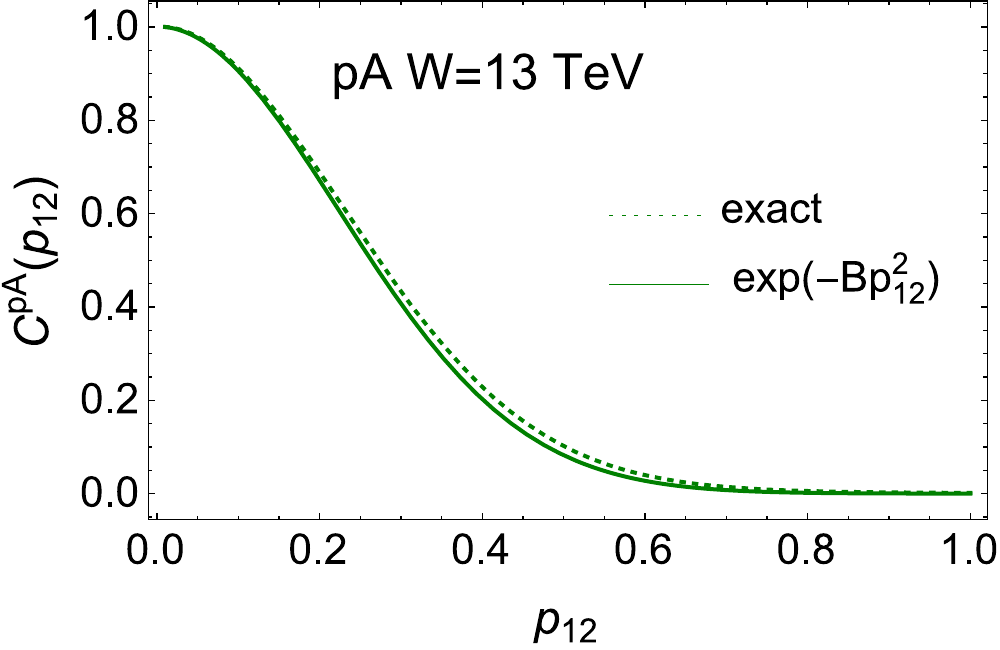}  &
 \includegraphics[width=8cm]{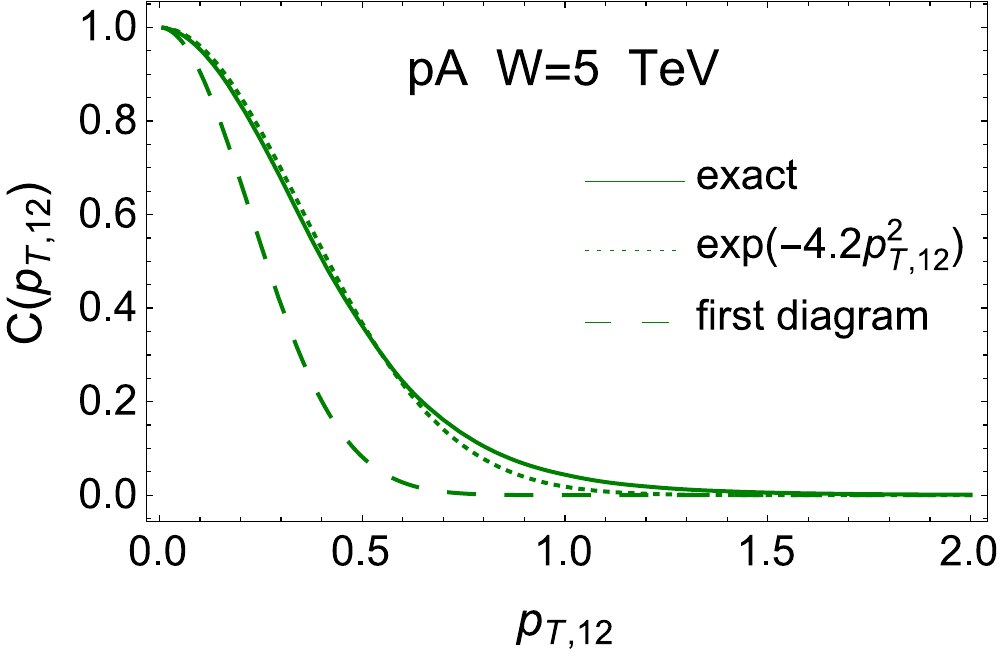}\\
      \fig{SA}-a&\fig{SA}-b\\
      \end{tabular}
           \caption{\protect\fig{SA}-a: comparison \protect\eq{1DCpA} 
  with the same equation where $ S_A^2\Lb k^2_T\Rb $ is replaced by
 $\delta\Lb k_T\Rb$. $R_A = 6.5 fm$ for  gold. $B = 10 \,GeV^{-2}$
 for proton at W =13 TeV.  \protect\fig{SA}-b: correlation function
 $C\Lb p_{T,12}\Rb$   for proton-lead scattering at W = 5 TeV in
 our model (see appendix C)   as it is defined in \eq{I2}, versus
 $p_{T,12} = |\vec{p}_{T1} - \vec{p}_{T2}|$.
      Dashed line corresponds to $\exp\Lb - B p^2_{T,12}\Rb$ with $B=4.2
 \,GeV^{-2}$, while the dotted line shows the dependence that we used in
 section 2 to calculate the first diagram: $\exp\Lb - B\,p^2_{T,12}\Rb$ 
 with $B = 10 \,GeV^{-2}$}
      \label{SA}
       \end{figure}

We expect that  $S_A\Lb k_T\Rb$ leads to small $k_T \sim 1/R_A$,   since
 the radius of nucleus is large. In \fig{SA}-a we compare  \eq{1DCpA} 
 with $\exp\Lb - B p^2_{T,12}\Rb$ which follows from \eq{1DCpA},
 replacing $ S_A^2\Lb k^2_T\Rb $  by $\delta\Lb k_T\Rb$. The agreement is
 impressive.

\begin{figure}[ht]
    \centering
    \begin{tabular}{ccc}
  \leavevmode
      \includegraphics[width=6cm]{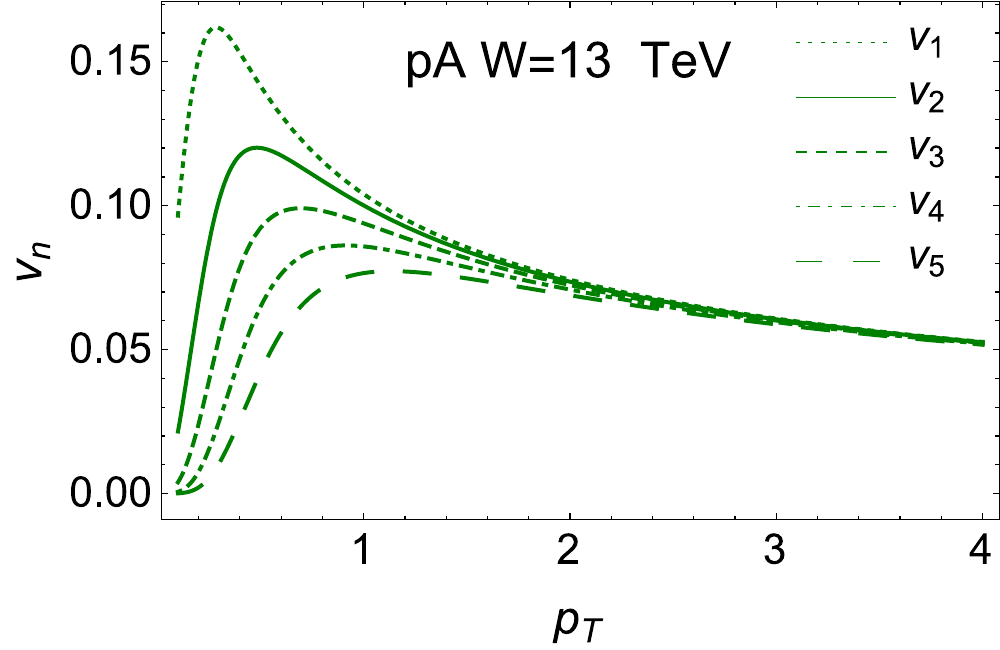} & 
 \includegraphics[width=6cm]{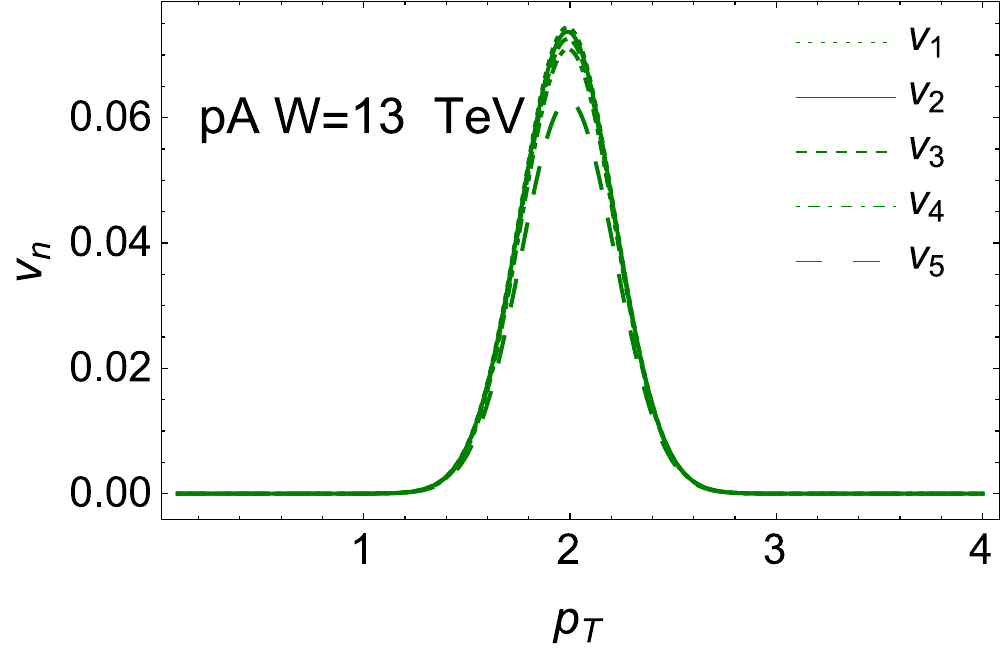} &
       \includegraphics[width=6cm]{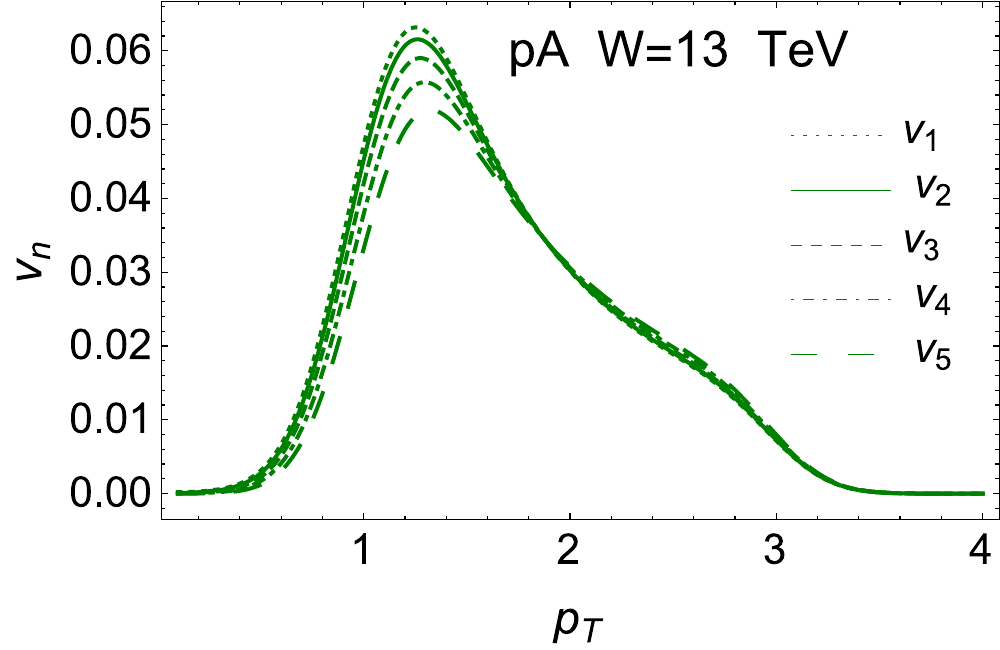}\\
       \fig{pamod}-a & \fig{pamod}-b & \fig{pamod}-c\\
          \includegraphics[width=6cm]{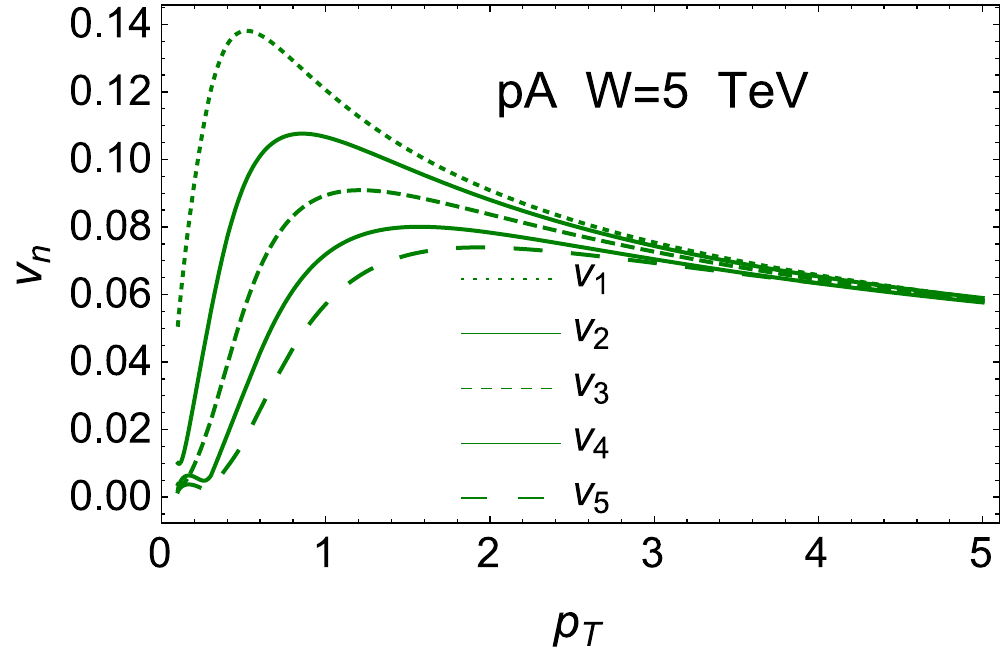} &
  \includegraphics[width=6cm]{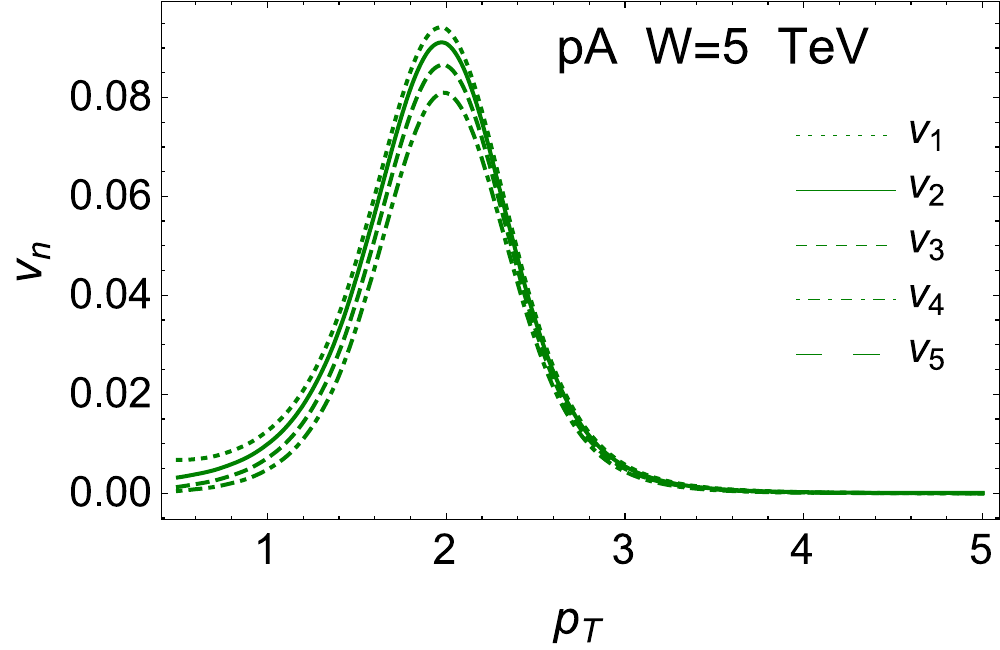} &
       \includegraphics[width=6cm]{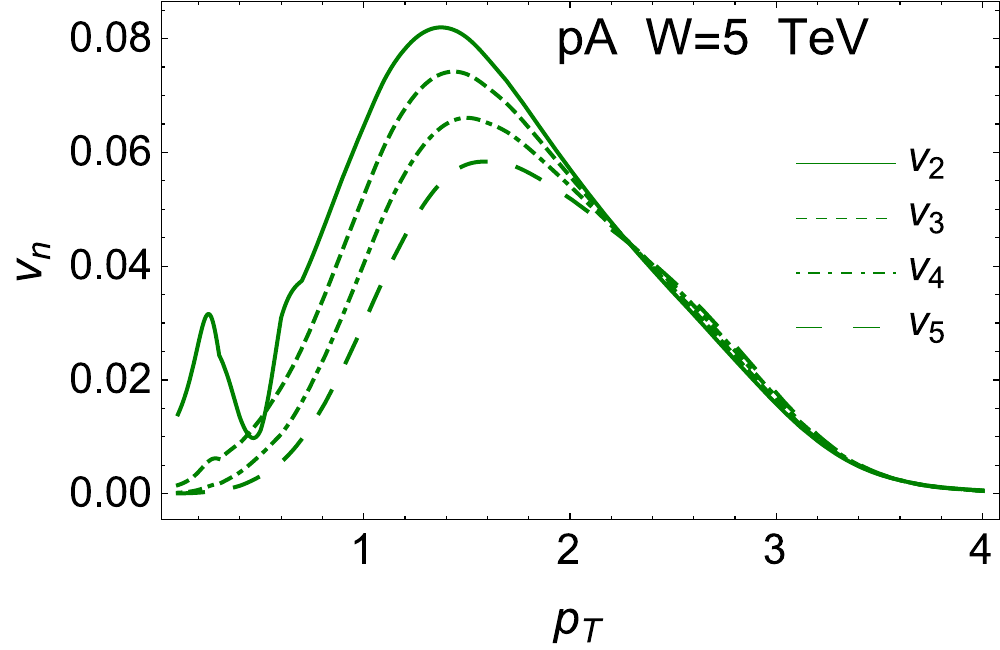}\\  
          \fig{pamod}-d & \fig{pamod}-e & \fig{pamod}-f\\            
       \end{tabular}     
           \caption{ $v_n$ versus $p_T$ for proton - gold
 (\fig{pamod}-a - \fig{pamod}-c) scattering at $ W = 13 \,TeV $
  and proton-lead  scattering at $W = 5\,TeV$ (\fig{pamod}-d -
 \fig{pamod}-c), using  \eq{VDNPP} and \eq{vn}. \fig{pamod}-a 
 and \fig{pamod}-d show $v_n$ that stem from \eq{vn}-1. 
In \fig{pamod}-b and \fig{pamod}-e  the estimates from \eq{vn}-2
 for $p^{\rm Ref}_T = 2 \,GeV$ are  plotted. \fig{pamod}-c and
 \fig{pamod}-f describe the same $v_n$ as in \fig{ppmod}-c but
 $ p^{\rm Ref}_T$ is taken in the interval $1 - 3 \,GeV$ as
 it is measured in Ref.\cite{ATLASPA}.}           
      \label{pamod}
       \end{figure}

\begin{figure}[ht]
    \centering
  \leavevmode
      \includegraphics[width=9cm]{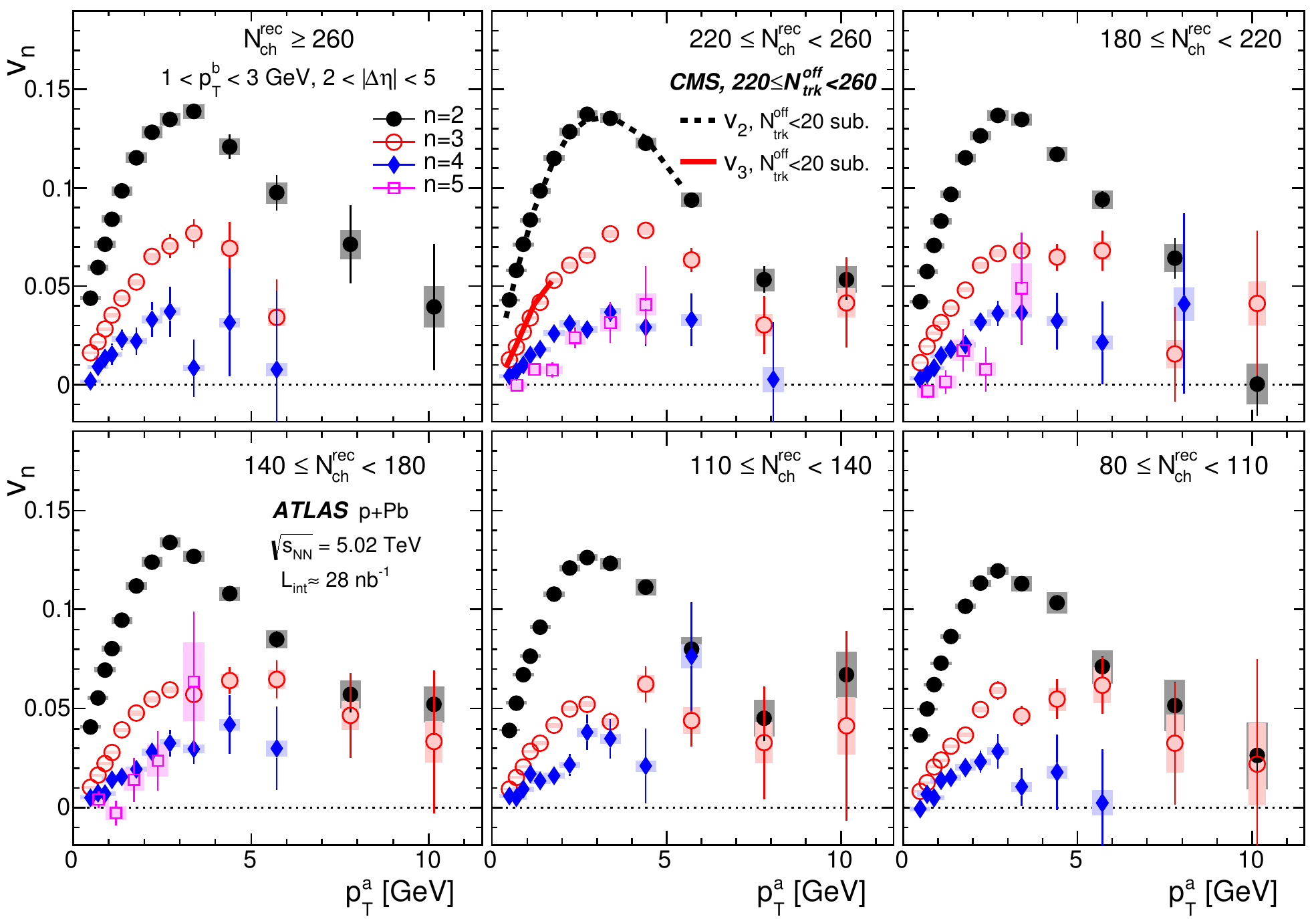}        
           \caption{ $v_n$ versus $p_T$ for proton - lead scattering
 at $W = 5\,TeV$ measured by ATLAS collaboration\cite{ATLASPP} (Fig.9
 from this paper)
        }           
      \label{paexp}
       \end{figure}

 
 In \fig{pamod} we plot the prediction for proton-gold scattering. One
 can see that the Bose-Einstein correlations generate large $v_n$ for
 $n\, \geq\,3$. Actually, we have several mechanisms (see, for example,
 review of Ref.\cite{KOLUREV}) for $v_n$ with even $n$, therefore, it
 is instructive to note that the simple estimates in this section lead
 to large $v_{2n - 1}$, larger than  has been measured \cite{ATLASPA}.
  It should be stressed that using  a more general approach which includes
 the diffractive production of small masses, as well as the shadowing
 corrections that lead to the shrinkage of diffractive peak, we obtain
 the predictions (see formulae in appendix C) which repeat the main
 features of our estimates in the simple model of \eq{1DCpA}. These
 calculations are plotted in \fig{pamod}-d - \fig{pamod}-f.
 In \fig{SA}-b  estimates for $C\Lb p_{T,12}\Rb$ in our model
 (see appendix C) are shown. One can see that  $C\Lb p_{T,12}\Rb
 $ are different, and the model gives a smaller interaction volume.
 However, all qualitative features turn out to be the same: larger
 interaction volume than for proton-proton scattering, $v_2$ is much
 smaller than the experimental value (see \fig{paexp}); $v_3$, $v_4$
 and even $v_5$ are close to the experimental values; and the value
 of the typical $p_T$ is about $1 \,GeV$ instead of $p_T = 3 - 4 \,GeV$
 in the experimental data.
 
  For nucleus-nucleus interaction $ C_{AA}\Lb R |\vec{p}_{T2} -  
 \vec{p}_{T1}|\Rb$  takes the form

   \bea \label{1DCAA}
 C_{AA}\Lb R |\vec{p}_{T2} -   \vec{p}_{T1}|\Rb\,\,&\propto&\,\,\int d^2 k_T
 \,  S_A^2  \Lb \Lb \vec{k}_T - \vec{p}_{T,12}\Rb^2\Rb \,S_A^2\Lb
 k^2_T\Rb\Bigg{/}     \int d^2 k_T \, S^2_A  \Lb \Lb k^2_T\Rb^2\Rb
 \,S_A^2\Lb k^2_T\Rb  \eea
 
\begin{figure}[ht]
    \centering
    \begin{tabular}{cccc}
  \leavevmode
      \includegraphics[width=4cm]{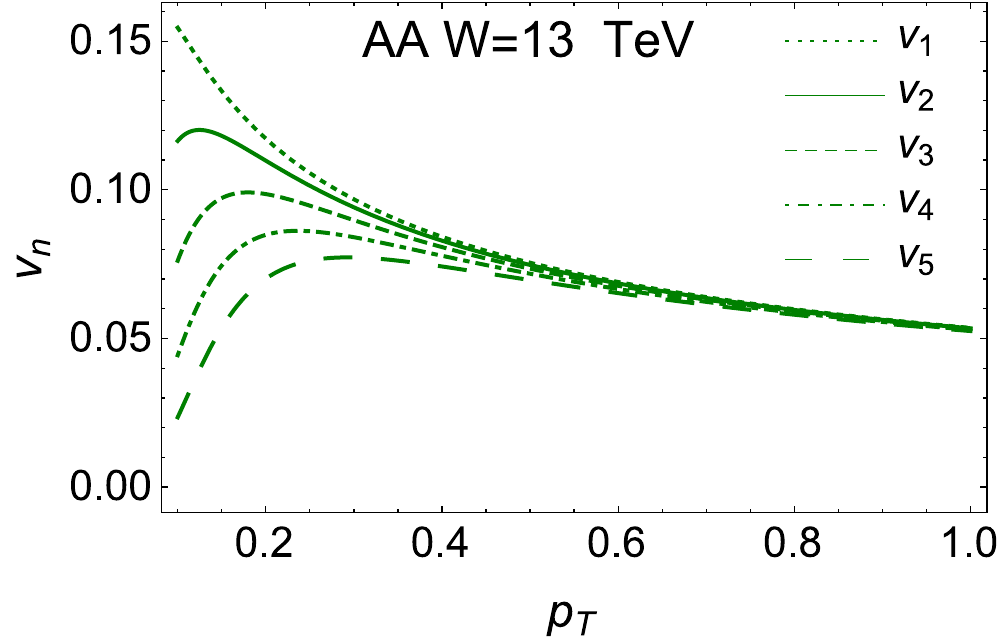} & 
 \includegraphics[width=4cm]{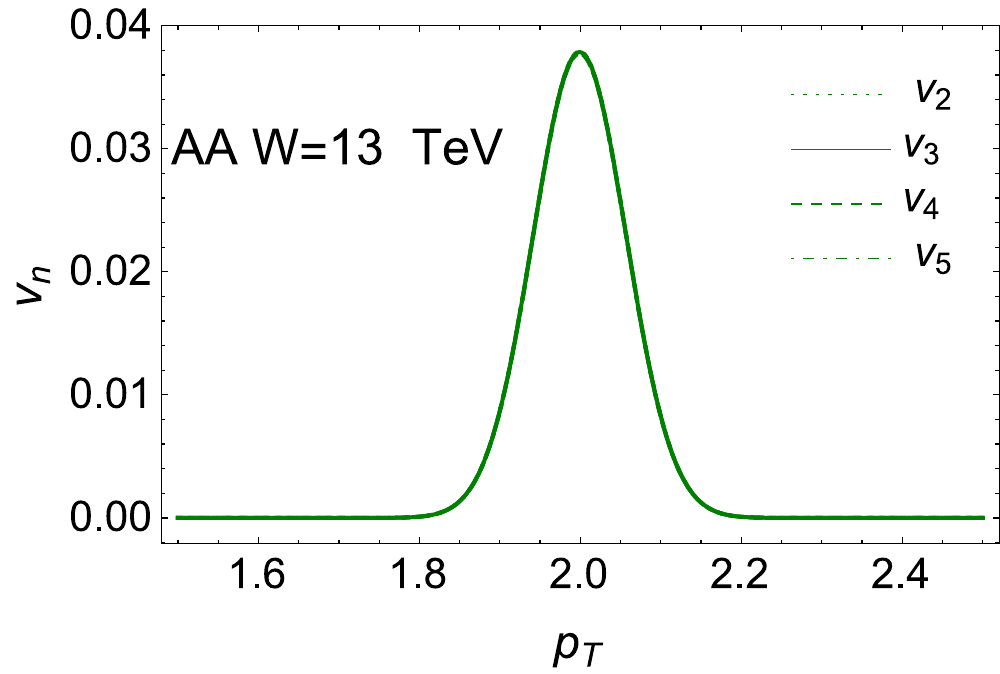} &
       \includegraphics[width=4cm]{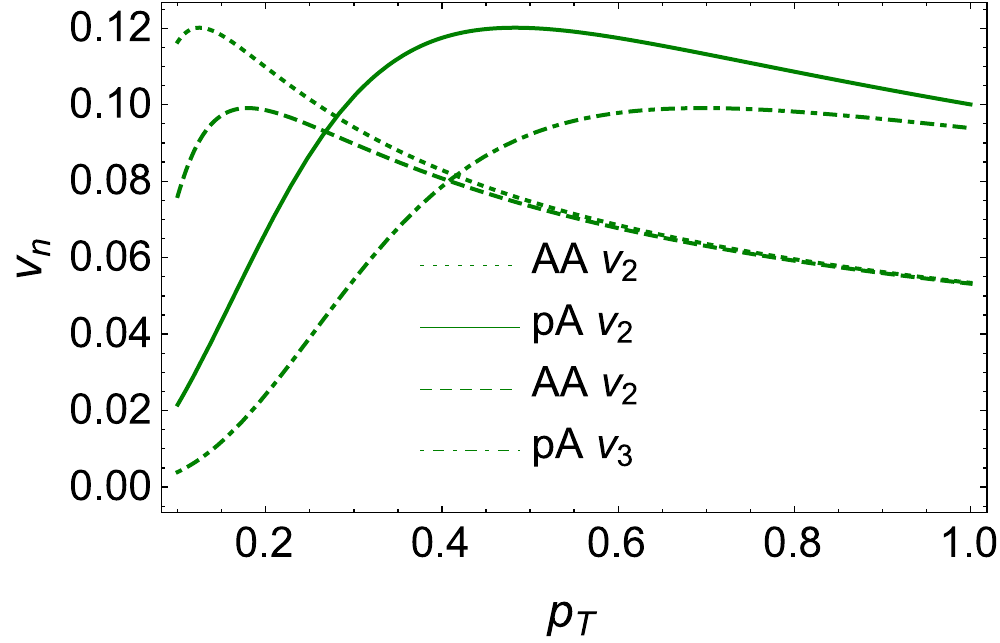}
 &\includegraphics[width=4cm]{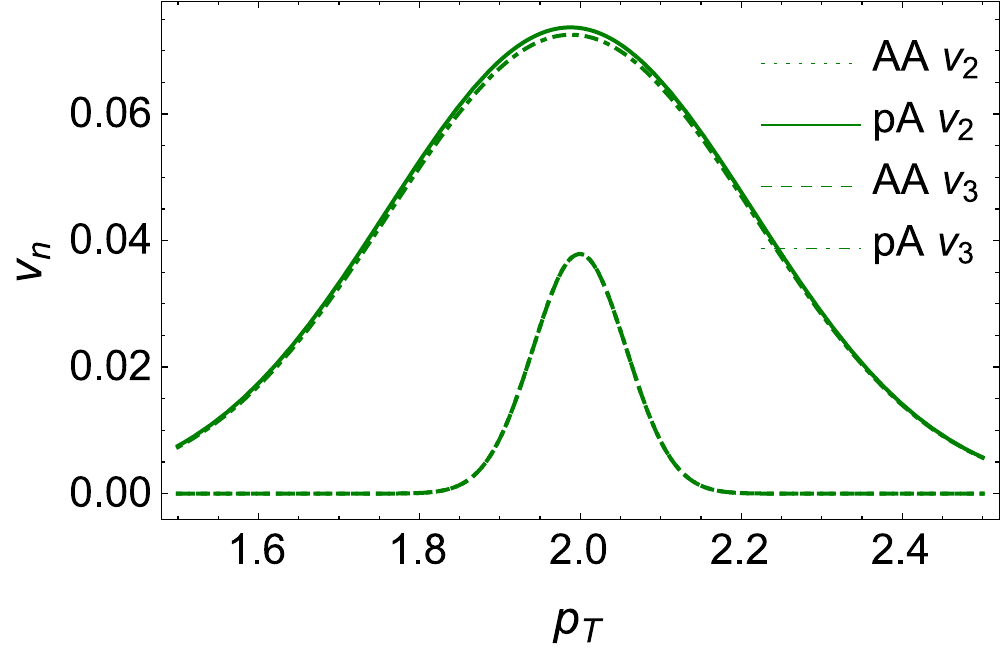}\\
       \fig{aamod}-a & \fig{aamod}-b & \fig{aamod}-c&\fig{aamod}-d\\
       \end{tabular}     
           \caption{ $v_n$ versus $p_T$ for   gold - gold
 scattering at $W = 13\,TeV$, using  \eq{VDNPP} and \eq{vn}. 
\fig{aamod}-a shows $v_n$ that stem from \eq{vn}-1. In \fig{aamod}-b
 the estimates from \eq{vn}-2 for $p^{\rm Ref}_T = 2 \,GeV$ is plotted.
\fig{aamod}c and \fig{aamod}-d describe the difference between
 proton-gold and gold-gold interactions.}           
      \label{aamod}
       \end{figure}

 \fig{aamod} shows $v_n$ for gold-gold scattering. One can see three
 major differences: $v_n$ values turns out to be smaller than for
 proton-nucleus scattering, especially when $p_{T2} = p^{\rm Ref}_T$ 
 differs from $p_{T1}$;  the momentum distribution is much narrower
 than for $p A$ scattering, and $v_n$ are the same for all $n$.
 
 Comparing \fig{ppmod}, \fig{pamod} and  \fig{aamod} we can conclude that
 the simplest estimates lead  to sufficiently large $v_n$ for both even
 and odd $n$, which are similar to those obtained in proton-proton and 
proton-nucleus
 collisions, but they are considerably smaller for the nucleus-nucleus 
case.
 Comparing these predictions with the experimental data of
 Refs.\cite{CMSPP,STARAA,PHOBOSAA,STARAA1,CMSPA,CMSAA,ALICEAA,ALICEPA,
ATLASPP,ATLASPA,ATLASAA} we see that the BE correlations should be taken
 into account in all three reactions, since they give sizable 
contributions.
\section{A  brief review of our model}
In this section we will give a brief review of our model which has
 been developed in our papers \cite{GLM2CH,GLMNI}. The advantage of
 the model is that it describes the experimental data on diffractive and
 elastic production\cite{GLM2CH}; the inclusive production \cite{GLMINCL}
  and large rapidity range (LRR) correlations \cite{GLMCOR}.

 As has been mentioned we need to build a  model which incorporates 
at 
least
 two non-perturbative phenomena: the correct large $b$ behaviour of the 
amplitude and the 
 hadron structure. These need to be incorporated so as  to reproduce 
in the framework of one approach, the main
 features of the experimental data, such as the increase of the 
interaction
 radius with energy, a sufficiently large cross section of diffraction
 production, as well as energy and multiplicity dependence of inclusive
 cross sections and two particle correlations.  On the other hand, we
 wish to include as much  information as possible from a theoretical
 approach based on QCD. 

\subsection{Theoretical input and `dressed' Pomeron Green function}
\begin{figure}[ht]
    \centering
  \leavevmode
      \includegraphics[width=16cm]{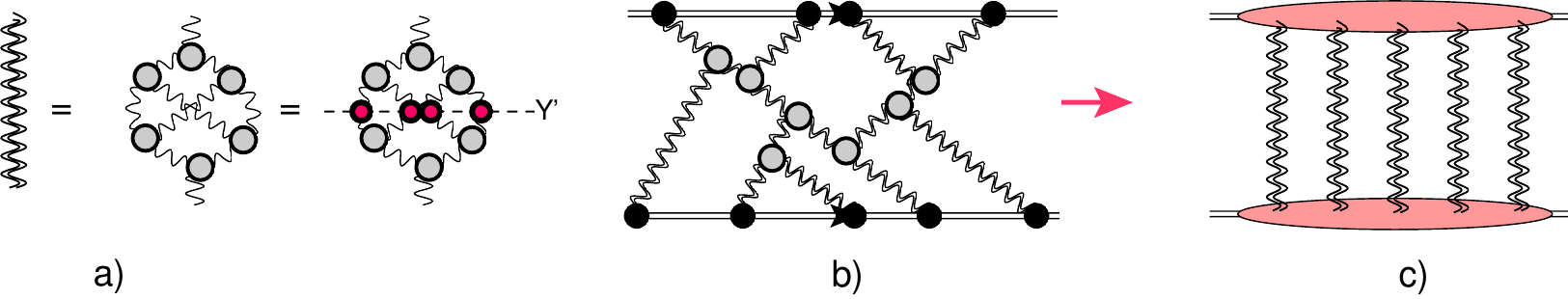}  
      \caption{\fig{amp}-a shows the set of the diagrams in the
 BFKL Pomeron calculus that produce the resulting (dressed) Green
 function of the Pomeron in the framework of high energy QCD. The red blobs
 denote the amplitude of dipole-dipole interaction at low energy.
 In \fig{amp}-b the net diagrams,    which   include
 the interaction of the BFKL Pomerons with colliding hadrons, are shown.
 The sum of the diagrams after integration
 over positions of $G_{3 \pom}$ in rapidity, reduces to \fig{amp}-c.  }
\label{amp}
   \end{figure}

  At the moment, the effective theory for QCD at high energies
 exists in two different formulations:  the CGC/saturation approach
 \cite{MV,MUCD,BK,JIMWLK}, and the BFKL Pomeron calculus 
\cite{BFKL,GLR,MUQI,MUPA,BART,BRN,KOLE,LELU,LMP,AKLL,AKLL1,LEPP}. 
 In building our model we rely on the BFKL Pomeron calculus, since
 the relation to diffractive physics is more evident in this approach.
  However, we are aware that the CGC/saturation approach gives a more 
general
 pattern\cite{AKLL,AKLL1}.  In Ref.\cite{AKLL1} it was proven that these
 two approaches are equivalent for
\beq \label{MOD1}
Y \,\leq\,\frac{2}{\Delta_{\mbox{\tiny BFKL}}}\,\ln\Lb
 \frac{1}{\Delta^2_{\mbox
{\tiny BFKL}}}\Rb
\eeq
where $\Delta_{\mbox{\tiny BFKL}}$ denotes the intercept of the BFKL 
 Pomeron. As we will see, in our model $ \Delta_{\mbox{\tiny BFKL}}\,
\approx\,0.2 - 0.25$    leading to $Y_{max} = 20 - 30$, which covers
 all accessible energies.
In addition in Ref.\cite{AKLL1} it is shown that  for such $Y$, we can
 safely use the Mueller-Patel-Salam-Iancu (MPSI) approach\cite{MPSI},
 which allows us to calculate the contribution to the resulting BFKL
 Pomeron Green function ( see \fig{amp}-a):
 \bea \label{GFMPSI}
 &&G^{\mbox{\tiny dressed}}_\pom\Lb Y, r, R; b \Rb\,\,=\\
 &&\,\,\int \prod_{i =1} d^2 r_i\, d^2 b_i\, d^2 r'_i\, d^2 b'_i\,
 N\Lb Y- Y', r, \{ r_i,b - b_i\}\Rb\,A^{\rm BA}_{\mbox{\tiny dipole-dipole}
}\Lb r_i, r'_i, \vec{b}_i - \vec{b'}_i\Rb
N\Lb  Y', R, \{ r'_i,b'_i\}\Rb\nn
\eea
where $A^{BA}_{\mbox{\tiny dipole-dipole}}$ is the dipole-dipole
 scattering amplitude in the Born approximation of perturbative
 QCD, and is shown in \fig{amp}-a by the red circles.   
 
 We need to find 
  the amplitude for the production of dipoles of size 
$r_i$ at
  impact parameters $b_i$.  This amplitude can be 
written as (see \fig{amp}-c)
    \bea \label{TI3}
&&N\Lb Y- Y', r, \{ r_i,b_i\}\Rb\,\,=\\
&&\,\,\sum^{\infty}_{n=1} \,\Lb - \,1\Rb^{n+1} \widetilde{C}_n\Lb
 \phi_0, r\Rb \prod^n_{i=1} G_\pom\Lb Y - Y';  r, r_i , b_i\Rb\,\,
=\,\,\sum^{\infty}_{n=1} \,\Lb - \,1\Rb^{n+1} \widetilde{C}_n\Lb
 \phi_0, r\Rb \prod^n_{i=1} G_\pom\Lb z - z_i\Rb\nn.
\eea
  $\widetilde{C}_n\Lb \phi_0, r\Rb  $ is shown as
 the  multi-Pomeron amplitudes ( pink ovals) in \fig{amp}-c. 
 
 The solution to the non-linear equation is of the following general form
 \beq \label{TI4}
N\Lb G_\pom\Lb \phi_0,z\Rb\Rb \,\,=\,\,\sum^{\infty}_{n=1} \,\Lb - 
\,1\Rb^{n+1} C_n\Lb \phi_0\Rb G_\pom^n\Lb \phi_0,z\Rb.
\eeq   
Comparing \eq{TI3} with \eq{TI4} we see 
\beq \label{TI5}
\widetilde{C}_n\Lb \phi_0, r\Rb\,\,\,=\,\,\,C_n\Lb \phi_0\Rb.
\eeq 
Coefficients $C_n$ can be found from the solution to the 
Balitsky-Kovchegov
 equation \cite{BK} in the saturation region
 (see Ref.\cite{LEPP}).
\beq \label{T16}
N^{\rm BK}\Lb G_\pom\Lb \phi_0,z\Rb\Rb \,\,=\,\,a\,\Lb 1
 - \exp\Lb -  G_\pom\Lb \phi_0,z\Rb\Rb\Rb\,\,+\,\,\Lb 1 - a\Rb
\frac{ G_\pom\Lb \phi_0,z\Rb}{1\,+\, G_\pom\Lb \phi_0,z\Rb},
\eeq
with $a$ = 0.65. \eq{T16} is a convenient parameterization of the
 numerical solution within accuracy better than 5\%.
Having $C_n$ we can calculate the Green function of the dressed BFKL
 Pomeron using \eq{GFMPSI} and the property of the BFKL Pomeron exchange:
\bea \label{POMTUN}
&&\frac{\as^2}{4 \pi} \,\,G_\pom\Lb Y - 0, r, R; b  \Rb\,=\,\\
&&\int d^2 r'
  d^2 b' \, d^2 r'' \,d^2 b'' \,G_\pom\Lb Y - Y', r, r', \vec{b}
 - \vec{b}^{\,'} \Rb \, \,
\,G_\pom\Lb Y' r'', R,  \vec{b}^{\,''} \Rb\,\,A^{\rm BA}_{\mbox{\tiny
 dipole-dipole}}\Lb r', r'', \vec{b''} - \vec{b'}\Rb\nn
\eea 
 
Carrying out the integrations in \eq{GFMPSI}, we obtain the Green
 function of the dressed Pomeron in the following form:
 
 \bea \label{G}
G^{\mbox{\tiny dressed}}\Lb T\Rb\,\,&=&\,\,a^2 (1 - \exp\Lb -T\Rb )  +
 2 a (1 - a)\frac{T}{1 + T} + (1 - a)^2 G\Lb T\Rb \nn\\
~~~&\mbox{with}&~~G\Lb T\Rb = 1 - \frac{1}{T} \exp\Lb \frac{1}{T}\Rb
 \Gamma\Lb 0, \frac{1}{T}\Rb
\eea
where $\Gamma\Lb s, z\Rb$ is the upper incomplete gamma function
 (see Ref.\cite{RY} formula {\bf 8.35}) and $T$ is  the BFKL Pomeron in the
 vicinity of the saturation scale
  \beq \label{T}
T\Lb r_\bot, s, b\Rb\,\,=\,\,\phi_0  \Lb r^2_\bot Q^2_s\Lb Y, b\Rb\Rb^{\bar
 \gamma}  
\eeq
  \subsection{Phenomenological assumptions and phenomenological parameters}
  The first phenomenological idea, is to fix the large impact parameter
 behaviour  by assuming that the saturation momentum depends on $b$ 
in the
 following way:
    \beq \label{QS}
Q^2_s\Lb b, Y\Rb\,\,=\,\,Q^2_{0s}\Lb b, Y_0\Rb\,e^{\lambda \,(Y - Y_0)}
\eeq
with
\beq \label{QS0}
Q^2_{0s}\Lb b, Y_0\Rb\,\,=\,\, \Lb m^2\Rb^{1 - 1/\bar \gamma}\,\Lb S\Lb b,
 m\Rb\Rb^{1/\bar{\gamma}} 
~~~~~~~S\Lb b , m \Rb \,\,=\,\,\frac{m^2}{2 \pi} e^{
 - m b}~~~\mbox{and}~~\bar \gamma\,=\,0.63
\eeq 

 We have introduced a new phenomenological parameter $m$ to
 describe the large $b$ behaviour. The $Y$ dependence  as well
 as  $r^2$ dependence, can be found from CGC/saturation approach 
\cite{KOLEB},
 since $\phi_0$ and $\lambda$ can be calculated in the leading order of
 perturbative QCD. However, since the higher order corrections turn out
 to be large \cite{HOCOR} we treat them as parameters to be fitted. $m$
 is non-perturbative parameter which determines the typical sizes of
 dipoles inside hadrons. As one can see from Table 1 from the fit 
 $m$ = 5.25 GeV, supporting our main assumption that we can
 apply the BFKL Pomeron calculus, based on perturbative QCD, to the
 soft interaction since $m \,\gg\,\mu_{soft}$ where $\mu_{soft}$ is
 the scale of soft interaction, which is of the order of the mass of
 pion or $\Lambda_{\rm QCD}$.
 
  Unfortunately, since the confinement problem is far from being 
solved, we have to assume
 a phenomenological approach for the structure of the colliding hadrons.
 We use a two channel model, which allows us to calculate the
 diffractive production in the region of small masses.
   In this model, we replace the rich structure of the 
 diffractively produced states, by a single  state with the wave 
function 
$\psi_D$, a la Good-Walker \cite{GW}.
  The observed physical 
hadronic and diffractive states are written in the form 
\beq \label{MF1}
\psi_h\,=\,\alpha\,\Psi_1+\beta\,\Psi_2\,;\,\,\,\,\,\,\,\,\,\,
\psi_D\,=\,-\beta\,\Psi_1+\alpha \,\Psi_2;~~~~~~~~~
\mbox{where}~~~~~~~ \alpha^2+\beta^2\,=\,1;
\eeq 

Functions $\psi_1$ and $\psi_2$  form a  
complete set of orthogonal
functions $\{ \psi_i \}$ which diagonalize the
interaction matrix $T$
\beq \label{GT1}
A^{i'k'}_{i,k}=<\psi_i\,\psi_k|\mathbf{T}|\psi_{i'}\,\psi_{k'}>=
A_{i,k}\,\delta_{i,i'}\,\delta_{k,k'}.
\eeq
The unitarity constraints take  the form
\beq \label{UNIT}
2\,\mbox{Im}\,A_{i,k}\left(s,b\right)=|A_{i,k}\left(s,b\right)|^2
+G^{in}_{i,k}(s,b),
\eeq
where $G^{in}_{i,k}$ denotes the contribution of all non 
diffractive inelastic processes,
i.e. it is the summed probability for these final states to be
produced in the scattering of a state $i$ off a state $k$. In \eq{UNIT} 
$\sqrt{s}=W$ denotes the energy of the colliding hadrons, and $b$ 
the 
impact  parameter.
A simple solution to \eq{UNIT} at high energies, has the eikonal form 
with an arbitrary opacity $\Omega_{ik}$, where the real 
part of the amplitude is much smaller than the imaginary part.
\beq \label{A}
A_{i,k}(s,b)=i \Lb 1 -\exp\Lb - \Omega_{i,k}(s,b)\Rb\Rb,
\eeq
\beq \label{GIN}
G^{in}_{i,k}(s,b)=1-\exp\Lb - 2\,\Omega_{i,k}(s,b)\Rb.
\eeq
\eq{GIN} implies that $P^S_{i,k}=\exp \Lb - 2\,\Omega_{i,k}(s,b) \Rb$, is 
the probability that the initial projectiles
$(i,k)$  reach the final state interaction unchanged, regardless of 
the initial state re-scatterings.
\par
  \subsection{Small parameters from the fit and the scattering amplitude}
  
 The first approach is to use the eikonal approximation for $\Omega$ in which
 \beq \label{EAPR}
 \Omega_{i,k}(r_\bot, Y - Y_0,b)\,\,=\,\int d^2 b'\,d^2 b''\,
 g_i\Lb \vec{b}'\Rb \,G^{\mbox{\tiny dressed}}\Lb T\Lb r_\bot,
 Y - Y_0, \vec{b}''\Rb\Rb\,g_k\Lb \vec{b} - \vec{b}'\ - \vec{b}''\Rb 
 \eeq 
 
 We propose a more general approach, which takes into account new
 small parameters, that come from the fit to the experimental data
 (see Table 1 and \fig{amp} for notations):
 \beq \label{NEWSP}
 G_{3\pom}\Big{/} g_i(b = 0 )\,\ll\,\,1;~~~~~~~~ m\,\gg\, m_1 
~\mbox{and}~m_2
 \eeq
 
 The second equation in \eq{NEWSP} leads to the fact that $b''$ in 
\eq{EAPR} is much
 smaller than $b$ and $ b'$,
  therefore, \eq{EAPR} can be re-written in
 a simpler form
 \bea \label{EAPR1}
 \Omega_{i,k}(r_\bot, Y - Y_0, b)\,\,&=&\,\Bigg(\int d^2 b''\,
G^{\mbox{\tiny dressed}}\Lb
 T\Lb r_\bot, Y - Y_0, \vec{b}''\Rb\Rb\Bigg)\,\int d^2 b' g_i\Lb
 \vec{b}'\Rb \,g_k\Lb
 \vec{b} - \vec{b}'\Rb \,\nn\\
 &=&\,\tilde{G}^{\mbox{\tiny dressed}}\Lb r_\bot, Y - Y_0\Rb\,\,
\int d^2 b' g_i\Lb
 \vec{b}'\Rb \,g_k\Lb \vec{b} - \vec{b}'\Rb 
\eea

Using the first small parameter of \eq{NEWSP}, we can see 
 that the main contribution stems from the net diagrams shown in \fig{amp}-b.
 The sum of these diagrams\cite{GLM2CH} leads to the following expression 
for $
 \Omega_{i,k}(s,b)$
 \bea \label{OMEGA}
\Omega\Lb r_\bot,  Y-Y_0; b\Rb~~&=& ~~ \int d^2 b'\,
\,\,\,\frac{ g_i\Lb\vec {b}'\Rb\,g_k\Lb\vec{b} -
 \vec{b}'\Rb\,\tilde{G}^{\mbox{\tiny dressed}}\Lb r_\bot, Y - Y_0\Rb
}
{1\,+\,G_{3\pom}\,\tilde{G}^{\mbox{\tiny dressed}}\Lb r_\bot, Y - Y_0\Rb\left[
g_i\Lb\vec{b}'\Rb + g_k\Lb\vec{b} - \vec{b}'\Rb\right]} ;\label{OM}\\
g_i\Lb b \Rb~~&=&~~g_i \,S_p\Lb b; m_i \Rb ;\label{g}
\eea
where
\beq \label{SB}
S_p\Lb b,m_i\Rb\,=\,\frac{1}{4 \pi} m^3_i \,b \,K_1\Lb m_i b \Rb
\eeq
\beq \label{GTILDE}
\tilde{G}^{\mbox{\tiny dressed}}\Lb r_\bot, Y -Y_0\Rb\,\,=\,\,\int d^2 b
 \,\,G^{\mbox{\tiny dressed}}\Lb T\Lb r_\bot, Y - Y_0, b\Rb\Rb
 \eeq
where $ T\Lb r_\bot, Y - Y_0, b\Rb$ is given by \eq{T}.

Note  that  $\tilde{G}^{\mbox{\tiny dressed}}\Lb Y - Y_0\Rb$ does not 
depend
 on $b$.  In all previous formulae, the value of the triple BFKL Pomeron
 vertex
 is known: $G_{3 \pom} = 1.29\,GeV^{-1}$.

\begin{table}[h]
\begin{tabular}{|l|l|l|l|l|l|l|l|l|l|}
\hline
model &$\lambda $ & $\phi_0$ ($GeV^{-2}$) &$g_1$ ($GeV^{-1}$)&$g_2$
 ($GeV^{-1}$)& $m(GeV)$ &$m_1(GeV)$& $m_2(GeV)$ & $\beta$& $a_{\pom
 \pom}$\\
\hline
 2 channel & 0.38& 0.0019 & 110.2&  11.2 & 5.25&0.92& 1.9 & 0.58 &0.21 \\
\hline
\end{tabular}
\caption{Fitted parameters of the model. The values are taken 
from Ref.\cite{GLM2CH}.}
\label{t1}
\end{table}

To simplify further discussion, we introduce the notation 

 \beq \label{NBK}
N^{BK}\Lb G^i_\pom\Lb r_\bot, Y,b \Rb\Rb \,\,=\,\,a\,\Lb 1
 - \exp\Lb -  G^i_\pom\Lb r_\bot, Y, b\Rb\Rb\Rb\,\,+\,\,\Lb 1 - a\Rb
\frac{ G^i_\pom\Lb  r_\bot, Y, b\Rb}{1\,+\, G^i_\pom\Lb r_\bot, Y, b\Rb},
\eeq 
 with $a = 0.65$ .
 \eq{NBK} is an analytical approximation to the numerical solution for  
the 
BK equation\cite{LEPP}. $G^{i}_\pom\Lb  r_\bot, Y; b\Rb \,=\,\,
 g_i\Lb b \Rb \,\tilde{G}^{\mbox{\tiny dressed}}\Lb r_\bot, Y - Y_0\Rb $.
 We recall that the BK equation sums the `fan'  diagrams.
 
 For the  elastic amplitude we have

\beq \label{EL}
a_{el}(b)\,=\,\Lb \alpha^4 A_{1,1}\,
+\,2 \alpha^2\,\beta^2\,A_{1,2}\,+\,\beta^4 A_{2,2}\Rb. 
\eeq 
 We will discuss   the inclusive production as well as LRR correlations 
 in appendix B.

 
 \section{Azimuthal angle correlation and the structure of the `dressed'
 Pomeron}
 As has been discussed, our model includes three dimensional scales:
 $m$,$m_1$ and $m_2$.
 $m_1$ and $m_2$ describe two typical sizes in the proton wave function,
 which could be associated with the distance between constituent quarks
 (size of proton) $R_p \sim 1/m_1$  and the size of the constituent quark
 $R_q \sim 1/m_2$ in the framework of the constituent quark model \cite{CQM}.
 The third scale: $m$, characterizes the impact parameter behaviour of the
 saturation scale, and is  intimately related to the structure of the 
dressed
 Pomeron in our model. In section 2 we discussed how two scales in the proton
 wave function  arise  in the BE correlations.
 Here, we would like  to show that the third scale leads to the BE 
correlations
 which can explain the values of $v_n$ observed experimentally. 
 
\begin{figure}[ht]
    \centering
  \leavevmode
      \includegraphics[width=17cm]{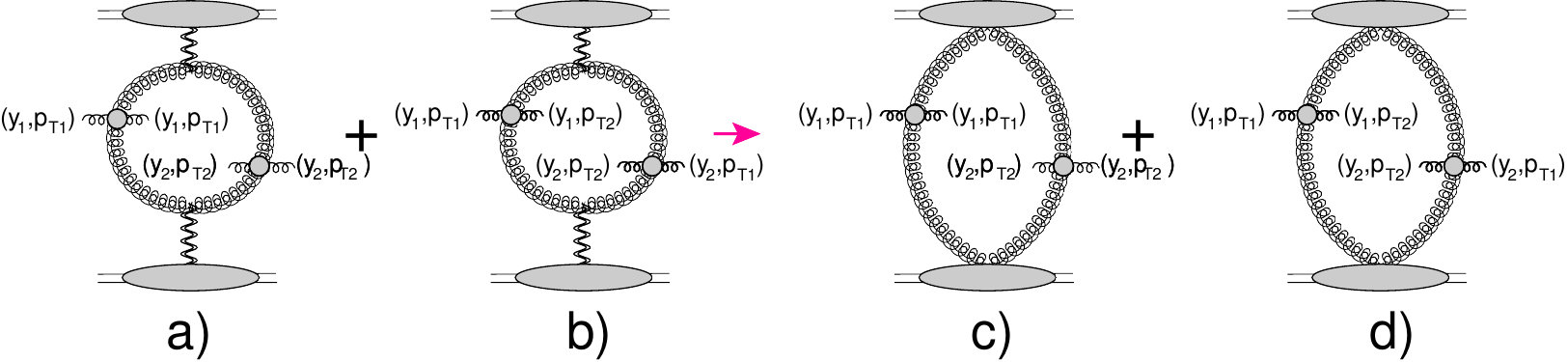}  
           \caption{ The Mueller diagrams for the BE correlation for
 the 'dressed' Pomeron. A blob denotes the vertex for gluon emission
 $a_{\pom \pom}$ (see \eq{PS2}).}           
      \label{enhdi}
       \end{figure}

 As we have discussed in section 3-A , the dressed Pomeron is the sum
 of enhanced diagrams (see \fig{amp}-a) which is given by \eq{G}. Therefore,
  the exchange of the dressed Pomeron generates the production of an 
infinite number
 of the parton showers and, in particular, two parton showers which generate
 the BE correlations as is 
 shown in \fig{enhdi}. Integration over rapidities of triple Pomeron
 vertices\cite{LMP} reduces the diagrams of \fig{enhdi}-a and \fig{enhdi}-b
 to the diagrams of \fig{enhdi}-c and \fig{enhdi}-d. We can calculate the
 probability to find two parton showers ($P_2$) inside of the dressed
 Pomeron expanding \eq{G}:
 \beq \label{PS1}
 P_2\,=\,(2 - 2 a + a^2/3)\,=\,0.91\,\,\,\,\,\,\mbox{for}\,\,\,\,\,\, a =0.65
 \eeq
 and the contribution of two parton showers production to the double
 inclusive cross section for the diagrams of \fig{enhdi}-a,
 is equal to
 \beq \label{PS2}
    \frac{d^2 \sigma}{d y_1 \,d y_2  d^2 p_{T1} d^2 p_{T2}}\,\,=\,\,a^2_{\pom
 \pom}\,P^2_2 \,\int d^2 k_T \, \, T\Lb k_T, Y- y_1\Rb \,T\Lb k_T,
  Y- y_2\Rb\, T\Lb k_T,  y_1\Rb \,T\Lb k_T,\  y_2\Rb   
\eeq
where $a_{\pom \pom}$ denotes the   Mueller vertex of gluon emission 
 (see \fig{enhdi}).  In our estimates 
 for the calculation of $v_n$, we do not need to know the probability
 $P_2$,  as well as the vertex $a_{\pom \pom}$, assuming that $a_{\pom
 \pom}$ is the same in \fig{enhdi}-a and in \fig{enhdi}-b. In \eq{PS2}
 all rapidities are in the  laboratory frame.

$T\Lb k_T, y\Rb$ is the Fourier image of $T\Lb b, y\Rb$ defined in
 \eq{T}-\eq{QS0} and it takes the form
\beq \label{PS3}
 T\Lb  k_T,  y \Rb \,\,=\,\,\phi_0\,\frac{1}{\Lb 1 + 
\frac{k^2_T}{m^2}\Rb^{3/2}}\,e^{\lambda \,\bar{\gamma}\Lb Y - Y_0\Rb}\
 \eeq
 
\begin{figure}[ht]
    \centering
  \leavevmode
  \includegraphics[width= 6cm]{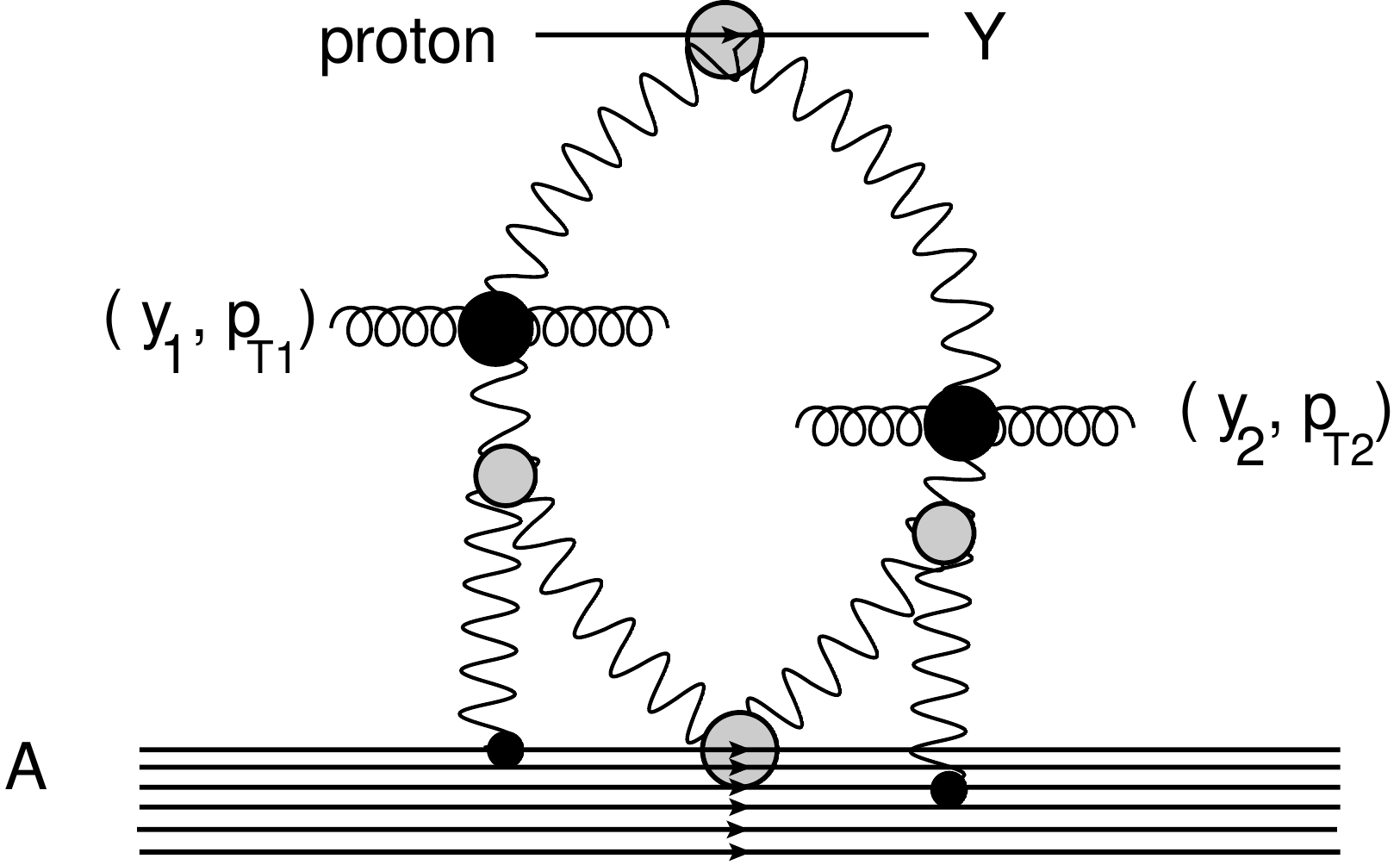}
  \caption{The Mueller diagrams for the BE correlation for the
 'dressed' Pomeron for proton-nucleus scattering. Black blob
 denotes the vertex for gluon emission $a_{\pom \pom}$ (see Eq.
 (41), the gray blob stands for the triple Pomeron vertex. }           
      \label{enhdipA}
       \end{figure}

\begin{figure}[ht]
    \centering
  \leavevmode
  \includegraphics[width= 11cm]{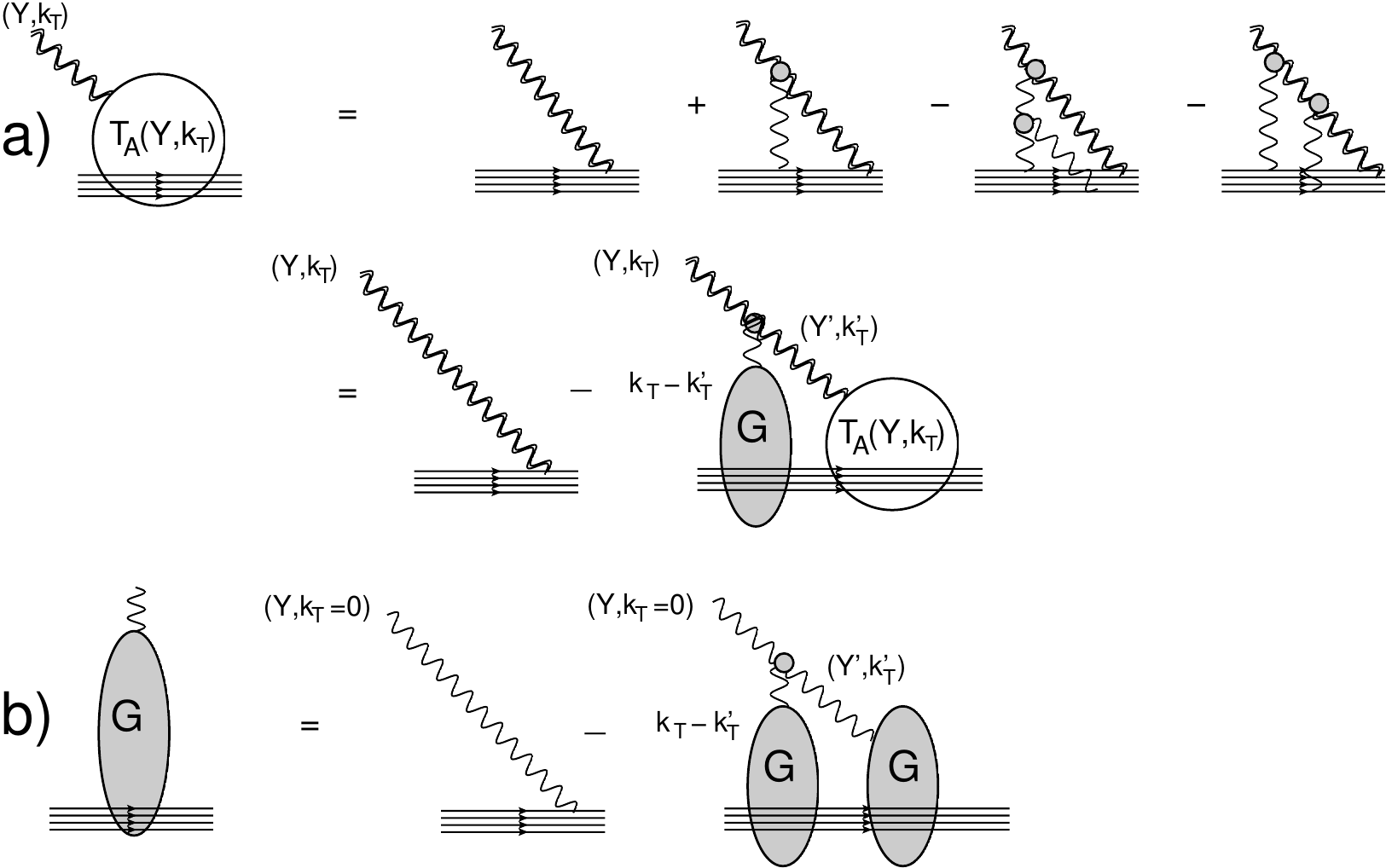}\\
  \caption{ Graphic form of equation for $T_A\Lb y, k_T\Rb$.
 The wavy double lines denote $T\Lb Y,k_T\Rb$ of \eq{PS3},
 while the wavy lines stand for $T_A \Lb y, k_T=0\Rb$. }           
      \label{enhdiA}
       \end{figure}

  For the interaction with nuclei, we need to take into account the 
interaction of
 the Pomeron with the nucleons inside the nucleus, as it is shown in
 \fig{enhdipA}. The equation for the resulting $T_A\Lb y, k_T\Rb$
 takes the form (see \fig{enhdiA}-a)
  \beq \label{PS4}
  T_A\Lb y, k_T\Rb  \,=\,  T\Lb y, k_T\Rb \,\,-\,\,\Gamma_{3 
\pom}\int^y_0\,d y'\,T\Lb y - y', k_T\Rb \int d^2 k'_T\, G_A\Lb y' \vec{k}
 - \vec{k}'\Rb \,T_A\Lb y', k'_T \Rb
  \eeq
 The triple Pomeron vertex  $\Gamma_{3\pom}$ will be
 calculated in our model below.

  The typical $| \vec{k} - \vec{k}'| \sim 1/R_A \,\ll\,1/m$ and, therefore,
 we can replace $  G_A\Lb y', \vec{k} - \vec{k}'\Rb \,$ by
  $\widetilde{G}_A\Lb y' \Rb \delta^{(2)}\Lb \vec{k} - \vec{k}'  \Rb$.  
Note
 that the normalization is such that the first diagram for
 $ \widetilde{G}_A = S_A\Lb b = 0\Rb  T\Lb y,k_T=0\Rb$,
 where $S_A\Lb b \Rb$ is defined in \eq{WS}.  After integration
 over $k'_T$,  \eq{PS4}
 reduces to the following equation
  
  \beq \label{PS5}
   T_A\Lb y, k_T\Rb  \,=\,  T\Lb y, k_T\Rb \,\,-\,\,\Gamma_{3 
\pom}\int^y_0\,d y'\,T\Lb y - y', k_T\Rb  \widetilde{G}_A\Lb y'
 \Rb \,T_A\Lb y', k'_T \Rb
  \eeq  
    For $\widetilde{G}_A$ we have the equation of \fig{enhdiA}-b, 
 which has the following analytical form:
  \beq \label{PS6}
  \widetilde{G}_A\Lb y \Rb  \,\,\,\,=\,\,\,S_A\Lb b=0\Rb \,T\Lb y, k_T=0 \Rb
  \,-\,\Gamma_{3 \pom}\int^y_0\,d y'\,\,T \Lb y - y', k_T\Rb 
 \widetilde{G}^2_A\Lb y' \Rb 
  \eeq  
  
      The solution to these two equations (\eq{PS5} and \eq{PS6}
 can be written as follows
      \beq \label{PS7} 
 T_A\Lb y,k_T\Rb\,\,=\,\,\frac{T\Lb y,k_T\Rb}{1 + \tilde{\Gamma}_{3 
\pom} \,S_A\Lb b=0 \Rb\,T\Lb y,k_T\Rb}
  \eeq
  where $\tilde{\Gamma}_{3 \pom} = \Gamma_{3 \pom}/\Lb \lambda
 \,\bar{\gamma}\Rb \,=\,P_2$.
  
   $T\Lb y, k_T\Rb$ has a physical meaning, of
 the BFKL amplitude in the vicinity of the saturation scale, where it has
 a geometric scaling behaviour \cite{GS}, and it depends on one variable 
$z
 = \ln\Lb r^2 Q^2_s(Y)\Rb$. For diagrams of \fig{enhdi} the typical 
  $r \sim 1/m_i $ and $z \to \lambda Y$.  It is well known that the main 
 contribution to the inclusive cross section stems from vicinity of the 
 saturation scale, since this cross section is proportional to $\nabla^2_r 
 N \Lb r, b; Y\Rb$, which tends to zero inside the saturation domain (see 
 \eq{A7}).  $N$ is the scattering amplitude of the dipole with size $r$. 
 The fact that we are dealing with the amplitude in the region where it 
 has geometric scaling behaviour, is the reason why a non-linear equation
of the BK type \cite{BK} is degenerate to one dimensional equations 
 (see \eq{PS5}
   - \eq{PS7}).
  
    Using \eq{PS2} we can calculate $C\Lb | \vec{p}_{T1} -
 \vec{p}_{T2}|\Rb$ for proton-proton scattering, given by \eq{I1}
 which is equal to
    \beq \label{PS8}
C_{\rm pp}\Lb | \vec{p}_{T1} - \vec{p}_{T2}|\Rb\,\,= \frac{1}{N^2_c - 1}
 \frac{    \,\int d^2 k_T \, \, T\Lb k_T, Y- y_1\Rb \,T\Lb k_T, 
 Y- y_2\Rb\, T\Lb \vec{k}_T  \,-\,\vec{p}_{T,12},  y_1\Rb \,T\Lb
  \vec{k}_T  \,-\,\vec{p}_{T,12},\  y_2\Rb  }{ \,\int d^2 k_T \,
 \, T\Lb k_T, Y- y_1\Rb \,T\Lb k_T,  Y- y_2\Rb\, T\Lb k_T,  y_1\Rb
 \,T\Lb k_T,\  y_2\Rb  } 
\eeq    

\begin{figure}[ht]
    \centering
  \leavevmode
  \includegraphics[width= 10cm]{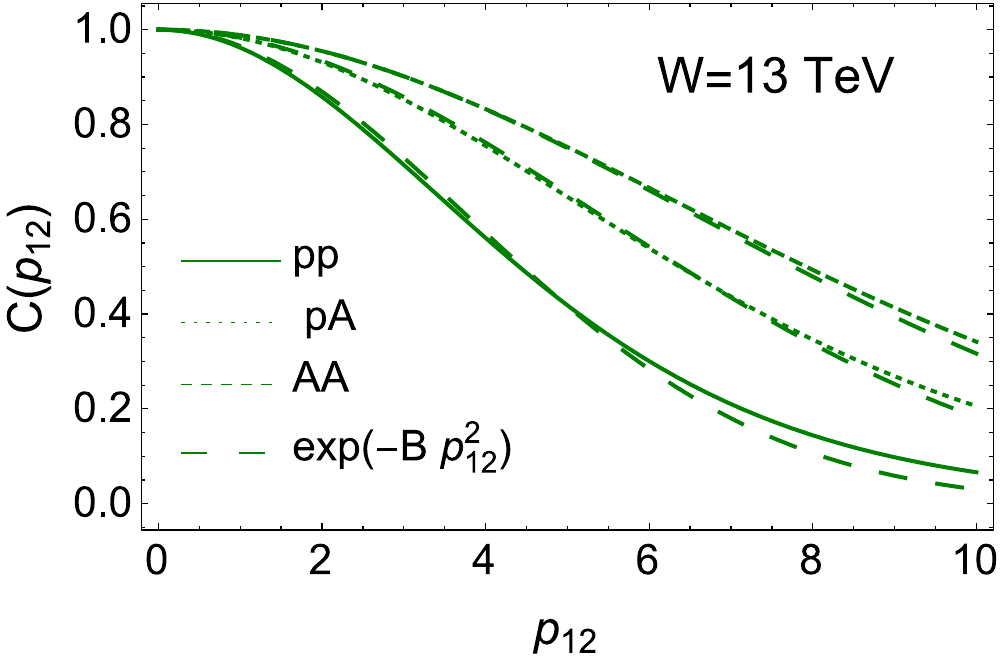}\\
  \caption{$ C\Lb | \vec{p}_{T1} - \vec{p}_{T2}| = p_{T,12}\Rb$,
 calculated using \eq{PS8} - \eq{PS10},  versus $p_{T,12}$ for
 three reactions: proton-proton, proton - lead  and lead - lead
 collisions at energy $W = 13\,TeV$. The long dashed curves
 correspond to  $\exp\Lb - B\, p^2_{T,12}\Rb$ with $B_{\rm pp}
 =0.035\,GeV^{-2}$, $B_{\rm pA} = 0.027\,Ge4V^{-2}$ and $B_{\rm AA} =
 0.022\,GeV^{-2}$.}           
      \label{C}
       \end{figure}

For proton-nucleus we have
    \beq \label{PS9}
C_{\rm pA}\Lb | \vec{p}_{T1} - \vec{p}_{T2}|\Rb\,\,= \frac{1}{N^2_c - 1}
 \frac{    \,\int d^2 k_T \, \, T_A\Lb k_T, Y- y_1\Rb \,T_A\Lb k_T, 
 Y- y_2\Rb\, T\Lb \vec{k}_T  \,-\,\vec{p}_{T,12},  y_1\Rb \,T\Lb 
 \vec{k}_T  \,-\,\vec{p}_{T,12},\  y_2\Rb  }{ \,\int d^2 k_T \, \,
 T\Lb k_T, Y- y_1\Rb \,T\Lb k_T,  Y- y_2\Rb\, T\Lb k_T,  y_1\Rb \,T\Lb k_T,\
  y_2\Rb  } 
\eeq
   and for nucleus - nucleus $C_{\rm AA}$ has the form:
      \beq \label{PS10}
C_{\rm AA}\Lb | \vec{p}_{T1} - \vec{p}_{T2}|\Rb\,\,= \frac{1}{N^2_c - 1}
 \frac{    \,\int d^2 k_T \, \, T_A\Lb k_T, Y- y_1\Rb \,T_A\Lb k_T,
  Y- y_2\Rb\, T_A\Lb \vec{k}_T  \,-\,\vec{p}_{T,12},  y_1\Rb \,T_A\Lb 
 \vec{k}_T  \,-\,\vec{p}_{T,12},\  y_2\Rb  }{ \,\int d^2 k_T \, \,
 T\Lb k_T, Y- y_1\Rb \,T\Lb k_T,  Y- y_2\Rb\, T\Lb k_T,  y_1\Rb \,T\Lb k_T,\
  y_2\Rb  } 
\eeq   
 The results of calculations for $C\Lb R_{\rm cor} p_{T,12}\Rb$ using
 \eq{PS8} - \eq{PS10} are plotted in \fig{C}, One can see that the radius
 of correlations  ($R_{\rm cor}^2 =B$)   turns out to be very small in
 comparison with the same radius in \fig{Cpp} and \fig{SA}. 
 
From $C\Lb R_{\rm cor} p_{T,12}\Rb$ we can calculate $v_n$ using \eq{I3}
 and \eq{vn}-1.
However,   $C\Lb R_{\rm cor} p_{T,12}\Rb$ shown in \fig{C}, are calculated
 for the production of gluon jets, while  experimentally $v_n$ are
 measured for a hadron. Following Ref.
\cite{LEREINC}  we explore the local parton-hadron duality(LPHD)
 suggested in Ref. \cite{LPHD}.

\begin{figure}[ht]
    \centering
    \begin{tabular}{ccc}
  \leavevmode
      \includegraphics[width=6cm]{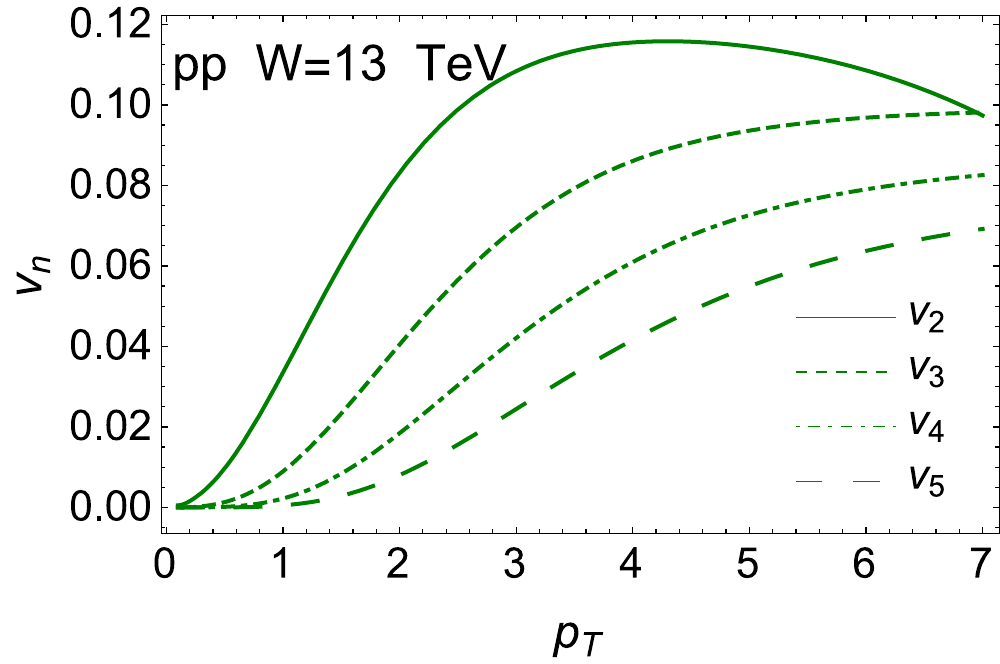} & 
 \includegraphics[width=6cm]{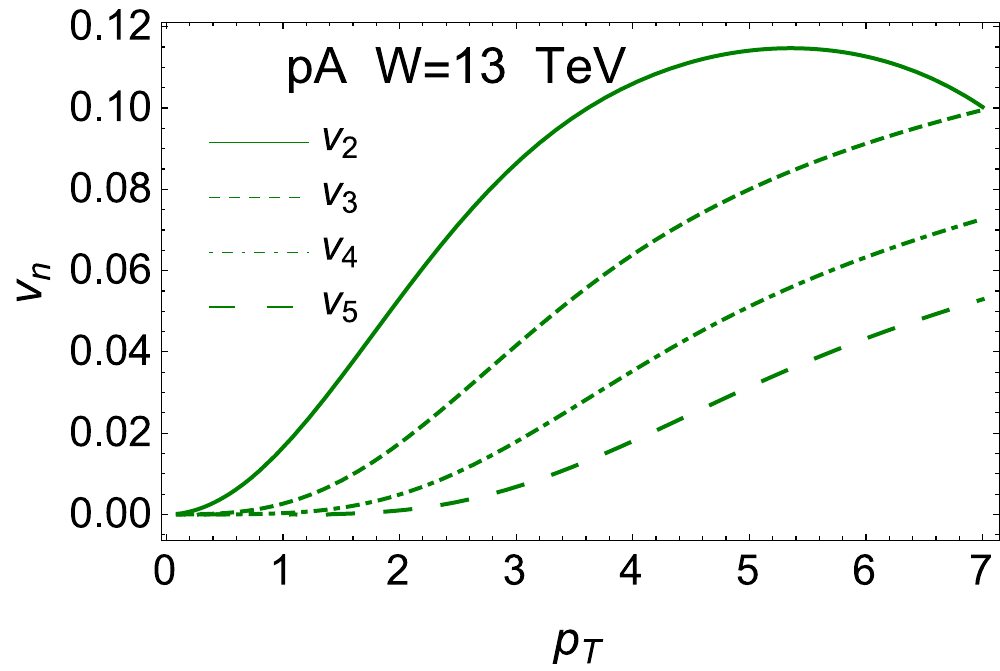} &
       \includegraphics[width=6cm]{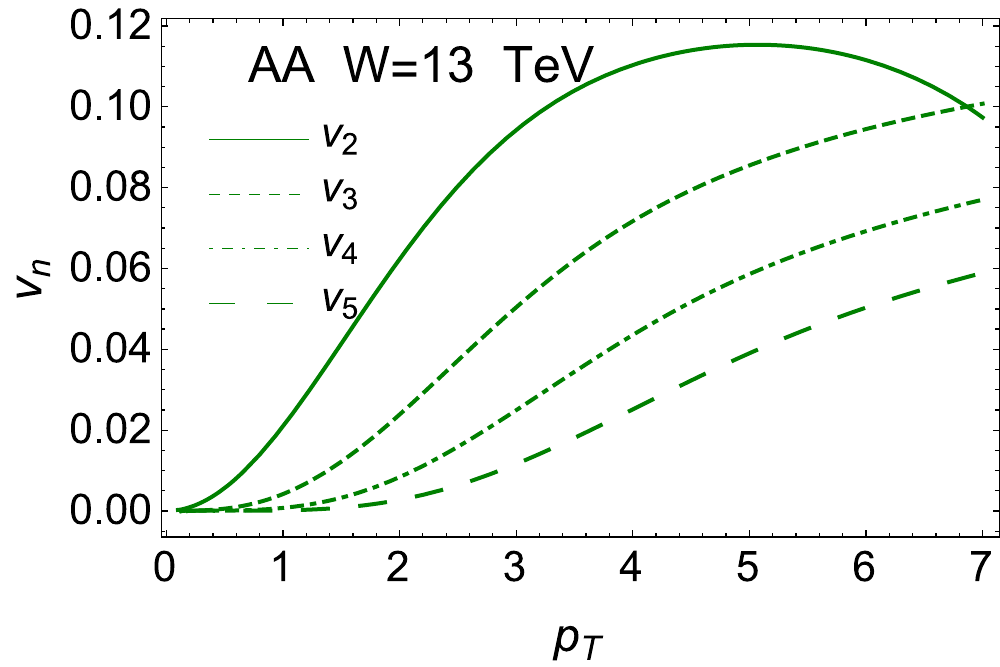}\\
       \fig{had}-a & \fig{had}-b & \fig{had}-c\\
       \end{tabular}     
           \caption{ $v_n$ versus $p_T$  at $W = 13\,TeV$ for proton-proton
 (\fig{had}-a), proton-lead (\fig{had}-b) and lead-lead (\fig{had}-c)
 scatterings, using  \eq{I3} and \eq{vn}-1. }           
      \label{had}
       \end{figure}

 In our approach the hadrons originate 
 from the decay of a gluon jet, and their transverse momenta are  
\beq \label{PS11}
\vec{p}_{\rm hadron,\, T}\,=\,z\,\vec{p}_{\rm jet,\,T}\,\,+\,\,
\vec{p}_{\rm intristic,\,T}
\eeq

where $z$ is the fraction of energy of the jet,  carried by the hadron.
 $\vec{p}_{\rm intristic,\,T}$   is the transverse momentum of the
 hadron in the mini-jet that has only longitudinal momentum.
 From \eq{PS11} we obtain that the average $p_T$ of hadrons is equal to
\beq \label{PS12}
\langle p_{\rm hadron,\, T}\rangle\,\,=\,\,\sqrt{z^2\,p^2_{\rm 
jet,\,T}\,\,+\,\,p^2_{\rm intristic,\,T}}
\eeq
In Ref.\cite{LEREINC} we  found that we need to take $z =0.5$
 and $p_{\rm intristic,\,T} = m_{\pi}$,  to describe the inclusive
 spectra of hadron at the LHC. Using \eq{PS12} we recalculate
 $v_n$ for a gluon jet to $v_n$ for hadrons, which are shown in
 \fig{had}. Comparing with the experimental data
  \cite{CMSPP,STARAA,PHOBOSAA,STARAA1,CMSPA,CMSAA,ALICEAA,ALICEPA,
ATLASPP,ATLASPA,ATLASAA}, and \fig{CppW}-b  and \fig{paexp},  one can
 see that we describe  the proton-proton scattering rather well,
 while for proton-nucleus we obtain $v_2$ which is smaller by 15- 20\%.
  For lead-lead collisions $v_2$ turns out to be two times smaller than
 the experimental value \cite{ATLASAA}. However, for central events
 with centrality  0 - 10\% measured $v_2$ is very close to our estimates.
 $v_n$ with $n \geq 3$ are larger than the experimental values.
  In general the $p_T$ distribution is wider than the experimental one.
 The LPHD approach and \eq{PS12} are very approximate, and we need to use 
a
 more advanced jet fragmentation function. Second, we need to add together 
the two
 mechanisms:  one discussed in this section and one discussed  in 
section 2.
 We need to include
a more advanced fragmentation function, 
   together with  more careful accounting
 of the emission vertex in QCD (see appendix A). We will
  consider  these in a future publication.

\begin{figure}[ht]
    \centering
    \begin{tabular}{cc}
  \leavevmode
      \includegraphics[width=7cm]{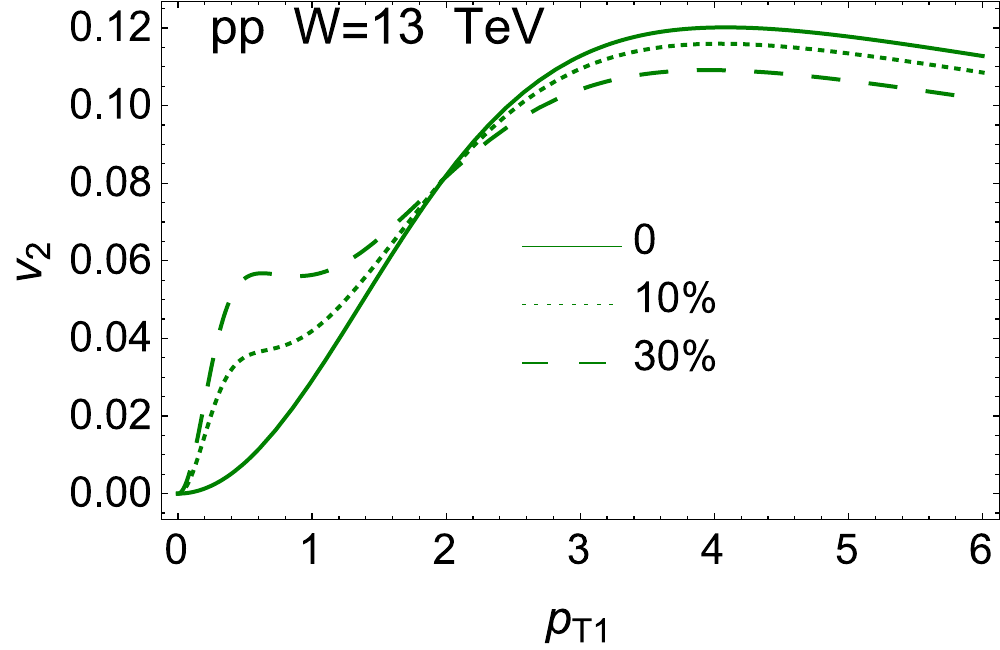}& 
 \includegraphics[width=7cm]{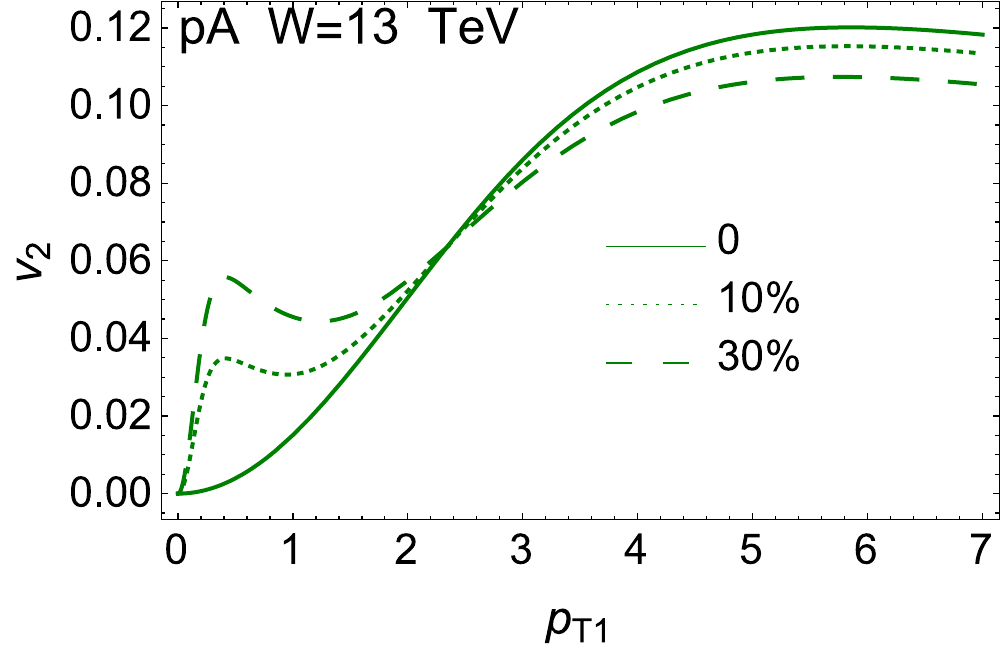}\\
       \fig{mix}-a & \fig{mix}-b\\
       \end{tabular}     
           \caption{ $v_2$ versus $p_T$  at $W = 13\,TeV$ for
 proton-proton (\fig{mix}-a)  and proton-lead (\fig{mix}-b)
 for the sum of two contributions: the `dressed' Pomeron structure
 and the diffractive production,  discussed in section 2. The percents
 indicate  the fraction of  diffractive production in the Pomeron 
structure.
   }           
      \label{mix}
       \end{figure}

 The estimates from our model show that the  mechanism 
 that has been discussed in section 2, yields about 10 -20\%
 of the contribution which we now consider.
 In \fig{mix} one can see how the sum of two mechanism occur
  in $v_2$.  One can see that the sum has a wider $p_T$  distribution 
and a smaller maximal value.
  For proton-proton collisions  both effects make predictions
 closer to the experimentally observed values of $v_2$ \cite{ATLASPP}. 

 
 One of the properties that has been violated in the estimates
 in section 2, was the factorization $r_n = 1$ where
 \beq \label{r}
r_n\,=\,\frac{V_{n \Delta}\Lb p_{T1}, p_{T2}\Rb}{\sqrt{V_{n
 \Delta}\Lb p_{T1},  p_{T1}\Rb\,V_{n \Delta}\Lb p_{T2},  p_{T2}\Rb}}\,=\,1
 \eeq
 
\begin{figure}[ht]
       \begin{tabular}{cccc}
\includegraphics[width=4.cm]{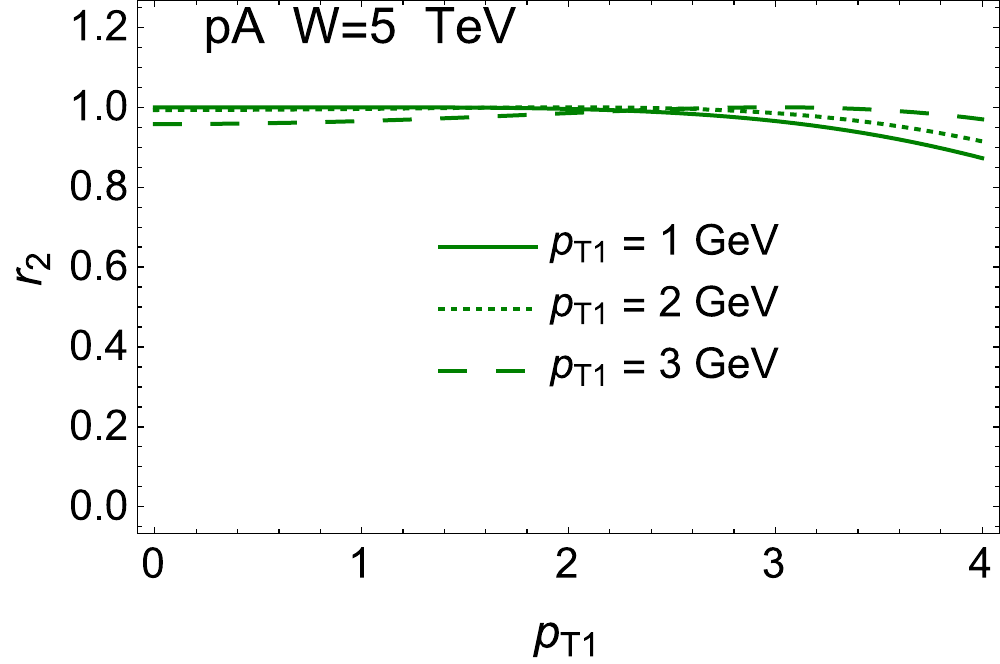} &  
\includegraphics[width=4.cm]{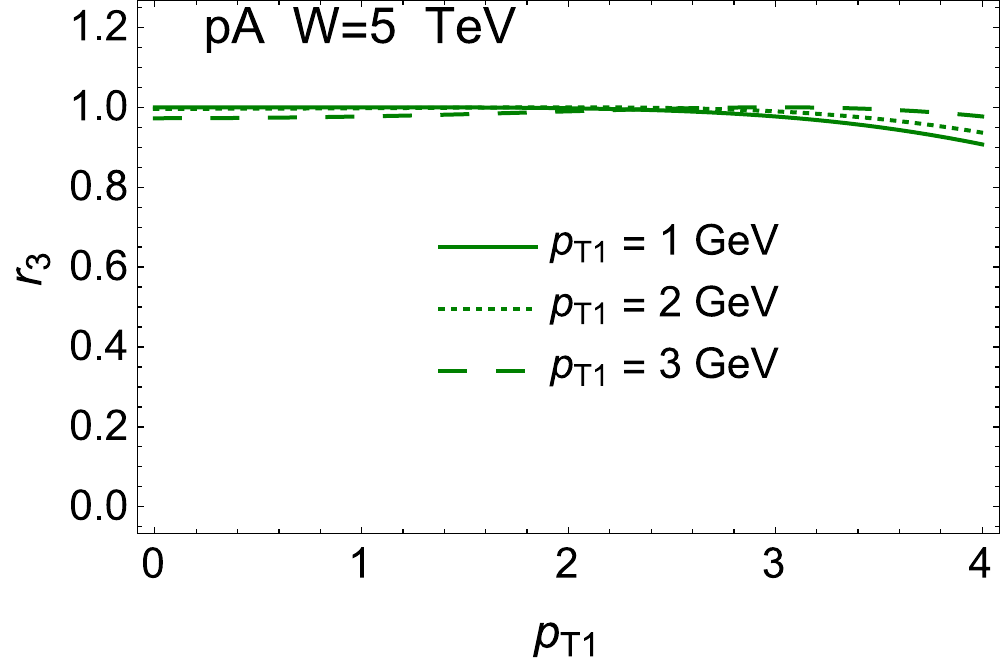} &
       \includegraphics[width=4.cm]{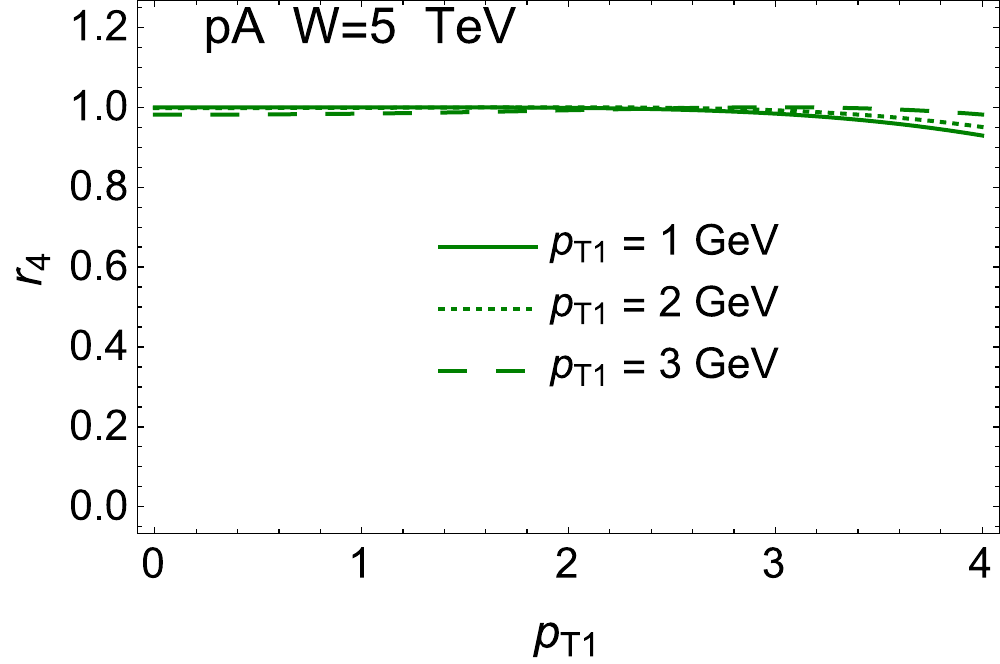}& 
\includegraphics[width=4.cm]{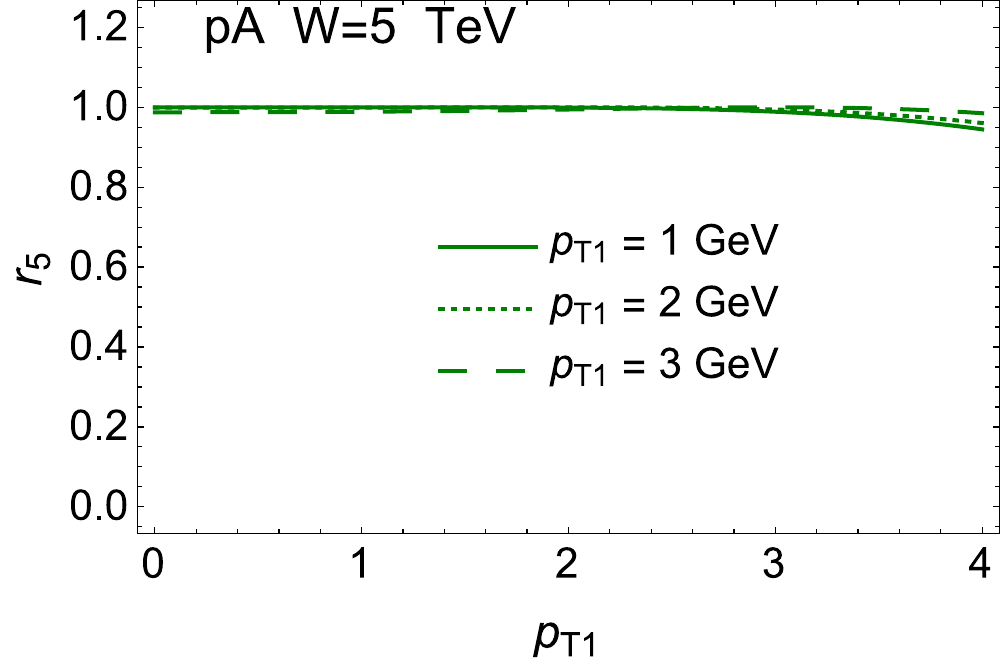}\\
       \fig{r1}-a & \fig{r1}-b & \fig{r1}-c &\fig{r1}-d\\
       \end{tabular}     
          \protect\caption{The ration $r_n$ versus $p_T$  at $W = 5\,TeV$
 for proton-lead  collisions.\, }           
      \label{r1}
       \end{figure}

 \fig{r1} shows that \eq{r} holds at least for $p_T \leq 4\,GeV$ in
 accordance with the experimental data (see Ref.\cite{ATLASPA}).

 \section{Conclusions}
In this paper we showed how three different dimensional  scales in high
 energy scattering, arise  in the Bose-Einstaein correlations
 that generates $v_n$,  for even and odd $n$. The first two scales are 
intimately
 related to the structure of the wave function of the hadron, and have an
  interpretation in the constituent quark model, as the distance
 between the constituent quarks and the size of the quark.  In a more
 formal way they characterize the size of the vertex of the BFKL Pomeron 
interaction with the hadron, and the typical size of the same vertex for
 the diffraction production, in the region of small mass. We 
demonstrated
 that these sizes lead to BE correlations which are large, but narrowly 
distributed in $p_T$ .   

 The third size is the value of the saturation momentum in the
 CGC/saturation approach, and has been used in the construction of
 our model for the high energy soft interactions. This size is 
incorporated
 in the structure of the `dressed' Pomeron in our model.
It turns out that this size leads to  values of $v_n$ which are close
 to the experimental values both for even and odd $n$, and they are 
broadly
 distributed in $p_T$.
In proton-proton scattering this mechanism is able to describe the
 experimental data both for even and odd $v_n$, while for proton-nucleus
 and nucleus-nucleus collisions we obtain smaller values of $v_2$:  
20-30\%
 smaller for proton-lead scattering, and two times smaller for lead-lead
 collisions. However, we would like to stress that for centrality 0-10\%,
 the structure of the Pomeron gives  values of  $v_n$ which are 
 very  close to the experimentally observed ones.

All estimates were made in the framework of our model for soft 
interactions
 which is based on CGC/saturation approach, but introduces non-perturbative
 parameters which describe the wave function of the hadron, and the large
 impact parameter behaviour of the saturation momentum. We describe in
 this model  the total, elastic and diffractive cross sections as well
 as the inclusive production and long  range rapidity correlations, and
 therefore, we trust that we can rely on the model when discussing the
 azimuthal angle correlations.

We demonstrated in this paper that BE correlations in the framework
 of CGC/saturation approach are able to explain  a substantial part
 if not the entire, experimental values of $v_n$ for both even and odd 
$n$.
 Therefore, we believe that is premature to conclude that the origin
of the observed long range
 rapidity correlations are  only due to  elliptic flow.


  {\bf Acknowledgements} 
   We thank our colleagues at Tel Aviv University and UTFSM for
 encouraging discussions. Our special thanks go to    
Carlos Cantreras, Alex Kovner and Michel  Lublinsky for
 elucidating discussions on the
 subject of this paper. This research was supported by the BSF grant   2012124, by    Proyecto Basal FB 0821(Chile) ,  Fondecyt (Chile) grant  1140842 and  by CONICYT grant PIA ACT1406.
  ~

 ~

 \appendix
 
 \section{}
 \subsection{BFKL contribution for the interference diagram}

 In this appendix we derive the BFKL contribution to  $ \frac{d  
 \sigma}{d y_1 d^2 p_{T1}}\Lb k_T, |  \vec{k_T} + \vec{p}_{T,12}|\Rb$
 given by \eq{1D2}.
       \begin{figure}[ht]
    \centering
  \leavevmode
      \includegraphics[width=3cm]{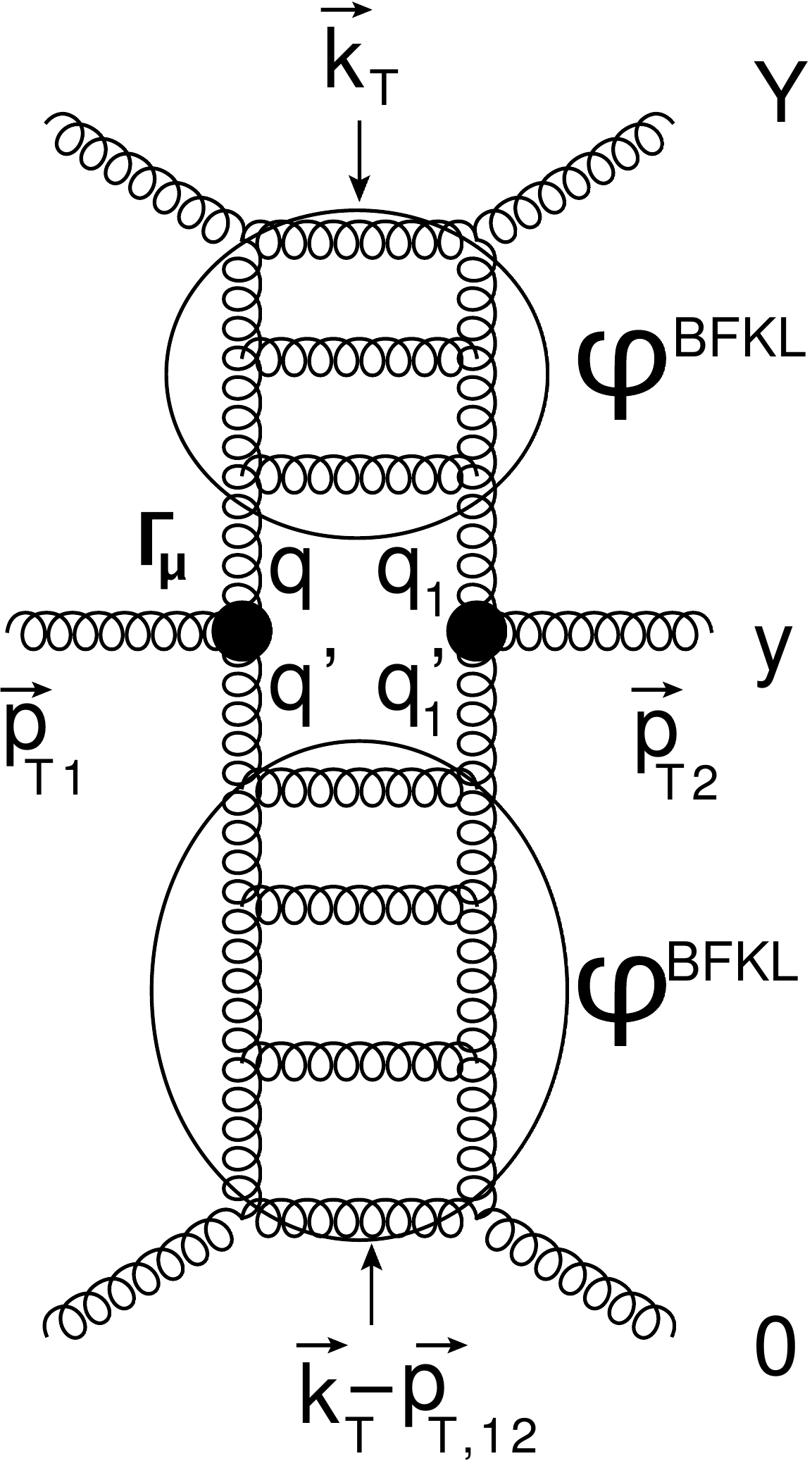} 
                 \caption{ The graphical representation of 
\protect\eq{1D2}.}   
        
      \label{bfkl}
       \end{figure}

 The Lipatov vertices $\Gamma_\mu\Lb q_T,p_{T1}\Rb$ and $\Gamma_\mu\Lb q_{T1},
p_{T2}\Rb$ have the form (see book \cite{KOLEB} for example):

   \beq \label{LIP}
   \Gamma_\mu\Lb q_T,p_{T1}\Rb\,\,=\,\,\frac{1}{p_{T1}^2}\Lb \,q^2_T\,
\vec{p}_{T1}   \,-\,\vec{q}_T\,p^2_{T1}\Rb;~~~~~~~~
     \Gamma_\mu\Lb q_{T1},p_{T2}\Rb\,\,=\,\,\frac{1}{p_{T2}^2}\Lb \,q^2_{T1}\,
\vec{p}_{T2}   \,-\,\vec{q}_{T1}\,p^2_{T2}\Rb;
      \eeq
      
      and
      
      \beq \label{A2}
         \Gamma_\mu\Lb q_T,p_{T1}\Rb\,  \Gamma_\mu\Lb q_{T1},p_{T2}\Rb
 \,\,=\,\,\frac{q^2_{T1}\,\Lb \vec{q}_T - \vec{p}_{T2}\Rb^2}{p^2_{T2}}
     \,+\,\frac{q^2_{T}\,\Lb \vec{q}_{T1} - \vec{p}_{T1}\Rb^2}{p^2_{T1}}
  \,-\,k^2_T\,-\,p^2_{T,12}\frac{q^2_T\,q^2_{T1}}{p^2_{T1}\,p^2_{T2}}
         \eeq
         
         where $\vec{p}_{T,12} \,=\,\vec{p}_{T1} - \vec{p}_{T2}$
 , $\vec{q}^{\,'}_T\,=\,\,\vec{q}_T - \vec{p}_{T1}$,  
$\vec{q}^{\,'}_{T1}\,
=\,\,\vec{q}_{T1}  - \vec{p}_{T2}$,         
         and  $\vec{q}_{T1} = \vec{q}_T - \vec{k}_T$.
       \eq{A2} can be re-written as
       \beq \label{A3}
      \Gamma_\mu\Lb q_T,p_{T1}\Rb\,  \Gamma_\mu\Lb q_{T1},p_{T2}\Rb \,\,=
\,\,\frac{q^2_{T1}\,\Lb \vec{q}^{\,'}_T + \vec{p}_{T,12}\Rb^2}{p^2_{T2}} 
    \,+\,\frac{q^2_{T}\,\Lb \vec{q}^{\,'}_{T1} - \vec{p}_{T,12}\Rb^2}
{p^2_{T1}}  \,-\,k^2_T\,-\,p^2_{T,12}\frac{q^2_T\,q^2_{T1}}{p^2_{T1}\,p^2_{T2}}
      \eeq
      $\phi^{\rm BFKL}$ satisfies the following equation:
      \beq \label{A4} 
      \frac{\partial \phi^{\rm BFKL}\Lb Y; q_T,k_T\Rb}{\partial \,Y}\,\,=
\,\,\bas \,\int \frac{d^2 q'_T}{\pi}\, K\Lb  k_T,q'_T\Rb\,\phi^{\rm BFKL}
\Lb Y; q'_T,k_T\Rb
      \eeq
      
      where
    \bea \label{AK}
      &&K\Lb k_T,q'_T\Rb\,\,=\\
      &&\,\,\Bigg(\frac{q^2_{T1}\,q'^2_T}{p^2_{T}}    
 \,+\,\frac{q^2_{T}\,q'^2_{T1}}{p^2_{T}}  \,-\,k^2_T\Bigg)\frac{1}{q'^2_T\,
q'^2_{T1}}\,   \,-\,\Bigg\{ q^2_T \int \frac{d^2 p_T}{p^2_T\,\Lb \vec{q}_T -
 \vec{p}_T\Rb^2}\,+\,q'^2_T \int \frac{d^2 p_T}{p^2_T\,\Lb \vec{q}^{\,'}_T -
 \vec{p}_T\Rb^2}  \Bigg\}\delta^{(2)}\Lb \vec{q}^{\,'}_t - \vec{q}_T\Rb\nn
      \eea    
      
            \eq{A4} is the BFKL equation  in the momentum representation, 
which
 has the following form in the coordinate representation\cite{LI,MUCD}:
  \bea \label{A5}
    &&  \frac{\partial N^{\rm BFKL}\Lb Y; x_{12}, b\Rb}{\partial \,Y}\,\,=\\
    &&\,\,\bas \,\int \frac{d^2 x_3}{\pi}\,\frac{x^2_{12}}{x^2_{13}\,x^2_{23}
 }\Big\{N\Lb Y; x_{13}, \vec{b} - \h \vec{x}_{23}\Rb\,+\,N\Lb Y; x_{23},
 \vec{b} - \h \vec{x}_{13}\Rb \,-\,            N\Lb Y;  x_{13}, b\Rb\Big\}\nn
      \eea
 where\cite{KOLEB,KOTUINC}
 \beq \label{AN}
    \frac{1}{\Lb \vec{q}_T + \h \vec{k}_T\Rb^2\,\Lb \h \vec{k}_T -
 \vec{q}_T\Rb^2}\phi^{\rm BFKL}\Lb q_T,k_T\Rb\,\,=\,\,\frac{2\,C_F}{\bas
 (2 \pi)^3}\int d^2 b\,d^2 x_{12}\,e^{i \vec{q}_T \cdot \vec{x}_{12}\,+\,i
 \vec{k}_T \cdot \vec{b}}
   \, N^{\rm BFKL}\Lb Y; x_{12}, b\Rb
         \eeq     
         
For diagrams \fig{2shiden}- c and \fig{2shiden}-d   $\vec{p}_{12} = 0$ 
  and  plugging \eq{AN}     in   \eq{1D2} we obtain that
\bea \label{A6}
&&\frac{d   \sigma}{d y_1 d^2 p_{T1}}\Lb k_T, k_T \Rb\,\,=\,\,\,\Lb\frac{2\,
C_F}{\bas (2 \pi)^3}\Rb^2\int d^2 x_{12} \,d^2 b\,d^2 b'\,
e^{i \vec{p}_{T1}\cdot \vec{x}_{12}\,+\,i \vec{k}_T \cdot \vec{b}} \\
&&\times 
\Bigg\{ \frac{1}{p_{T1}^2}\Lb\Lb\Lb  \h \vec{\nabla}_b \,+\,
\vec{\nabla}_{x_{12}}\Rb^2 N^{\rm BFKL}\Lb Y - y; x_{12}, 
\vec{b} - \vec{b}'\Rb\Rb\,\Lb
\Lb  \h \vec{\nabla}_{b'} \,-\,\vec{\nabla}_{x_{12}}\Rb^2\,N^{\rm BFKL}\Lb 
 y; x_{12},  \vec{b}'\Rb\Rb\,\right.\nn\\
&&\left.+\,\Lb\Lb  \h \vec{\nabla}_b \,-\,\vec{\nabla}_{x_{12}}\Rb^2 N^{\rm
 BFKL}\Lb Y - y; x_{12}, \vec{b} - \vec{b}'\Rb\Rb\,\Lb
\Lb  \h \vec{\nabla}_{b'} \,+\,\vec{\nabla}_{x_{12}}\Rb^2\,N^{\rm BFKL}\Lb 
 y; x_{12},  \vec{b}'\Rb\Rb\Rb\nn\\
&& \,-\,\Lb\vec{\nabla}_b  N^{\rm BFKL}\Lb Y - y; x_{12}, \vec{b} -
 \vec{b}' \Rb \Rb \cdot \Lb \vec{\nabla}_{b'}N^{\rm BFKL}\Lb y; x_{12},
 \vec{b}'\Rb\Rb\Bigg\}\nn
\eea
    \eq{A6} in the limit $k_T \to 0$,     degenerates to the expression
 for the inclusive cross section which has the elegant form derived in
 Ref.\cite{KOTUINC}
    \bea \label{A7}
&&\frac{d   \sigma}{d y_1 d^2 p_{T1}}\Lb k_T=0, k_T=0 \Rb\,\,=\\
&&\Lb\frac{2\,C_F}{\bas (2 \pi)^3}\Rb^2 \frac{1}{p^2_{T1}}\,
 \int d^2 x_{12}\,e^{i \vec{p}_{T1} \cdot x_{12}}\Lb\nabla^2_{x_{12}}\int
 d^2 b\, N^{\rm BFKL}\Lb Y -   y; x_{12},  \vec{b}\Rb\Rb\,\Lb
\nabla^2_{x_{12}}\int d^2 b'\, N^{\rm BFKL}\Lb   y; x_{12}, 
 \vec{b}'\Rb\Rb\nn
\eea

The interesting feature of \eq{A6} and \eq{A7} is, that they remain  
correct, if we replace $2 N^{\rm BFKL}$ by  $ N_G\,=\,2\, N\, - \,N^2$,
 where $N$ is  the solution of the Balitsky-Kovchegov equation\cite{BK}.
 Inside  the saturation domain where $N \to 1$, both equations lead to
 negligible contributions. In other words, in both equations the main
 contributions stem from the vicinity of the saturation scale, where
 $x^2_{12}\,Q^2_s  \approx\,1$.

The solution for  the scattering amplitude of two dipoles $r_1$ and $r_2$ 
 to \eq{A5} is known\cite{LI}
\bea \label{GENSOL}
&&N_\pom\Lb r_1,r_2; Y, b \Rb\,\,= \\
&&\,\,\sum_{n=0}^{\infty}
\int \frac{d \gamma}{2\,\pi\,i}\,\phi^{(n)}_{in}(\gamma; r_2)
\,\,d^2\, R_1 \,\,d^2\,R_2\,\delta(\vec{R}_1 - \vec{R}_2 - 
\vec{b})\,
e^{\omega(\gamma, n )\,Y}
\,E^{\gamma,n}\Lb r_1, R_1\Rb\,E^{1 - \gamma,n}\Lb r_2, R_2 \Rb\nn
\eea
 where  the  functions $\phi^{(n)}_{in}(\gamma; r_2)$ are determined
 by the initial conditions at low energies and 

\beq \label{OMEGA}
\omega(\gamma, n)\,\,=\,\,\bas \chi(\gamma, n)\,\,  =\,\,\bas \Lb 2 
\psi\Lb 1\Rb \,-\,\psi\Lb \gamma + |n|/2\Rb\,\,-\,\,\psi\Lb 1 
 - \gamma + |n|/2\Rb\Rb;
\eeq
where $~\psi\Lb \gamma\Rb \,\,=\,\,d \ln \Gamma\Lb \gamma\Rb/d
 \gamma $ and $\Gamma\Lb \gamma\Rb$ is  Euler gamma function. 
Functions $ E^{n, \gamma} \Lb \rho_{1a},\rho_{2a}\Rb$ are given
 by the following equations.
\begin{align}\label{EFUN}
  E^{n, \gamma} \Lb \rho_{1a},\rho_{2a}\Rb \,=\, \Lb
  \frac{\rho_{12}}{\rho_{1a} \, \rho_{2a}}\Rb^{1 - \gamma + n/2}
  \, \Lb \frac{\rho^*_{12}}{\rho^*_{1a} \, \rho^*_{2a}}
  \Rb^{1 - \gamma - n/2},
\end{align}
In \eq{EFUN} we use  complex numbers to characterize the point
 on the plane
\begin{align}\label{COMNUM}
  \rho_i = x_{i,1} + i \, x_{i,2};\,\,\,\,\,\,\, \rho^*_i = x_{i,1} - 
 i \, x_{i,2}
\end{align}
where the indices $1$ and $2$ denote  two transverse axes. Note that
\beq \label{NOT}
\rho_{12}\,\rho^*_{12}\,\,=\,\,r^2_i ;~~~~~~\rho_{1 a}\,\rho^*_{1 a}\,=
\,\Lb\vec{R}_i\,-\,\frac{1}{2}\vec{r}_{i}\Rb^2~~~~~~\rho_{2 a}\,\rho^*_{2a}\,
=\,\Lb\vec{R}_i\,+\,\frac{1}{2}\vec{r}_{i}\Rb^2
\eeq
At large values of $Y$, the main contribution stems from the first term 
with $n =0$.  For this term, \eq{EFUN} can be re-written in the form
\beq \label{E}
E^{\gamma,0}\Lb r_i,R_i\Rb \,\,=\,\,\left( \,\frac{r^2_{i}}{(\vec{R}_i
\,+\,\frac{1}{2}\vec{r}_{i})^2\,\,
(\vec{R}_i\,-\,\frac{1}{2}\vec{r}_{i})^2}\,\right)^{1 - \gamma}\,\,.
\eeq

The integrals over $R_1$ and $R_2$ were taken in Refs.\cite{LI,NAPE}
 and at $n=0$ we have
\bea \label{H}
&&H^\ga\Lb w, w^*\Rb\,\,\equiv\,\,\int d^2\,R_1\,E^{\ga,0}\Lb r_{1},R_1\Rb\,
 E^{1 - \ga, 0}\Lb r_{2}, \vec{R}_1
\,-\,\vec{b}\Rb\,= \\
&&
\,\frac{ (\gamma - \h)^2}{( 
\gamma (1 - \gamma)
)^2} \Big\{b_\ga\,w^\gamma\,{w^*}^\gamma\,F\Lb\gamma, \gamma, 2\gamma, w\Rb\,
F\Lb\gamma, \gamma, 2\gamma, w^*\Rb\nn\\
&&
\,+   b_{1 - \ga} w^{1 -
\gamma}{w^*}^{1-\gamma}
F\Lb 1 - \gamma, 1 -\gamma, 2 - 2\gamma, w\Rb\,F\Lb 1 - \gamma,1 -\gamma,2 
-2\gamma, w^*\Rb \Big\}\nn\\
&&\xrightarrow{b\,\gg\,r_1\,\, \mbox{\small and/or} \,\,r_2}\,\,
\,\frac{ (\gamma - \h)^2}{( 
\gamma (1 - \gamma)
)^2} \Big\{b_\ga\,w^\gamma\,{w^*}^\gamma
\,+  b_{1 - \ga} w^{1 -
\gamma}{w^*}^{1-\gamma}\Big\}\,=\,\,\frac{ (\gamma - \h)^2}{( 
\gamma (1 - \gamma)
)^2} \Big\{b_\ga\,\Lb \frac{r^2_1\,r^2_2}{b^4}\Rb^\gamma
\,+  b_{1 - \ga} \Lb \frac{r^2_1\,r^2_2}{b^4}\Rb^{1- \gamma}\Big\}\nn
\eea
where $F$ is hypergeometric function \cite{RY}. In  \eq{H}
 $w\,w^* $ and $b_{\ga}$ are equal
\beq \label{W}
w\,w^*\,\,=\,\,\frac{r^2_{1}\,r^2_{2}}{\Lb\vec{b} - \h\Lb\,\vec{r}_{1}\,
- \,\vec{r}_{2}\Rb\Rb^2
\,\Lb\vec{b} \,+\, \h \Lb\,\vec{r}_{1} \,- \,\vec{r}_{2}\Rb\Rb^2};~~~~~~~~
b_{\ga} \, = \, \pi^3 \, 2^{4(1/2 - \ga)} \, \frac{\Gamma \Lb\ga \Rb}{
\Gamma \Lb 1/2 - \ga \Rb}
  \, \frac{\Gamma \Lb 1 - \ga  \Rb}{\Gamma \Lb 1/2 + \ga \Rb}.
\eeq
Therefore, at large $b$, $N^{\rm BFKL}$ decreases as a power of $b$ 
which violates the Froissart theorem \cite{KOWE}. At present, as
 has been mentioned above, we cannot suggest a modification of the
 equation of the CGC/saturation approach  in which the correct\cite{FROI}
 exponential behaviour at large $b$ 
would be incorporated. So we doomed to build a model. We discussed our 
model 
 in section 3.
\subsection{Born diagrams}
\begin{figure}[ht]
 \includegraphics[width=10cm]{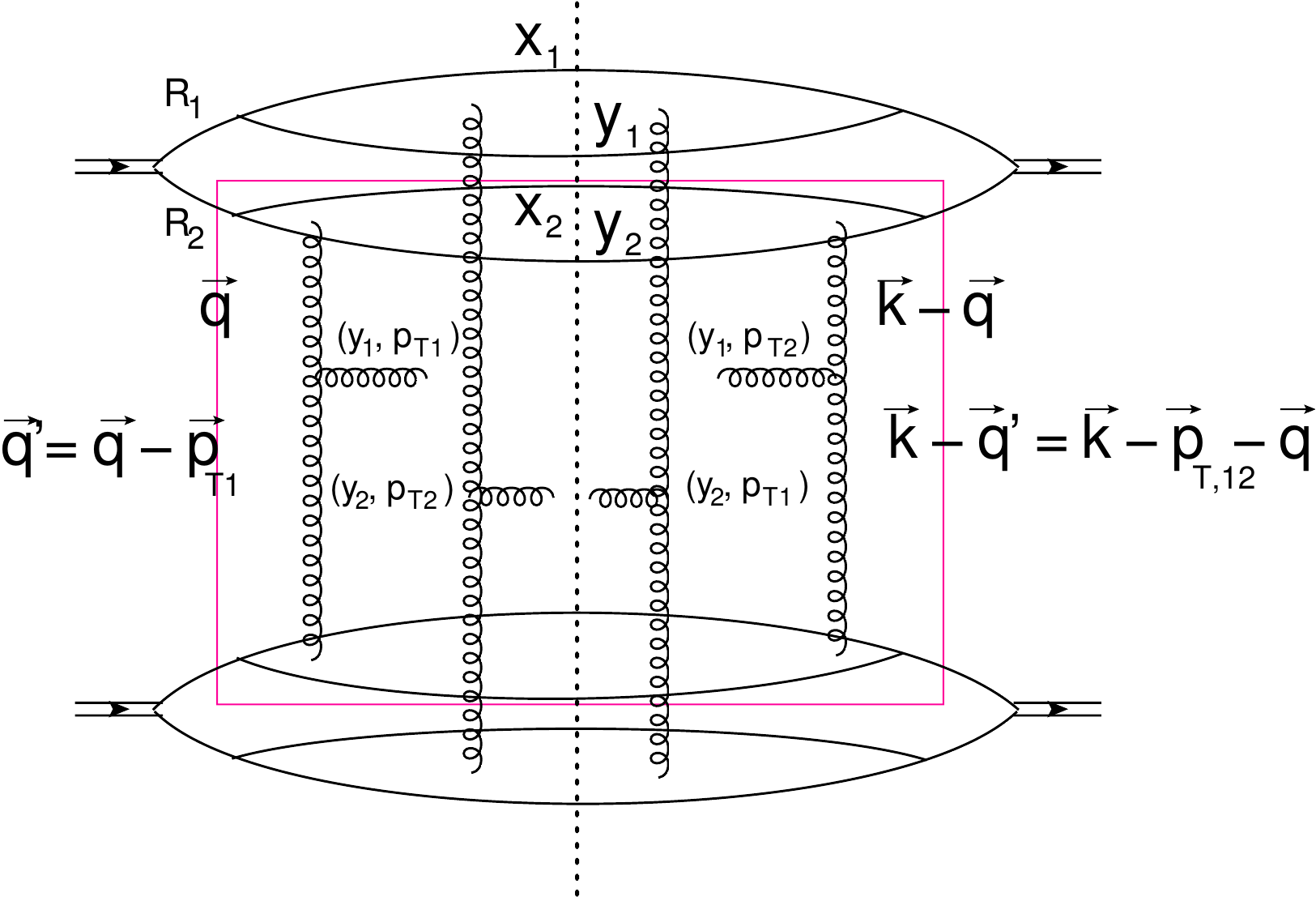}
   \protect\caption{The Born interference diagram  for production
 of two identical gluons with rapidities:$y_1$ and $y_2$ and
 transverse momenta $\vec{p}_{T1}$ and $\vec{p}_{T2}$.$\vec{R}_i =
 \h \Lb  \vec{x}_i +\vec{y}_i\Rb$, $\vec{p}_{T,12} = \vec{p}_{T1} -
 \vec{p}_{T2}$. Red rectangle shows function $\Phi\Lb \vec{k},
\vec{p}_{T1},\vec{p}_{T2}\Rb$(see text). }
\label{ba2}
   \end{figure}

The spirited discussions with our colleagues, showed us that  it 
would be benificial  to add a
 general discussion of the BFKL contribution, by  calculating  of the
 first Born diagrams for the production of two identical gluons that have
 rapidities $y_1$ and $y_2$, and carry momenta
 $\vec{p}_{T1}$ and $\vec{p}_{T2}$. These diagrams
 are shown in \fig{ba2} for the scattering of the
 bound states of two  oniums  (two dipoles). Such a
 model for the scattering systems allows us to use
 the perturbative QCD approach, and has the analogy
 in the simplest bound system: deuteron.

 The two onium bound state is described by the wave function $\Psi\Lb 
\vec{R}_1
 - \vec{R}_2\Rb$, where $R_i $ is the coordinate of the onium which is
 equal $\vec{R}_i = \h \Lb  \vec{x}_i +\vec{y}_i\Rb$ where $\vec{x}_i$
 and $\vec{y}_i$ are coordinates of quark and antiquark in the onium (see
 \fig{ba2}).
We introduce two new functions that describe the form factor of our bound
 state ($G\Lb q \Rb$), and the interaction of two gluons with the onium:
\bea \label{A10}
G\Lb q\Rb \,&=&\,\int d^2 R \,|\Psi\Lb R\Rb|^2\, \,e^{i\,\vec{q} \cdot
 \vec{R}}~~~\mbox{with}~~~\vec{R} = \vec{R}_1 - \vec{R}_2;\nn\\
\phi_{\rm onium}\Lb q, k\Rb\,&=&\,2\,\int d^2 r_i \,|\psi_{\rm onium}\Lb
 r_i \Rb|^2\,\,e^{i\,\h \vec{k}\cdot \vec{r}_i}
\Lb 1  + e^{i \vec{q}\cdot\vec{r}_i}\Rb~~~~~\mbox{with}~~~~~\vec{r}_i \,=\,
 \vec{x}_i \,-\,\vec{y}_i;
\eea
The contribution of the diagram  of \fig{ba2} can be written as\footnote{
 We omit all numerical factors as well as $\bas^6$.}
\beq \label{A11}
\sigma_{\rm interference}\,\propto\,\int \frac{d^2 k}{4 \pi^2}\,G\Lb k\Rb
 \,G\Lb \vec{k} + \vec{p}_{T,12}\Rb\,I^2\Lb \vec{k},\vec{p}_{T1},\vec{p}_{T2}
\Rb 
\eeq
where 
\bea \label{A12}
I\Lb \vec{k},\vec{p}_{T1},\vec{p}_{T2}\Rb\,&\propto&\,\int \frac{d^2 q}{4\,
 \pi^2}\,\,\phi_{\rm onium}\Lb \vec{k},\vec{q}\Rb \,\phi_{\rm onium}\Lb
 \vec{k}\, -\,\vec{p}_{T,12} ,\vec{q}\Rb \,\nn\\
&\times&
\Bigg\{\frac{1}{q^2\,\Lb \vec{k} - \vec{q}\Rb^2} \, 
  \Gamma_\mu\Lb q_T, p_{T1}\Rb\,  \Gamma_\mu\Lb q_{T1}, p_{T2}\Rb\,\,\frac{1}
{\Lb \vec{q} - \vec{p}_{T1}\Rb^2\,\Lb \vec{k} - \vec{p}_{T,12} - 
\vec{q}\Rb^2}\Bigg\}
\eea
where $  \Gamma_\mu\Lb q_T, p_{T1}\Rb\,  \Gamma_\mu\Lb q_{T1},
 p_{T2}\Rb$ is given by \eq{A2} and \eq{A3}.

One can see from \eq{A11} and  \eq{I12} that the typical $q \approx 
1/r$,
 where $r$ is he size of the onium , while  the typical values of $k
 \propto 1/R$, where  $R$ is the size of the bound state. Assuming that
 $R \gg r$,  we  see that $k\, \ll \,q$.  Anticipating $p_{T,12} \propto
 1/R$, we can reduce
the contribution of the interference diagram to the following form:
\beq \label{A13}
\sigma_{\rm interference}\,\propto\,\frac{1}{p^2_{T1} \,p^2_{T,2}}\int
 \frac{d^2 k}{4 \pi^2}\,G\Lb k\Rb \,G\Lb \vec{k} + \vec{p}_{T,12}\Rb\Bigg(\int
 \frac{d^2 q}{4 \pi^2}\frac{1}{q^2\,\Lb \vec{q} - \vec{p}_{T1}\Rb^2}\Bigg)^2
\eeq
In \eq{A13} we assume that $p_{T1} \approx p_{T2}$ and one can see that
 $p_{T,12}$ from this equation is indeed of the order of $1/R$,  being much
 smaller than $p_{Ti}$ if they are of the order of $1/r$. For $1/R\,\ll\,
 p_{Ti} \,\ll\,1/r$ we need to take  $ \Gamma_\mu\Lb q_T, p_{T1}\Rb\, 
 \Gamma_\mu\Lb q_{T1}, p_{T2}\Rb\,\,=\,\,\Lb \frac{1}{p^2_{T1}}\,+\,
\frac{1}{p^2_{T2}}\Rb\, \frac{1}{q^4}$.

\section{BE correlations in the model: diffractive production in
 the small mass region. }
 
\subsubsection{Inclusive production}

\begin{figure}[ht]
    \centering
  \leavevmode
      \includegraphics[width=10cm,height=4cm]{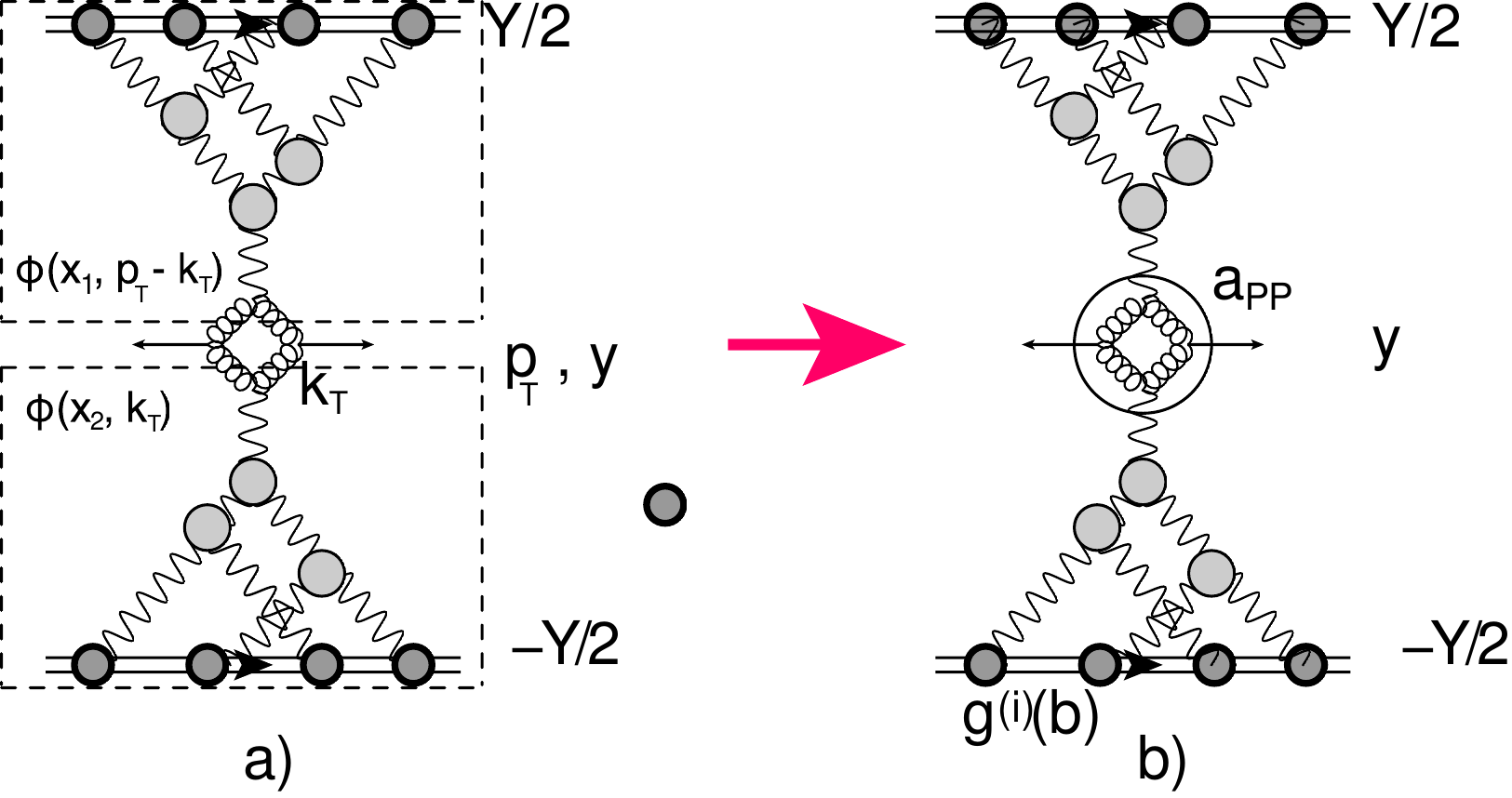}  
      \caption{The graphic representation of \protect\eq{CGCF}
 (see \fig{inclgen}-a).For the sake of simplicity all other indices
   in $\phi\Lb x_1,p_T - k_T\Rb$ and $\phi\Lb x_2,k_T\Rb$ are omitted.
 The wavy lines denote the BFKL Pomerons, while the helical lines 
illustrate
 the gluons. In \fig{inclgen}-b  the Mueller diagram for  inclusive
 production is shown. }
\label{inclgen}
   \end{figure}

 The inclusive production in the framework of the CGC/saturation
 approach comprises two stages: the gluon mini-jet productions and
 the decay of this mini-jet into hadrons.
 For mini-jet production, we
 use the  $k_T$ factorization  formula, which has been proven in  Ref.
 \cite{KOTUINC}  in the framework of the CGC/saturation approach 
(see appendix for details).

\beq \label{CGCF}
\frac{d \sigma}{d y \,d^2 p_T}\,\,=\,\,\frac{2 \pi 
\bas}{C_F}\frac{1}{p^2_T}\,\int
 d^2 k_{T}\,\,\phi^{h_1}_G\Lb x_1;\vec{k}_T\Rb\,\phi^{h_2}_G\Lb x_2;
 \vec{p}_T
 -\vec{k}_T\Rb
\eeq
where $\phi^{h_i}_G$ denotes the probability to find a gluon that
 carries the fraction $x_i$  of energy with $k_T$ transverse 
momentum,
 and $ \bas \,= \,\as N_c/\pi$,  with the number of colours equal to
 $N_c$. $\h Y + y \,=\,\ln(1/x_1)$ and $ \h Y - y =\ln(1/x_2)$.
 $\phi^{h_i}_G$ is the solution of the Balitsky-Kovchegov(BK)
 \cite{BK} non-linear evolution equation, and can be viewed as
 the sum of `fan' diagrams of the BFKL Pomeron interactions,  shown
 in \fig{inclgen}. 

 In our model the sum of `fan' diagrams is given by \eq{T16}. 
Assuming that the main contribution to 
 $$\frac{d \sigma}{d y } \,=\,\int d^2 p_T\,\frac{d \sigma}{d y \,d^2 p_T} $$
 stems from $p_T \,\leq\,Q_s$, we obtain the following formula:
 \bea \label{INCF}
   \frac{d \sigma}{d y}\,\,&=&\,\,\int d^2 p_T\,\frac{d
 \sigma}{d y \,d^2 p_T}\,\,=\,\,a_{\pom \pom}\,\ln\Lb W/W_0\Rb\Bigg(
 \alpha^2 \,In^{(1)}\Lb \h Y + y\Rb \, +\beta^2 \, In^{(2)}\Lb
 \h Y + y\Rb\Bigg)\nn\\
 &\times& \Bigg(
 \alpha^2 \,In^{(1)}\Lb \h Y - y\Rb \, +\beta^2 \, In^{(2)}\Lb
 \h Y - y\Rb\Bigg)
   \eea
   where
 \bea \label{IN}
&&In^{(i)}\Lb y\Rb\,=\,\int d^2 b \,\,N^{BK}\Lb g^{(i)}\,S\Lb m_i,
 b\Rb \,\tilde{G}_\pom\Lb y\Rb\Rb~~~~~~~\mbox{or}~~~~~~ In^{(i)}\Lb y
 \Rb = I^{\rm BK}_i\Lb y, Q_T=0\Rb;\nn\\
 && ~~~\mbox{with}~~~
  I^{\rm BK}_i\Lb y, Q_T\Rb\,=\,\int d^2 b \,e^{i \vec{b}
 \cdot \vec{Q}_T}\,\,N^{BK}\Lb g^{(i)}\,S\Lb m_i,
 b\Rb \,\tilde{G}_\pom\Lb y\Rb\Rb  
  \eea
where  $\tilde{G}_\pom\Lb y\Rb$  and $N^{BK}$ have been  defined in
 \eq{GTILDE} and in \eq{T16}, respectively.
 Regarding the factor in front of \eq{INCF}  i.e. $ \ln\Lb
 W/W_0\Rb$, where $W = \sqrt{s}$ is the energy of collision in c.m. frame,
 and $W_0$ is the  value of energy from which we  start our approach.
  One can see that \eq{CGCF}
is divergent in the region of small $p_T \,<\,Q_s$. Indeed,
 in this region $\phi$'s in \eq{CGCF} do not depend on $p_T$,
 since $k_T \approx\,Q_s \,>\,p_T$, and the integration over $p_T$
 leads to $\ln\Lb Q^2_s/m^2_{soft}\Rb$, where $m_{soft}$ is the
 non-perturbative scale,  that includes the confinement of quarks
 and gluons ($m_{soft} \sim \Lambda_{QCD})$. 
 
 \subsubsection{LRR correlations}
 In our previous paper \cite{GLMCOR}, we showed that in the framework
 of our model that has been described above, the main source of the long
 range rapidity correlation, is the correlation  between two parton 
showers.  In other words, it was shown that 
the contribution to the correlation function from enhanced and
 semi-enhanced diagrams, turns out to be negligibly small.

\begin{figure}[ht]
    \centering
  \leavevmode
      \includegraphics[width=15cm,height=4cm]{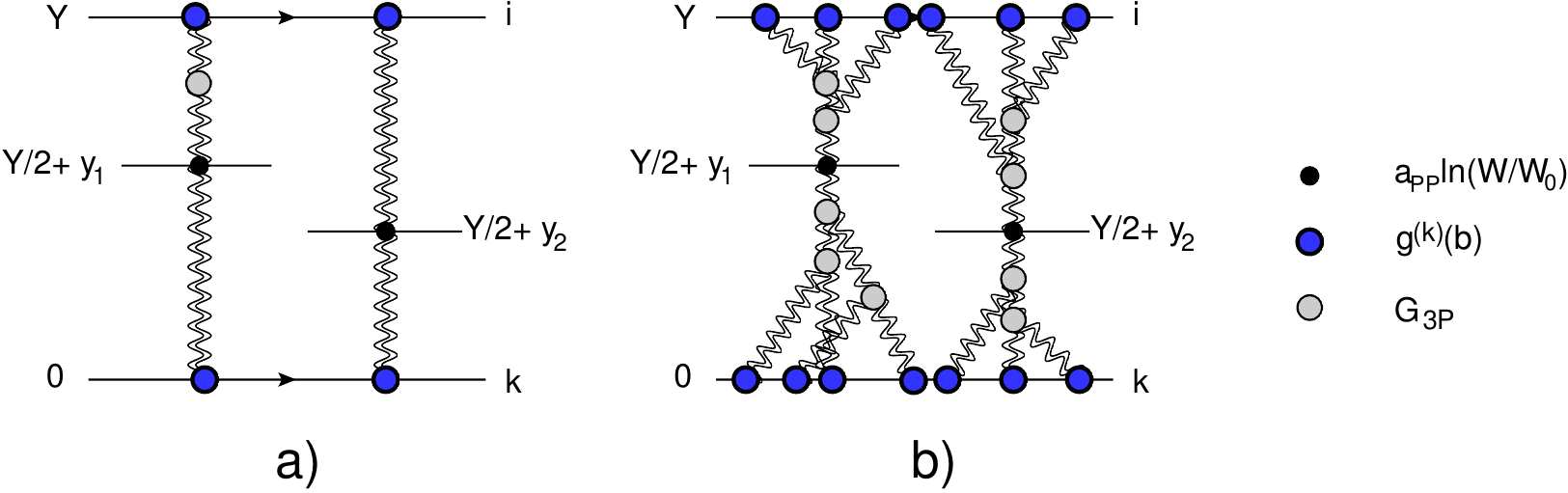}  
      \caption{The  Mueller diagram for the  rapidity correlation between
 two particles produced in two parton showers. \protect\fig{cor2sh}-a  shows
 the first Mueller diagram, while \protect\fig{cor2sh}-b indicates the 
structure
 of general diagrams.
  The double  wavy lines describe the dressed  BFKL Pomerons.
 The blobs stand for the vertices as shown in the legend.}
\label{cor2sh}
   \end{figure}

  The appropriate   Mueller diagrams are shown
 in \fig{cor2sh}. 
 Examining this diagram, we see that the contribution to the double
 inclusive cross section, differs from the product of two single inclusive
 cross sections.  This difference generates the rapidity correlation
 function, which is defined as
 \beq \label{RCF}
  R\Lb y_1, y_2\Rb\,\,=\,\ \frac{\frac{1}{\sigma_{in}}
 \frac{d^2 \sigma}{d y_1\,d y_2}}{\frac{1}{\sigma_{in}}
 \frac{d \sigma}{d y_1} \, \frac{1}{\sigma_{in}} \frac{d
 \sigma}{d y_2}  }\,\,-\,\,1
  \eeq


  There are two reasons for the difference between the 
 double inclusive cross section due to production of 
 two parton showers, and the products of inclusive cross
 sections: the first, is that in the 
expression for the double
 inclusive cross section, we integrate  the product of the single
 inclusive inclusive cross sections, over $b$ or $Q_T$(see \fig{inclgen}-a
 and \eq{1D1}). The second, 
is that the 
summation 
 over $i$ and $k$  for  the product of single  inclusive cross
 sections, is for fixed $i$ and $k$(see \fig{inclgen}-a).

Introducing the  following new function, enables us to write the 
 analytical expression for the double inclusive cross section: 
\bea \label{2SH1}
I^{(i,k)}\Lb y, b \Rb\,\,&=&\,\,\tilde a_{\pom \pom} \,\int d^2 b'
 \,\,N^{BK}\Bigg( g^{(i)}\,S\Lb m_i,b'\Rb
 \tilde{G}^{\mbox{\tiny dressed}}\Lb r_\bot = 1/m, \h Y + y
 \Rb\Bigg)\,\nn\\
 &\times& N^{BK}\Bigg( g^{(k)}\,S\Lb m_k,
 \vec{b} - \vec{b}^{\,'}\Rb \tilde{G}^{\mbox{\tiny dressed}}\Lb
 r_\bot=1/m,\h Y- y \Rb\Bigg)
\eea

 where $\tilde{a}_{\pom \pom} = a_{\pom \pom} \,\ln\Lb W/W_0\Rb$.

 Using \eq{2SH1} we can write the double inclusive cross section in
 two equivalent  forms
 \bea 
 &&\frac{d^2 \sigma^{\mbox{\tiny 2 parton showers}}}{ d
 y_1\,\,d y_2}\,\,=\,\,\int d^2p_{1T}\,d^2 p_{2T}\frac{d^2
 \sigma^{\mbox{\tiny 2 parton showers}}}{ d
 y_1\,\,d y_2\,d^2 p_{1T}\,d^2 p_{2T}}\,\,=\,\,\int d^2 b \,\,\Bigg\{ \alpha^4
 \,I^{(1,1)}\Lb y_1, b \Rb\,I^{(1,1)}\Lb y_2, b \Rb\nn\\
 && + \alpha^2\,\beta^2\,\Lb  I^{(1,2)}\Lb y_1, b \Rb
\,I^{(1,2)}\Lb y_2, b \Rb\,\,+\,\,I^{(2,1)}\Lb y_1,
 b \Rb\,I^{(2,1)}\Lb y_2, b \Rb \Rb\,\,+\,\,\beta^4\,I^{(2,2)}\Lb
 y_1, b \Rb\,I^{(2,2)}\Lb y_2, b \Rb \Bigg\}\label{2SH2} \\
&&=\frac{ \tilde{a}^2_{\pom \pom}}{p^2_{T1}\,p^2_{T2}}\,
\int d^2 Q_T \Bigg( \alpha^2  I^{\rm BK}_1\Lb \h Y +y_1;
 Q_T\Rb I^{\rm BK}_1\Lb\h Y +y_2; Q_T\Rb\,+\,\beta^2\, 
 I^{\rm BK}_2\Lb\h Y +y_1; Q_T\Rb I^{\rm BK}_2\Lb\h Y +y_2;
 Q_T\Rb\Bigg) \nn\\
 &&\times\, \Bigg( \alpha^2  I^{\rm BK}_1\Lb \h Y -y_1;
 Q_T\Rb I^{\rm BK}_1\Lb\h Y -y_2; Q_T\Rb\,+\, \beta^2\, I^{\rm BK}_2\Lb\h
 Y -y_1; Q_T\Rb I^{\rm BK}_2\Lb\h Y -y_2; Q_T\Rb\Bigg) \,\nn\\
 &&\equiv\, \,\,\frac{ \tilde{a}^2_{\pom \pom}}{p^2_{T1}\,p^2_{T2}}
\,\int d^2 Q_T \,F^{\rm BK}_{12}\Lb \h Y + y_1, \h Y + y_2; Q_T\Rb 
\,\,F^{\rm BK}_{12}\Lb \h Y - y_1, \h Y - y_2; Q_T\Rb  \label{2SH2Q}
\eea

where  $F^{\rm BK}_{12}$ is equal to 

\beq \label{FBL12}
F^{\rm BK}_{12}\Lb Y_1, Y_2; Q_T\Rb\,=\,\alpha^2  I^{\rm BK}_1\Lb  Y_1;
 Q_T\Rb I^{\rm BK}_1\Lb Y_2; Q_T\Rb\,+\, \beta^2\, I^{\rm BK}_2\Lb Y_1; Q_T\Rb I^{\rm BK}_2\Lb Y_2; Q_T\Rb
 \eeq

Recall that all rapidities are in the c.m. frame.

{\boldmath{  \subsubsection{$v_n$ for proton-proton collisions}}}
 Using \eq{2SH2Q}, \eq{I2} can be re-written in the following form
 
 \bea \label{PP1}
&&   \frac{d^2 \sigma}{d y_1 \,d y_2  d^2 p_{T1} d^2 p_{T2}}\,\,=\,\, 
\frac{ \tilde{a}^2_{\pom \pom}}{p^2_{T1}\,p^2_{T2}}\,\int d^2 k_T \,
F^{\rm BK}_{12}\Lb \h Y + y_1, \h Y + y_2; k_T\Rb \,\,F^{\rm BK}_{12}\Lb
 \h Y - y_1, \h Y - y_2; k_T\Rb \\
   &&+\,\,\frac{ \tilde{a}^2_{\pom \pom}}{N^2_c - 1}\,\frac{1}{4}\Lb
 \frac{1}{p^2_{T1}}\,+\,\frac{1}{p^2_{T2}}\Rb^2
   \int d^2 k_T \,F^{\rm BK}_{12}\Lb \h Y + y_1, \h Y + y_2; \vec{k}_T 
\Rb \,\,F^{\rm BK}_{12}\Lb \h Y - y_1, \h Y - y_2;\vec{k}_T + \vec{p}_{T,12}
 \Rb\nn
   \eea
   
   In \eq{PP1} we neglected  the contribution $\propto p^2_{T,12}$ in the
  vertex of gluon emission in \fig{bfkl} (see 
Appendix A1), as well as 
the dependence of the BFKL Pomeron on the momentum transfer.  
The small size of both quantities 
   stem from the fact that in our model,  $k_T$ dependence
 in \eq{PP1} is determined by the proton structure and the typical 
$k_T \sim m_1 {\rm}$ or $m_2$ (see Table 1), while typical transverse
 momentum in the BFKL Pomeron is about $Q_s$ or $m$, and it is much
 larger than $m_1$ or $m_2$.

   ~


\section{Hadron-nucleus interaction in the model}
 
In the case of the  hadron-nucleus interaction the general formula of 
\eq{2SH2Q}
 can be re-written in the form\cite{GLMPA}
  \bea 
 &&\frac{d^2 \sigma^{\mbox{\tiny 2 parton showers}}}{ d
 y_1\,\,d y_2}\,\,=\,\,\int d^2p_{1T}\,d^2 p_{2T}\frac{d^2
 \sigma^{\mbox{\tiny 2 parton showers}}}{ d
 y_1\,\,d y_2\,d^2 p_{1T}\,d^2 p_{2T}}\,\,= \\
 &&=\frac{ \tilde{a}^2_{\pom \pom}}{p^2_{T1}\,p^2_{T2}}\,\int
 d^2 Q_T \Bigg( \alpha^2  I^{\rm BK}_1\Lb \h Y +y_1; Q_T\Rb
 I^{\rm BK}_1\Lb\h Y +y_2; Q_T\Rb\,+\,\beta^2\, 
 I^{\rm BK}_2\Lb\h Y +y_1; Q_T\Rb I^{\rm BK}_2\Lb\h Y +y_2; Q_T\Rb\Bigg) \nn\\
 &&\times\, I^{\rm BK}_A\Lb\h Y -y_1; Q_T\Rb\,I^{\rm BK}_A\Lb\h Y -y_2;
 Q_T\Rb\ \,\nn\\
 &&\equiv\, \,\,\frac{ \tilde{a}^2_{\pom \pom}}{p^2_{T1}\,p^2_{T2}}\,
\int d^2 Q_T \,F^{\rm BK}_{12}\Lb \h Y + y_1, \h Y + y_2; Q_T\Rb \,
\,F^{\rm BK}_{A}\Lb \h Y - y_1, \h Y - y_2; Q_T\Rb  \label{2SH2AQ}
\eea 
 where 
 \beq \label{IA}
  I^{\rm BK}_A\Lb y,Q_T\Rb\,\,=\,\,\int d^2 b e^{i \vec{b} \cdot \vec{Q}_T}
 \,N^{\rm BK}\Lb \Lb \alpha^2 In^{(1)}\Lb y\Rb \,+\,\beta^2\,In^{(2)}\Lb
 y\Rb\Rb
  S_A\Lb b \Rb\Rb
  \eeq
  where $In^{(i)}$ are defined in \eq{IN} and $S_A\Lb b \Rb$ is the
 nucleus Wood-Saxon distribution\cite{WS} given by 

 \beq \label{WS}
S_A\Lb b \Rb\,\,=\,\ \int^\infty_{- \infty} d z \,\frac{\rho_0}{1 \,+\,
 \exp\Big(  \frac{\sqrt{z^2 + b^2} - R_A}{h}\Big)}
~~~~~~~~\mbox{where}~~~~~~ \int d^2 b \,S_A\Lb b \Rb\,=\,A
\eeq 
For  gold we have $R_A = 6.38 \,fm$ and $h = 0.535 \,fm$, while for
lead we have $R_A=6.68 \,fm$ and $h=0.546\,fm$ \cite{WS}.
  In \eq{2SH2AQ} we have taken into account that the typical
 impact parameters in hadron-hadron interaction are much smaller
 than the radius of nucleus ($R_A$). Indeed,  the typical  $b$
 in hadron-hadron collisions  are  $\alpha'_{eff} Y$ or less,
 where $\alpha'_{eff}$ is the effective slope of the BFKL
 Pomeron trajectory, which occurs in our model as a result
 of  shadowing corrections.

Using \eq{2SH2AQ} we can re-write \eq{PP1} for proton-proton in 
the following form for proton-nucleus scattering

\bea \label{PA1}
&&   \frac{d^2 \sigma}{d y_1 \,d y_2  d^2 p_{T1} d^2 p_{T2}}\,\,=\,\,
 \frac{ \tilde{a}^2_{\pom \pom}}{p^2_{T1}\,p^2_{T2}}\,\int d^2 k_T 
\,F^{\rm BK}_{12}\Lb \h Y + y_1, \h Y + y_2; k_T\Rb \,\,F^{\rm BK}_{A}\Lb
 \h Y - y_1, \h Y - y_2; k_T\Rb \\
   &&+\,\,\frac{ \tilde{a}^2_{\pom \pom}}{N^2_c - 1}\,\frac{1}{4}\Lb 
\frac{1}{p^2_{T1}}\,+\,\frac{1}{p^2_{T2}}\Rb^2
   \int d^2 k_T \,F^{\rm BK}_{12}\Lb \h Y + y_1, \h Y + y_2; \vec{k}_T
 + \vec{p}_{T,12}\Rb \,\,F^{\rm BK}_{A}\Lb \h Y - y_1, \h Y - y_2;\vec{k}_T
  \Rb\nn
   \eea
   

   and for nucleus-nuclues scattering we have
   \bea \label{AA1}
&&   \frac{d^2 \sigma}{d y_1 \,d y_2  d^2 p_{T1} d^2 p_{T2}}\,\,=\,\,
 \frac{ \tilde{a}^2_{\pom \pom}}{p^2_{T1}\,p^2_{T2}}\,\int d^2 k_T 
\,F^{\rm BK}_{12}\Lb \h Y + y_1, \h Y + y_2; k_T\Rb \,\,F^{\rm BK}_{A}
\Lb \h Y - y_1, \h Y - y_2; k_T\Rb \\
   &&+\,\,\frac{ \tilde{a}^2_{\pom \pom}}{N^2_c - 1}\,\frac{1}{4}\Lb
 \frac{1}{p^2_{T1}}\,+\,\frac{1}{p^2_{T2}}\Rb^2
   \int d^2 k_T \,F^{\rm BK}_{A}\Lb \h Y + y_1, \h Y + y_2; \vec{k}_T
 + \vec{p}_{T,12}\Rb \,\,F^{\rm BK}_{A}\Lb \h Y - y_1, \h Y - y_2;\vec{k}_T 
 \Rb\nn
   \eea

\end{document}